\documentclass[aps,prx,twocolumn,reprint,amsmath,amssymb,superscriptaddress,floatfix,footinbib,longbibliography, tikz, dvipsnames]{revtex4-1}
\usepackage{natbib}

\usepackage{amsmath}
\usepackage{amssymb}
\usepackage{amsthm}
\usepackage{physics}

\usepackage{dsfont}
\DeclareMathAlphabet{\mbb}{U}{BOONDOX-ds}{m}{n} 

\usepackage{array}
\usepackage{multirow}
\usepackage{enumerate}
\usepackage{bm}
\usepackage{cancel}
\usepackage{nicefrac}

\usepackage{graphicx}
\usepackage{empheq}
\usepackage{float}
\usepackage{tabularx}

\newcolumntype{C}{>{\centering}X}

\usepackage{standalone}
\usepackage{tikz}
\usepackage{pgfplots}
\pgfplotsset{compat=1.5}
\usetikzlibrary{arrows}
\usetikzlibrary{decorations.markings}   
\usetikzlibrary{patterns}
\usetikzlibrary{decorations.pathmorphing}

\usepackage{xcolor}
\usepackage{url}
\usepackage[colorlinks]{hyperref}
\definecolor{python_green}{RGB}{43,160,43}
\definecolor{python_blue}{RGB}{32,119,180}
\definecolor{python_orange}{RGB}{255,127,15}
\hypersetup{%
        plainpages=true,
        breaklinks=true,
        hypertexnames=false,
        pageanchor=true,
        colorlinks=true,
        linkcolor={python_blue},
        citecolor={python_orange},
        urlcolor={python_blue},
        anchorcolor={black}
      }
\usepackage[all]{hypcap} 

\usepackage{mleftright} 

\newcommand{\figref}[1]{\mbox{Fig.~\ref{#1}}}

\newcommand{\secref}[1]{\mbox{Sec.~\ref{#1}}}

\newcommand{\appref}[1]{\mbox{Appendix~\ref{#1}}} 
\renewcommand{\eqref}[1]{\mbox{Eq.~(\ref{#1})}}
\newcommand{\eqsref}[2]{\mbox{Eqs.~(\ref{#1})--(\ref{#2})}}

\newcommand{\figpanel}[2]{Fig.~\hyperref[#1]{\ref*{#1}(#2)}}
\newcommand{\figpanels}[3]{Fig.~\hyperref[#1]{\ref*{#1}(#2)--(#3)}}
\newcommand{\figpanelNoPrefix}[2]{\hyperref[#1]{\ref*{#1}(#2)}}

\newcommand{\Hc}{\text{H.c.}}
\newcommand{\R}{\mathbb{R}}

\newcommand{\N}{\mathbb{N}}
\newcommand{\e}{\mathrm{e}}

\newcommand{\Hint}{H_\text{int}}
\newcommand{\bathfreq}{\omega_B}

\newcommand{\detuning}{\Delta}

\newcommand{\ncpts}{P}
\newcommand{\overlap}{\cap}

\newcommand{\al}{\alpha}
\newcommand{\be}{\beta}

\newcommand{\de}{\delta}

\newcommand{\ka}{\kappa}
\newcommand{\lm}{\lambda}
\newcommand{\om}{\omega}

\newcommand{\pvec}[1]{\vec{#1}\mkern2mu\vphantom{#1}}

\newcommand{\Ga}{\Gamma}
\newcommand{\GT}{\Gamma^T}

\newcommand{\De}{\Delta}
\newcommand{\Lm}{\Lambda}

\AtBeginDocument{%
    \newwrite\bibnotes
    \def\bibnotesext{Notes.bib}
    \immediate\openout\bibnotes=\jobname\bibnotesext
    \immediate\write\bibnotes{@CONTROL{REVTEX41Control}}
    \immediate\write\bibnotes{@CONTROL{%
    apsrev41Control,author="08",editor="1",pages="0",title="0",year="1"}}
     \if@filesw
     \immediate\write\@auxout{\string\citation{apsrev41Control}}%
    \fi
}%

\begin{document}

\title{Avoiding decoherence with giant atoms in a two-dimensional structured environment}
\date{\today}

\author{Emil Raaholt Ingelsten}
\affiliation{Department of Microtechnology and Nanoscience, Chalmers University of Technology, 412 96 Gothenburg, Sweden}

\author{Anton Frisk Kockum}
\affiliation{Department of Microtechnology and Nanoscience, Chalmers University of Technology, 412 96 Gothenburg, Sweden}

\author{Ariadna Soro}
\email{soro@chalmers.se}
\affiliation{Department of Microtechnology and Nanoscience, Chalmers University of Technology, 412 96 Gothenburg, Sweden}


\begin{abstract}

Giant atoms are quantum emitters that can couple to light at multiple discrete points. Such atoms have been shown to interact without decohering via a one-dimensional waveguide. 
Here, we study how giant atoms behave when coupled to a two-dimensional square lattice of coupled cavities, an environment characterized by a finite energy band and band gaps. 
In particular, we describe the role that bound states in the continuum (BICs) play in how giant atoms avoid decoherence. By developing numerical methods, we are able to investigate the dynamics of the system and show the appearance of interfering BICs within a single giant atom, as well as oscillating BICs between many giant atoms. In this way, we find the geometric arrangements of atomic coupling points that yield protection from decoherence in the two-dimensional lattice.
These results on engineering the interaction between light and matter may find applications in quantum simulation and quantum information processing.

\end{abstract}

\maketitle


\section{Introduction}

In the past decade, a new paradigm of quantum emitters has been increasingly attracting interest: so-called giant atoms (GAs)~\cite{FriskKockum2021}.
These atoms, which may be artificial, earn their name by breaking the dipole approximation: the assumption that atoms are \emph{small} compared to the wavelength of the field they interact with.
\emph{Giant} atoms instead couple to light (or other bosonic fields) at several discrete points, which can be spaced wavelengths apart.
The interference between emission and absorption through these coupling points then leads to a plethora of remarkable features, such as frequency-dependent decay rates and Lamb shifts~\cite{Kockum2014, Vadiraj2021}, waveguide-mediated decoherence-free interaction~\cite{FriskKockum2018a, Kannan2020, Carollo2020, Soro2022, Soro2023, Du2023b}, and oscillating bound states~\cite{Guo2020, Guo2020a, Terradas-Brianso2022, Noachtar2022, Lim2023}.

Since 2014, several experimental demonstrations of GAs have been achieved, both with superconducting qubits coupled to surface acoustic waves~\cite{Gustafsson14, Aref2016, Manenti2017, Noguchi2017, Satzinger2018, Moores2018, Bolgar2018, Sletten2019, Bienfait2019, Andersson2019, Bienfait2020, Andersson2020} and to microwave waveguides~\cite{Kannan2020, Vadiraj2021, Joshi2023}, and several other implementations have been proposed~\cite{Gonzalez-Tudela2019, Du2022}.
Recently, giant-atom physics has also been explored beyond the atomic paradigm in giant molecules~\cite{Guimond2020, Gheeraert2020, Zhang2021, Yin2022a, Kannan2023} or giant spin ensembles~\cite{Wang2022}.
However, most studies to date (both theoretical and experimental) have focused on GAs coupled to one-dimensional reservoirs: most commonly to continuous waveguides~\cite{Kockum2014, Guo2017, FriskKockum2018a, Karg2019, Kannan2020, Guo2020, Guo2020a, Ask2020, Vadiraj2021, Du2021, Feng2021, Cai2021, Soro2022, Yin2022, Noachtar2022, Chen2022, Du2022a, Joshi2023, Du2023a, Santos2023, Wang2024, Du2023b, Zhou2023, Gu2023, Xu2024}, but recently also to structured ones~\cite{Longhi2020, Zhao2020, Wang21, Longhi2021, Yu2021a, Vega2021, Wang2021, Xiao2022, Du2022, Cheng2022, Zhang2022, Soro2023, Lim2023, Du2023, Du2023c, Bag2023, Jia2024, Gao2023}.

\begin{figure}[t]
    \centering
    \includegraphics[width=\columnwidth]{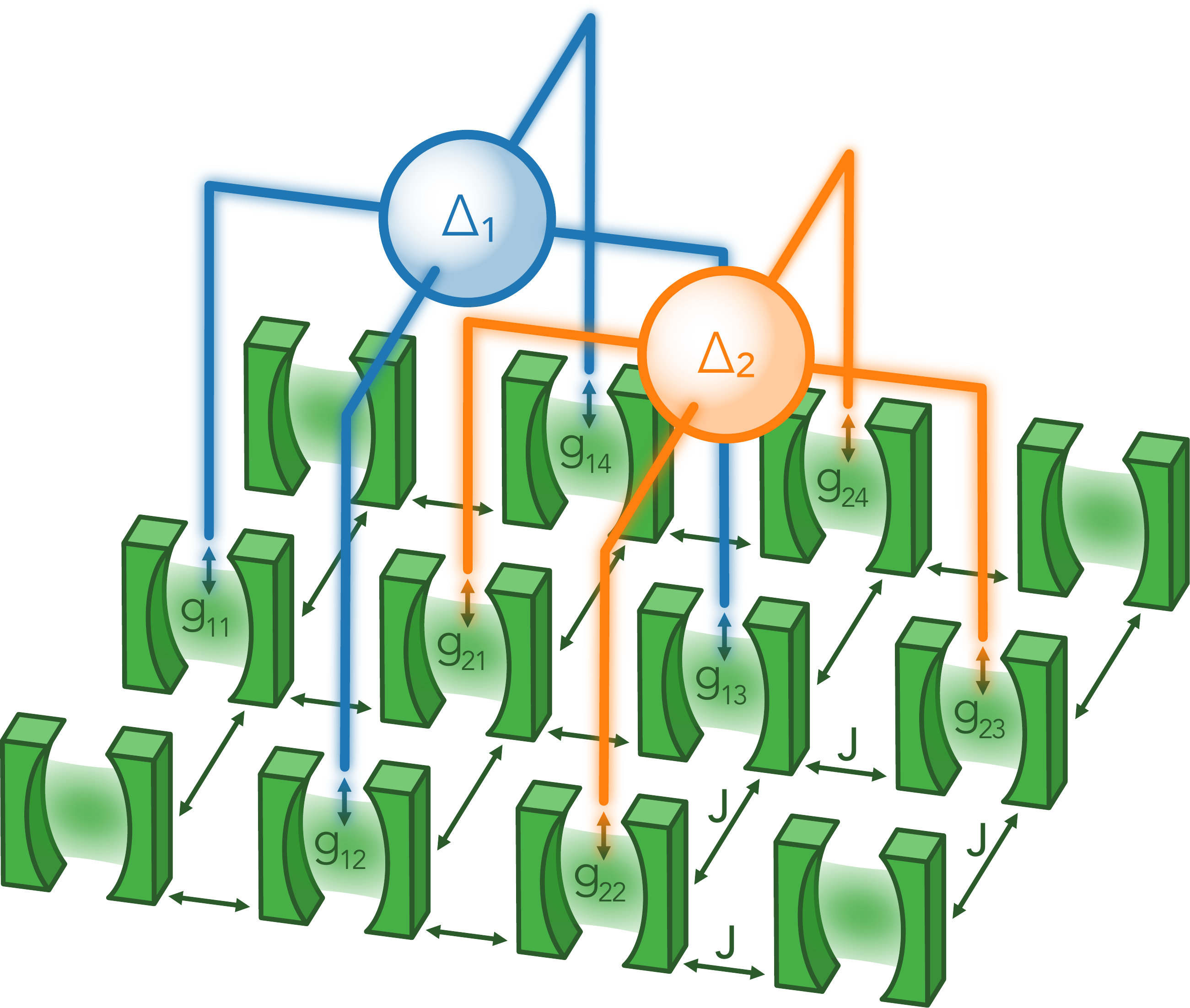}
    \caption{Two giant atoms in a braided configuration coupled to a 2D structured bath.
    The bath is modeled as a square lattice of $N\times N$ cavities with resonance frequency $\bathfreq$ and nearest-neighbor coupling strength $J$.
    The atoms are two-level systems, with transition frequencies $\omega_i$ detuned from the middle of the band by $\detuning_i=\omega_i-\bathfreq$ ($i=1,2$). They are coupled to the cavities with coupling strength $g_{ip}$ at each coupling point, where $i$ refers to the atom and $p$ to the connection point. 
    }
    \label{fig:setup}
\end{figure}

Here, we instead study GAs coupled to a two-dimensional (2D) structured environment, modeled as a square lattice of coupled cavities with nearest-neighbor interaction (see \figref{fig:setup}).
This environment has been studied in depth in relation to small atoms~\cite{Gonzalez-Tudela2017, Gonzalez-Tudela2017a, Galve2018, Yu2019, Feiguin2020, Ruks2022, Vega2023, Windt2024}, and there are additional works on small atoms coupled to other 2D~\cite{Gonzalez-Tudela2018a, Bienias2022, Ruks2022, Vega2023, DeBernardis2023, Tecer2023} or higher-dimensional~\cite{Gonzalez-Tudela2018, Feiguin2020, Garcia-Elcano2020, Garcia-Elcano2023} structured environments.
However, GAs in a 2D square lattice have only received limited attention so far: in Ref.~\cite{Gonzalez-Tudela2019}, the focus was on a particular proposal for an experimental implementation and on engineering unconventional emission patterns from a single GA; in Ref.~\cite{Vega2023}, the study revolved around harnessing topologically protected propagating modes.

To describe the dynamics of the GAs when they are tuned to the band of the 2D bath, we develop numerical methods (split-operator approach) and use complex-analysis techniques (resolvent formalism).
We focus on ways the GAs can avoid decoherence.
In doing so, we find that bound states in the continuum~\cite{vonNeumann1929, Hsu2016, Kang2023, Xu2023} (BICs) arise in certain geometries and make it possible for GAs to exhibit both subradiance and decoherence-free interaction. 
We also show different interference patterns of the BICs enclosed by eight coupling points of a single GA.
Moreover, by studying the bound-state dynamics, we identify decoherence-free interaction~\cite{FriskKockum2018a, Kannan2020} as a many-atom analogue of previously reported oscillating BICs~\cite{Guo2020, Guo2020a, Terradas-Brianso2022, Noachtar2022, Lim2023}.
Finally, we explore many-GA configurations that exploit decoherence-free interactions for potential applications in quantum simulation and computing, such as chains with pairwise interactions, triads with all-to-all interaction, and grids with effective long-range interactions.

This article is structured as follows. In \secref{sec:theory}, we present and discuss a Hamiltonian model for GAs coupled to a 2D square lattice of coupled cavities.
In \secref{sec:numerics}, we develop an efficient numerical method to simulate the dynamics of this system.
The results from the simulations are shown in \secref{sec:results}, where we focus on ways in which GAs become protected from decoherence and delve into the characterization of BICs.
In particular, we investigate the subradiance of a single GA in \secref{sec:1GA}, the decoherence-free interaction between two GAs  in \secref{sec:2GAs}, and other configurations involving many GAs in \secref{sec:manyGAs}.
We conclude in \secref{sec:conclusion} with a summary and an outlook.
Moreover, we include \appref{app:resolvent}, where we show the resolvent-formalism techniques that support the results of this article, and \appref{app:numerics}, where we provide additional details for the numerical method described in \secref{sec:numerics}.


\section{Theoretical framework}
\label{sec:theory}

We start by deriving a Hamiltonian model for the setup shown in \figref{fig:setup}. Both the setup and the Hamiltonian are inspired by those presented in Refs.~\cite{Gonzalez-Tudela2017, Gonzalez-Tudela2019, Soro2023}.

The structured 2D reservoir we consider can be described as a square lattice of $N\times N$ cavities with resonance frequencies $\bathfreq$ and nearest-neighbor couplings $J$.
Taking the separation of adjacent cavities as the unit of length, we characterize the position of each cavity with a coordinate vector $\vec{n}=(n_x, n_y)$, where $n_x, n_y \in[0,N-1]$, and label each corresponding cavity annihilation operator as $a_{\vec{n}}$. The resulting bath Hamiltonian in real space, rotating at frequency $\bathfreq$, reads ($\hbar = 1$ throughout this article)
\begin{equation}
    H_B = -J \sum_{\expval{\vec{n},\vec{m}}} \mleft( a_{\vec{n}}^\dagger a_{\vec{m}} + \Hc \mright),
    \label{eq:H_B_real}
\end{equation}
where $\expval{\vec{n},\vec{m}}$ denotes summation over all pairs of neighboring cavities at $\vec{n}$ and $\vec{m}$, and H.c.~denotes the Hermitian conjugate.

The Hamiltonian in \eqref{eq:H_B_real} can be diagonalized by introducing periodic boundary conditions and the operators in momentum space
\begin{equation}
a_{\vec{n}}=\frac{1}{N}\sum_{\vec{k}} a_{\vec{k}}\; e^{-i\vec{k}\cdot\vec{n}},
\end{equation}
where $\vec{k}=(k_x, k_y)$ is the wave vector, with $k_x,k_y \in \{-\pi, \dots, \pi-\frac{2\pi}{N}\}$.
In that basis,
\begin{equation}
    H_B = \sum_{\vec{k}} \omega(\vec{k})a_{\vec{k}}^\dagger a_{\vec{k}},
    \label{eq:H_B_momentum}
\end{equation}
with
\begin{equation}
\omega(\vec{k}) = -2J(\cos k_x + \cos k_y).
\label{eq:dispersion}
\end{equation}

The dispersion relation in \eqref{eq:dispersion}, although describing a fairly simple 2D structured bath, gives rise to some very interesting properties.
First and foremost, it results in an energy band in the range $\omega(\vec{k}) \in [-4J, 4J]$. 
Within this band, the energy dispersion varies widely: it is isotropic close to the band edges, but becomes highly anisotropic at the band center [i.e., at $\omega(\vec{k})=0$]~\cite{Gonzalez-Tudela2017}.
This is easy to see from the definition of group velocity: at $\omega(\vec{k})=0$,
\begin{equation}
\vec{v}_g = \eval{\vec{\nabla}\omega(\vec{k})}_{\omega(\vec{k})=0} = 2J \sin k\mqty[1\\ \pm 1]
\label{eq:group-velocity}
\end{equation}
for any $k\in \{-\pi, \dots, \pi-\frac{2\pi}{N}\}$.
At the band center, excitations can thus only propagate along two orthogonal diagonals, which we hereafter refer to as the $\smqty[\;\;\,1\\ \pm1]$ diagonals.
Furthermore, note that $\vec{v}_g$ vanishes for $k = \{0,\pm \pi\}$, which yields a divergent density of states, i.e., a type of Van Hove singularity present in many 2D structured baths~\cite{Gonzalez-Tudela2017, Gonzalez-Tudela2017a}.

We now consider $M$ two-level GAs, with transition frequencies $\omega_i$, detuned from the middle of the band by $\detuning_i=\omega_i-\bathfreq$. The bare Hamiltonian of these GAs (rotating at $\bathfreq$) is
\begin{equation}
    H_A = \sum_{i=1}^M \detuning_i \sigma^+_i \sigma^-_i,
    \label{eq:H_A}
\end{equation}
where $\sigma^\pm_i$ denote the atomic ladder operators.
If the $i$th GA couples to the bath at $P_i$ points, the interaction of the atoms with the bath under the rotating-wave approximation (RWA) can be described by
\begin{equation}
\begin{aligned}
    \Hint &= \sum_{i=1}^M \sum_{p=1}^{P_i} g_{ip} \mleft(a_{\vec{n}_{ip}} \sigma^+_i + \Hc \mright) \\
    &=\sum_{i=1}^M \sum_{p=1}^{P_i} \frac{g_{ip}}{N} \sum_{\vec{k}} \mleft(e^{-i\vec{k}\cdot\vec{n}_{ip}} a_{\vec{k}}\sigma^+_i + \Hc \mright),
    \label{eq:H_int}
\end{aligned}
\end{equation}
where $\vec{n}_{ip}$ denotes the position of the cavity which interacts with the $p$th coupling point of the $i$th atom.
The coupling strength between the atom and the bath at this point is $g_{ip}$, which we assume to be real for convenience.
Although we leave complex coupling strengths out of the scope of this manuscript, we note that they are experimentally achievable~\cite{Roushan2017, Gonzalez-Tudela2019, Zhang2021, Joshi2023} and can be used to engineer chiral emissions and interactions~\cite{Lodahl2017, Roushan2017, Gonzalez-Tudela2019, Zhang2021, Chen2022, Wang2022, Wang2024, Joshi2023}.

With the definitions above, the total Hamiltonian of the system is
\begin{equation}
H=H_B + H_A + \Hint .
\label{eq:H}
\end{equation}
We note that this model is valid in the single-excitation subspace, as well as under the assumption of coupling to a single polarization of light and a single bosonic band.
We also work in the continuum limit (i.e., $N\to\infty$ in the analytics, and $tJ \ll N/2$ in the simulations), which allows us to disregard effects arising from the finite size of the bath. 
Additionally, we neglect couplings to other reservoirs by assuming that the losses induced by such couplings occur at a significantly lower rate than the relevant dynamics we study.
Lastly, we assume weak coupling strengths  $g_{ip} < J$ and $g_{ip} \ll \omega_i, \bathfreq\; \forall i, p$, in order to comply with the RWA we applied in \eqref{eq:H_int}.

This theoretical model may be used to describe cold atoms coupled to photonic crystals~\cite{Hood2016, Yu2019} or optical lattices~\cite{Krinner2018, Stewart2020, Gonzalez-Tudela2019}, as well as superconducting qubits coupled to microwave photonic crystals~\cite{Liu2017, Sundaresan2019, Harrington2019, Kollar2019, Carusotto2020} or superconducting metamaterials~\cite{Mirhosseini2018, Indrajeet2020, Kim2021, Ferreira2021, Scigliuzzo2022, Zhang2023, Jouanny2024}.
Due to the intricate nature of the setup (2D resonator lattice \emph{and} multiplicity of atoms \emph{and} multiplicity of coupling points per atom arranged in a nontrivial manner), the experimental realization of cold atoms in the GA regime requires excellent control of a dynamical state-dependent optical lattice~\cite{Gonzalez-Tudela2019}.
Such implementation remains elusive to date, and we thus consider our setup to be most readily implementable with superconducting qubits.
We note that, in such a case, arranging for the multiple coupling points may be aided by flip-chip technology~\cite{Rosenberg2017, Rahamim2017, Kosen2022}.


\section{Numerical methods}
\label{sec:numerics}

The aforementioned Van Hove singularities [see \secref{sec:theory}, after \eqref{eq:group-velocity}] in the middle of the band and at the band edges introduce branch cuts that make it hard to analytically calculate how the exact atomic and bath populations evolve over time.
While we provide derivations for those quantities in the \appref{app:resolvent}, we rely heavily on numerical simulations in order to study the dynamics of the system.
In this section, we present the numerical methods we develop and use in this work.

We base our numerical method on a so-called \textit{split-operator} approach~\cite{Press2007}, which has been used in previous studies of small and giant atoms coupled to structured environments~\cite{Gonzalez-Tudela2017, Soro2023}.
Essentially, this method is based on splitting the full Hamiltonian in \eqref{eq:H} into the bath part $H_B$ [diagonal in Fourier space, \eqref{eq:H_B_momentum}], the atomic part $H_A$ [diagonal in real space, \eqref{eq:H_A}], and the interaction part $H_\text{int}$ [\eqref{eq:H_int}], and evolving the system by repeatedly applying the approximate time-evolution operator
\begin{equation}
    \tilde{U}(\Delta t) = \mathcal{F}^{-1} U_B(\Delta t) \mathcal{F} U_A(\Delta t),
\end{equation}
where
\begin{equation}
    U_A(\Delta t) = e^{ -i \qty(H_A + H_\text{int}) \Delta t}, \quad U_B(\Delta t) = e^{ -i H_B \Delta t },
\end{equation}
and $\mathcal{F}$ denotes a Fourier transform, implemented in practice as a fast Fourier transform (FFT).
While this is not equivalent to applying the exact time-evolution operator $U(t) = \exp[ -i \qty(H_B + H_A + H_\text{int} ) t ]$, it is accurate to $\mathcal{O}(\Delta t^3)$, provided that the copies of $\tilde{U}(\Delta t)$ are sandwiched between an initial $U_A\qty(-\Delta t / 2)$ and a final $U_A\qty(\Delta t / 2)$~\cite{Hatano2005}.

Since $U_B$ is diagonal in Fourier space, it is trivial to compute.
The same does not hold for $U_A$ in real space due to the presence of $H_\text{int}$.
However, for small atoms, it has been previously used~\cite{Gonzalez-Tudela2017} that, as long as no cavity couples to multiple atoms, $U_A$ can be calculated with a computational complexity linear in $M$ and $N^d$, where $d$ is the dimensionality of the bath (in our case: $d=2$).
The reason for this complexity can most easily be seen by examining the structure of $H_A + \Hint$.

Defining the state vector for our system such that the first $M$ elements correspond to the (bare) atomic excited states and the following $N^2$ elements correspond to the states where one of the cavities is excited (and the atoms are in their ground states), we can write the combined atomic and interaction Hamiltonian as a block matrix:
\begin{equation}
    H_A + \Hint = \begin{bmatrix}
        D & \Ga \\
        \GT & \mbb{0}
    \end{bmatrix}.
\label{eq:H_A+H_int}
\end{equation}
Here, $D$ is a diagonal $M \times M$ matrix containing the detunings $\De_i$ for the different atoms, $\Ga$ is an $M \times N^2$ matrix containing all terms related to atom-bath interaction, and $\mbb{0}$ is the $N^2 \times N^2$ zero matrix.
For the case of small atoms, where atom $i$ couples to a single cavity with coupling strength $g_i$,
\begin{equation}
    \qty[\Ga]_{in} = \begin{cases}
        g_i & \text{if atom $i$ couples to cavity $n$} \\
        0 & \text{otherwise}.
    \end{cases}
\end{equation}
Assuming each cavity only couples to a single atom, no interaction occurs between atoms without taking $H_B$ into account.
Thus, each atom $i$ is isolated and can be modeled using an effective Hamiltonian
\begin{equation}
    H_i = \begin{bmatrix}
        \Delta_i & g_i \\
        g_i & 0
    \end{bmatrix}.
\end{equation}
In other words, to apply $U_A$ to a state, one only needs to evaluate $M$ different $2 \times 2$ matrix exponentials:
\begin{equation}
    U_i(\De t) = \exp( -i H_i \De t ).
\end{equation}
The matrix elements of $U_i$ can then be used to apply the same time evolution as the one caused by $U_A$ to atom $i$ and its coupled cavity. 
This is significantly cheaper than computing $U_A$ directly, which would entail computing an $(M + N^2) \times (M + N^2)$ matrix exponential.
In turn, this would have a computational cost scaling like $(M+N^2)^3$, since the cost of exponentiating an $n \times n$ matrix in general scales like $n^3$ using state-of-the-art algorithms \cite{AlMohy2010, Moler2003}.

As described in \secref{sec:theory}, the most general GA case involves each atom $i$ coupling to $P_i$ resonators with coupling strengths $\qty{g_{ip}}_{p=1}^{P_i}$.
As long as there are still no cavities coupling to multiple atoms, the small-atom method described above can be generalized to handle GAs by simply extending the size of the effective Hamiltonian with an additional row and column for each added coupling point, such that
\begin{equation}
    H_i = \begin{bmatrix}
        \Delta_i & \pvec{g}_{\!i} \\
        \pvec{g}_{\!i}^T & \mbb{0}
    \end{bmatrix},
\end{equation}
where $\pvec{g}_{\!i}$ is a $1 \times P_i$ row matrix with elements $\qty[ \pvec{g}_{\!i} ]_p = g_{ip}$.
This Hamiltonian can then be used completely analogously to how $H_i$ is used in the small-atom case.

However, as we show in \appref{app:numerics}, we can generalize the small-atom method to handle GAs without increasing the size of the effective Hamiltonian. We can thus not only apply $U_A$ to a state at a cost linear in $M$ and $N^2$ for GAs, just as has previously been possible for small atoms, but the cost can be made linear in $P_i$.
In contrast, the cost of exponentiating a $(1 + P_i) \times (1 + P_i)$ matrix directly would scale as $(1+P_i)^3$. While this difference is not very significant for most of the configurations examined in this study, the $P_i$-linear version of the method would be significantly faster at modeling more complicated setups, e.g., ones based on reverse design like those described in Ref.~\cite{Gonzalez-Tudela2019}.
Note that our generalized method works for baths of arbitrary dimensionality, but we focus mainly on the 2D case in this paper.

The $P_i$-linear method for computing the elements of $U_A$ is based on using an effective Hamiltonian
\begin{equation}
    H_i = \begin{bmatrix}
        \Delta_i & G_i \\
        G_i & 0
    \end{bmatrix},
\label{eq:H_i-GA}
\end{equation}
where
\begin{equation}
    G_i = \sqrt{\sum_{p=1}^{P_i} g_{ip}^2}
    \label{eq:G_i}
\end{equation}
is the effective coupling strength of atom $i$. If the atom couples equally strongly to each coupling point, i.e., $g_{ip} = g_i$ for every $p$, \eqref{eq:G_i} reduces to
\begin{equation}
    G_i = \sqrt{P_i} g_i.
\label{eq:G_i_equal}
\end{equation}

Since the $\Ga$ block in \eqref{eq:H_A+H_int} is now more complicated than in the small-atom case, the elements of $U_A$ are more complicated than simply being copies of the elements in $U_i$ when using our $2\times2$ effective Hamiltonian [\eqref{eq:H_i-GA}].
In fact, as shown in \appref{app:numerics}, again assuming that no cavities couple to multiple atoms, $U_A$ can be written as a block matrix
\begin{equation}
    U_A = \begin{bmatrix}
        \mathds{1}_M + D_0 & D_1 \Ga \\
        \GT D_1 & \mathds{1}_{N^2} + \GT D_2 \Ga
    \end{bmatrix},
    \label{eq:UA_and_Dl}
\end{equation}
where $D_{0,1,2}$ are diagonal matrices that can be computed from the elements of $U_i$. Specifically, the matrix elements of $U_A(\Delta t)$ can be expressed as follows:
\begin{widetext}
\begin{equation}
\begin{dcases}
    \qty[\mathds{1}_M + D_0]_{ij} = \de_{ij} \qty[{U}_i]_{11}, \\
    \qty[D_1 \Ga]_{in} = \qty[\GT D_1]_{ni} = \begin{dcases}
        \frac{g_{i p_n}}{G_i} \qty[{U}_i]_{12} & \text{if cavity $n$ couples to atom $i$ at point $p_n$} \\
        0 & \text{otherwise},
    \end{dcases} \\
    \qty[\mathds{1}_{N^d} + \GT D_2 \Ga]_{mn} = \begin{dcases}
        \de_{mn} + \frac{g_{i p_m} g_{i p_n}}{G_i^2} \qty( \qty[{U}_i]_{22} - 1 ) & \text{if cavities $m,n$ couple to atom $i$ at $p_m, p_n$} \\
        \de_{mn} & \text{otherwise}.
    \end{dcases} 
\end{dcases}
\label{eq:UA_els}
\end{equation}
\end{widetext}
Applying the map encoded by $U_A$ can thus be done, to accuracy $\mathcal{O}(\Delta t^3)$, with computational cost linear in $M$, $N^2$, and $P_i$ for GAs.
Doing so moves the bottleneck of the split-operator algorithm to the computation of the time evolution associated with $H_B$.
Specifically, the bottleneck becomes the two FFTs performed in each time step, which have complexity $\mathcal{O}\qty(N^2 \log N)$~\cite{Cooley1965}, since the computation and application of $U_B$ in the bath eigenbasis only has complexity $\mathcal{O}(N^2)$.

As discussed in Ref.~\cite{Gonzalez-Tudela2017}, the computational cost of the time-evolution algorithm can be reduced further by introducing an additional approximation based on going to the continuum limit, discretizing frequency space and exploiting the periodicity of $\om(\pvec{k})$.


\section{Avoiding decoherence}
\label{sec:results}

In this Section, we explore ways in which GAs can avoid relaxing into the bath.
Therefore, we focus on the case $\detuning / J  = 0$, where the atoms are tuned to the middle of the band and thus can only emit along two orthogonal diagonals, as shown in \secref{sec:theory} [see the discussion around \eqref{eq:group-velocity}].
This restriction in emission directions makes it relatively easy to engineer interference such that emission to the bath from different atoms, or from different coupling points belonging to the same atom, cancels completely~\cite{Gonzalez-Tudela2017, Gonzalez-Tudela2017a, Galve2018, Gonzalez-Tudela2019, Windt2024}. 

\begin{figure}[t]
    \centering
    \includegraphics[width=0.45\columnwidth]{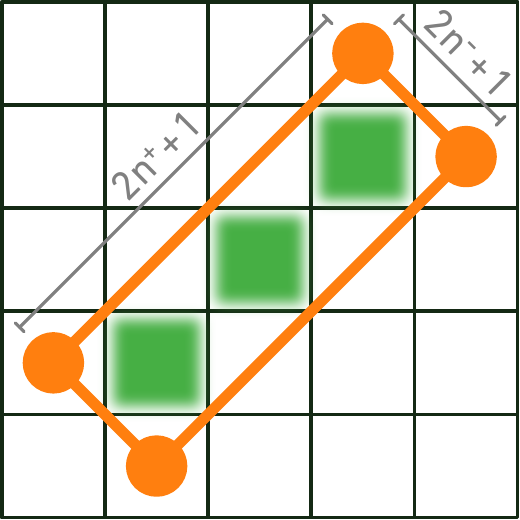}
    \caption{A perfectly subradiant giant atom and its populated bound state in the continuum. The grid denotes the lattice of coupled cavities, with each square corresponding to one lattice site. The orange dots linked by a solid line denote the four coupling points of the giant atom. Note that the coupling points are separated by odd distances along each diagonal: $2n^{+}+1=3$ in the $\smqty[1\\1]$ direction and  $2n^{-}+1=1$ in the $\smqty[\;\;\,1\\-1]$ direction. The green shading of lattice sites shows the evenly distributed photonic population of the bound state in the continuum, enclosed by the coupling points of the giant atom.}
    \label{fig:subradiance+bic}
\end{figure}


\subsection{A single giant atom --- subradiance}
\label{sec:1GA}

We typically refer to an ensemble of atoms that radiate to their environment at a slower rate than that given by Fermi's golden rule as \textit{subradiant}~\cite{Dicke1954, Gross1982}.
In the case of GAs, it makes sense to use that term not only about collective emission, but also for a single atom, as the interference between its coupling points is analogous to the interference between many small atoms.
A \textit{perfectly subradiant} GA is one which does not decay into the bath.
In this subsection, we first show how a GA can be perfectly subradiant with four coupling points, and how this subradiance is connected to a BIC.
We then derive an analytical expression for this BIC and expand the discussion to various setups with more than four coupling points.

\subsubsection{Subradiance and bound state in the continuum for a giant atom with four coupling points}

In a 2D square lattice, a single GA can be perfectly subradiant if it has at least four coupling points.
This is because it needs two points to interfere destructively and cancel the emission along each of the diagonals.
Assuming all points couple with the same strength, achieving destructive interference requires the coupling points to be separated by an odd distance along each of the diagonals (see \figref{fig:subradiance+bic}), since such a separation makes the excitations acquire a phase shift equal to an odd multiple of $\pi$ when traveling between the coupling points.
Conversely, an even distance between coupling points will lead to superradiance.

The interference between coupling points causing the subradiance of a GA is built through a \textit{bound state in the continuum} (BIC).
This is an eigenstate of the full Hamiltonian whose energy lies in the continuum, but does not interact with the propagating modes.
The BIC is thus a \emph{dressed state} of the system, whose photonic part is as a standing wave of localized excitations between the coupling points of the giant atom (see \figref{fig:subradiance+bic}).
The existence of such BICs has been previously reported for giant atoms in 1D~\cite{Soro2023, Lim2023, Zhang2023a} and 2D~\cite{Gonzalez-Tudela2019} structured environments, although often under different names, such as trapped emission or real pole.

\begin{figure}[t]
    \centering
    \includegraphics[width=\columnwidth]{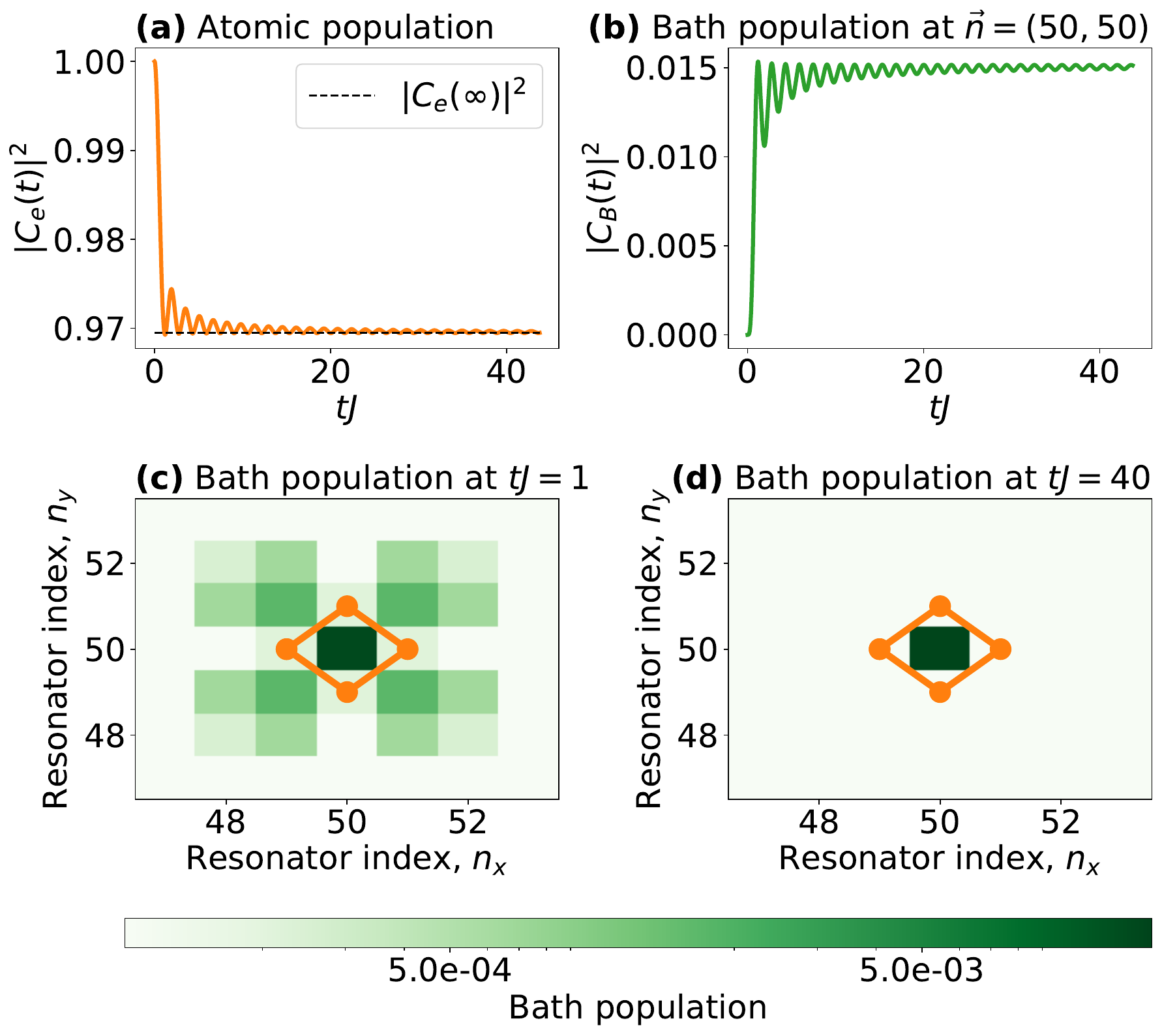}
    \caption{Subradiant dynamics of a giant atom with 4 coupling points (orange markers connected by lines), all with coupling strength $g/J=0.125$ ($G/J=0.25$). The atom is tuned to the middle of the band ($\detuning/J=0$) in a lattice of $100\times100$ coupled resonators. (a) Population of the atom, starting in the excited state and saturating at $\abs{C_e(\infty)}^2$. (b) Population of the photonic part of the bound state, which corresponds to the only cavity enclosed by the coupling points of the atom. (c,d) Population of the bath in real space, at different times $tJ = 1, 40$.
    \label{fig:subradiant_dynamics}}
\end{figure}

The role of the BIC is clearly seen in the dynamics of the system (see \figref{fig:subradiant_dynamics}).
Starting with the atom in its bare excited state $\ket{e}$, we first observe a short decay that corresponds to the atom populating the photonic part of the BIC $\ket{B}$.
During this time, which is the time it takes an excitation to travel through the bath and reach the other coupling points, there is also a small leakage into the environment, owing to the fact that the interference necessary for the subradiance has not yet been built.
The excitation leaked away from the atom is approximately equal to the excitation trapped in the photonic part of the BIC.
This occurs because each coupling point should give off half of its emissions in the direction of another coupling point, and half along the diagonals pointing outward.
After this decay, both the atom and the trapped excitation reach a nonradiative steady state: the BIC or dressed state.
Note that in some experimental platforms, the leakage to the bath can be avoided by driving the dressed state (BIC) instead of the bare atomic excited state.

From resolvent formalism (see \appref{app:resolvent}), we know that the atomic steady-state population is given by~\cite{Gonzalez-Tudela2017, Soro2023}
\begin{equation}
\begin{aligned}
\abs{C_e(\infty)}^2 &= \abs{\lim_{t\to\infty} \Res[\frac{e^{-izt}}{z-\detuning-\Sigma_e(z)}]}^2 \\
&=  \abs{\frac{1}{1 -\eval{\partial_z \Sigma_e(z)}_{z=\detuning}}}^2,
\end{aligned}
\end{equation}
where $z$ is the energy of the BIC and $\Sigma_e(z)$ is the self-energy of the atom.
In particular, for a GA tuned to the middle of the band ($\detuning / J = 0$) with $\ncpts=4$ coupling points and area $(2n^{+}+1)\times(2n^{-}+1)$,
\begin{equation}
    -\eval{\partial_z \Sigma_e(z)}_{z=0} = \frac{g^2 }{J^2}(2n^{+}+1)(2n^{-}+1). 
    \label{eq:partial_Sigma_e}
\end{equation}
In fact, it can be shown that in the single-excitation regime, the self-energy (and thus the time evolution) of a GA with coupling strength $g = g_0/\sqrt{4}$ is exactly the same as that of the state $\ket{+}=\frac{1}{2}(\ket{eggg} + \ket{gegg} + \ket{ggeg} + \ket{ggge})$ of a set of four small atoms with coupling strength $g_0$ coupled to the same points.
This fact is quite intuitive from the derivation of the effective coupling strength and Hamiltonian shown in \secref{sec:numerics} [see discussion around \eqsref{eq:H_i-GA}{eq:G_i_equal}].

\subsubsection{An analytical expression for the dressed excited state}

Above we described the time evolution of the \textit{bare} excited state of a perfectly subradiant GA, but here we derive a concise analytic expression for the \textit{dressed} excited state, i.e., the eigenstate of the full Hamiltonian remaining stationary as $t \to \infty$.
We consider the simplest perfectly subradiant configuration, with four coupling points surrounding a single photonic BIC peak, as seen in \figref{fig:subradiant_dynamics}.
Choosing the photonic BIC peak as the origin of our coordinate system and using notation where $\ket{\pvec{n}}$ is the state with an excitation present at the cavity with coordinate vector $\pvec{n}$, we define
\begin{align}
    \ket{B} &= \ket{\begin{bmatrix} 0 \\ 0 \end{bmatrix}}, \\
    \ket{C} &= \mleft( \ket{\begin{bmatrix} 1 \\ 0 \end{bmatrix}} + \ket{\begin{bmatrix} -1 \\ 0 \end{bmatrix}} + \ket{\begin{bmatrix} 0 \\ 1 \end{bmatrix}} + \ket{\begin{bmatrix} 0 \\ -1 \end{bmatrix}} \mright) / 2.
\end{align}
Inspired by what we observe in our numerical simulations in \figref{fig:subradiant_dynamics}, we make the following ansatz for the dressed excited state:
\begin{equation}
    \ket{e'} = \al \ket{e} + \be \ket{B},
\end{equation}
where $\ket{e}$ is the bare excited state and $\al,\be \in \mathbb{C}$. Applying $H$ from \eqref{eq:H} to this ansatz yields
\begin{equation}
\begin{aligned}
    H \ket{e'} &= \al H \ket{e} + \be H \ket{B} = \qty{\detuning/J = 0} = \\
    &= 2 \mleft( \al g - \be J \mright) \ket{C}.
\end{aligned}
\end{equation}
Since $\ket{e'}$ is not proportional to $\ket{C}$, the only way for $\ket{e'}$ to be an eigenstate of the full Hamiltonian is if $\be/\al = g/J$, making $\ket{e'}$ have eigenenergy $0$.
Normalization then gives that the dressed eigenstate for this perfectly subradiant GA is
\begin{equation}
    \ket{e'} = \frac{1}{\sqrt{1 + g^2/J^2}} \mleft( \ket{e} + \frac{g}{J} \ket{B} \mright).
\end{equation}

This same derivation can be shown to hold for any perfectly subradiant configuration with only slight modifications.
The main thing to note is that, in the general $\detuning/J = 0$ case, the $\ket{B}$ state needs to be defined with an alternating phase, related to how taking an odd number of steps along a diagonal in the lattice makes an excitation acquire a phase shift of $\pi$.
For example, for the configuration in \figref{fig:subradiance+bic}, one needs to use
\begin{equation}
    \ket{B} = \mleft( \ket{\begin{bmatrix} 1 \\ 1 \end{bmatrix}} - \ket{\begin{bmatrix} 0 \\ 0 \end{bmatrix}} + \ket{\begin{bmatrix} -1 \\ -1 \end{bmatrix}} \mright) / \sqrt{3},
    \label{eq:B_alt_phase}
\end{equation}
so that $H \ket{B}$ only involves the cavities at the coupling points.
In this case, when we have three photonic BIC peaks, the dressed excited state becomes
\begin{equation}
    \ket{e'} = \frac{1}{\sqrt{1 + 3g^2/J^2}} \qty( \ket{e} + \sqrt{3} \frac{g}{J} \ket{B} ).
\end{equation}

Finally, we note that these derivations can be extended to yield configurations of coupling points that avoid decoherence even when $\detuning/J \neq 0$.
The BICs in such cases have a nature that is less atomic and more photonic [i.e., $\abs{\braket{e}{e'}}_{\detuning/J=0} > \abs{\braket{e}{e'}}_{\detuning/J\neq 0}$], in agreement with what has been observed in 1D~\cite{Soro2023}.

\begin{figure}[t]
    \centering
    \includegraphics[width=\columnwidth]{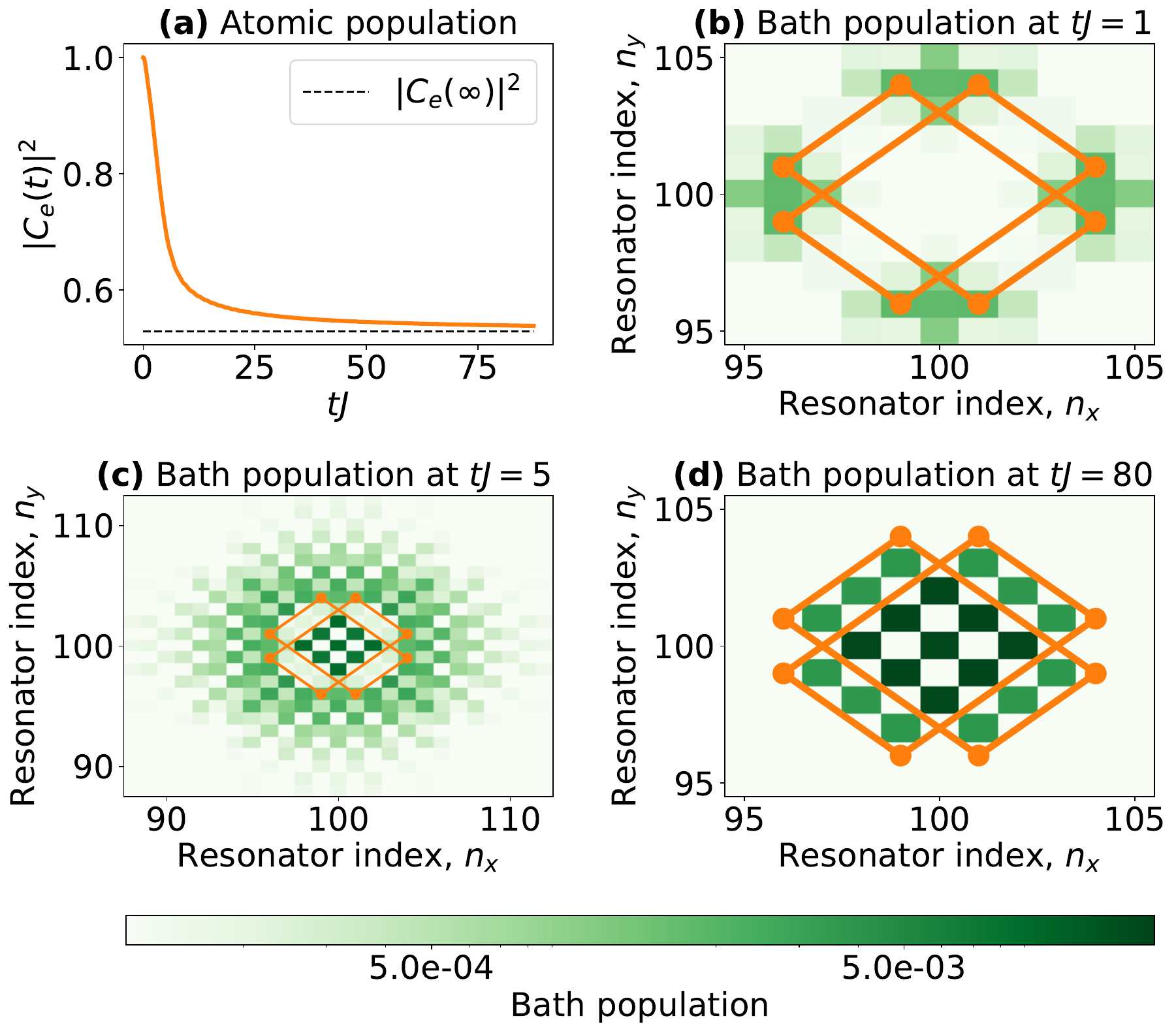}
    \caption{Subradiant dynamics of a giant atom with 8 coupling points (orange markers connected by lines), all with coupling strength $g/J=0.088$ ($G/J=0.25$). The atom is tuned to the middle of the band ($\detuning/J=0$) in a lattice of $200\times200$ coupled resonators. (a) Population of the atom, starting in the excited state and saturating at $\abs{C_e(\infty)}^2$. (b,c,d) Population of the bath in real space, at different times $tJ=1, 5, 80$. Panel (d) shows a constructive interference between the photonic part of the bound states in the continuum of the two subradiant sets of coupling points.}
    \label{fig:8cPts_constructive}
\end{figure}

\begin{figure}[t]
    \centering
    \includegraphics[width=\columnwidth]{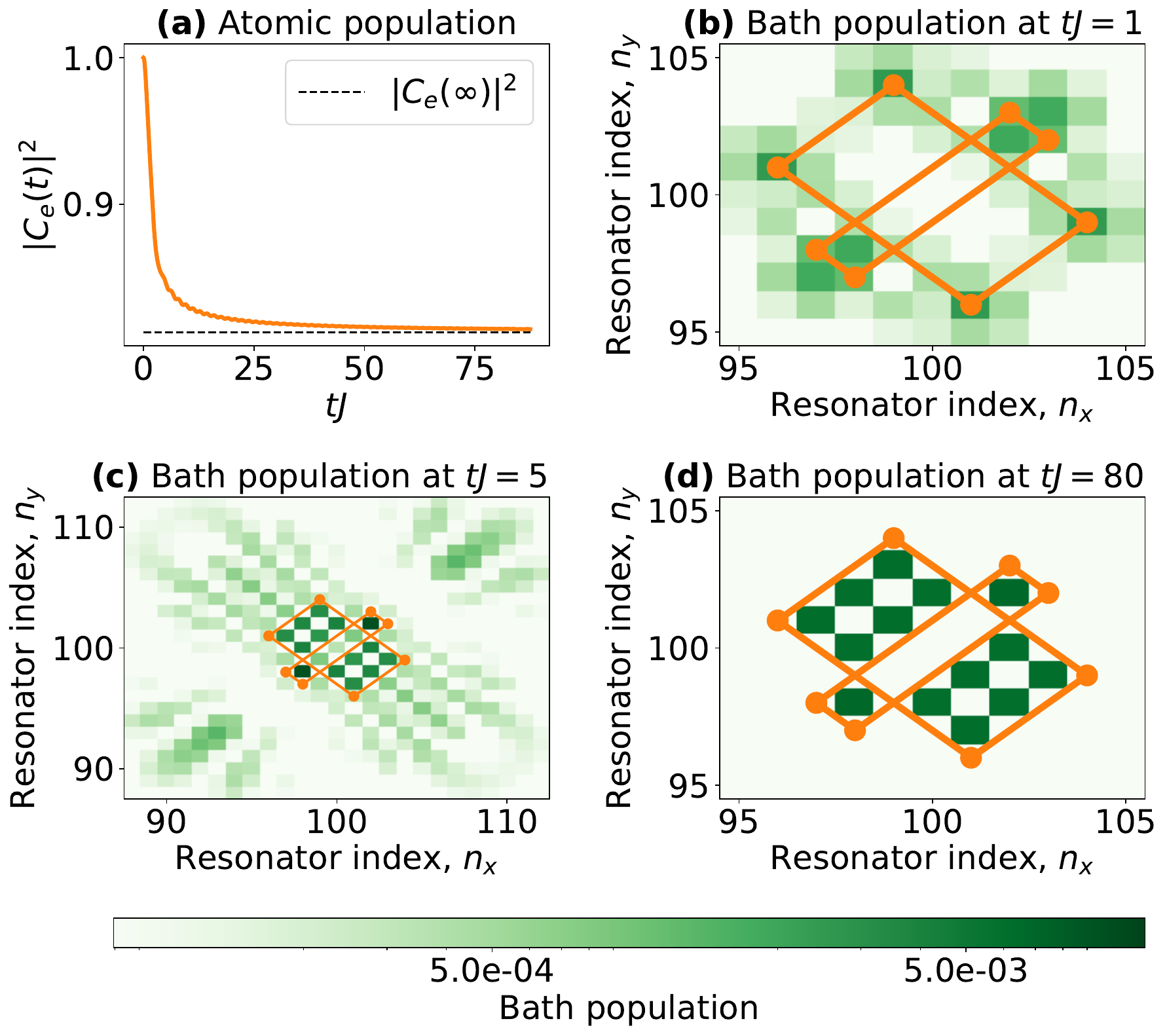}
    \caption{Subradiant dynamics of a giant atom with 8 coupling points (orange markers connected by lines), all with coupling strength $g/J=0.088$ ($G/J=0.25$). The atom is tuned to the middle of the band ($\detuning/J=0$) in a lattice of $200\times200$ coupled resonators. (a) Population of the atom, starting in the excited state and saturating at $\abs{C_e(\infty)}^2$. (b,c,d) Population of the bath in real space, at different times $tJ=1, 5, 80$. Panel (d) shows a destructive interference between the photonic part of the bound states in the continuum of the two subradiant sets of coupling points.}
    \label{fig:8cPts_destructive}
\end{figure}

\begin{figure}
    \centering
    \includegraphics[width=\columnwidth]{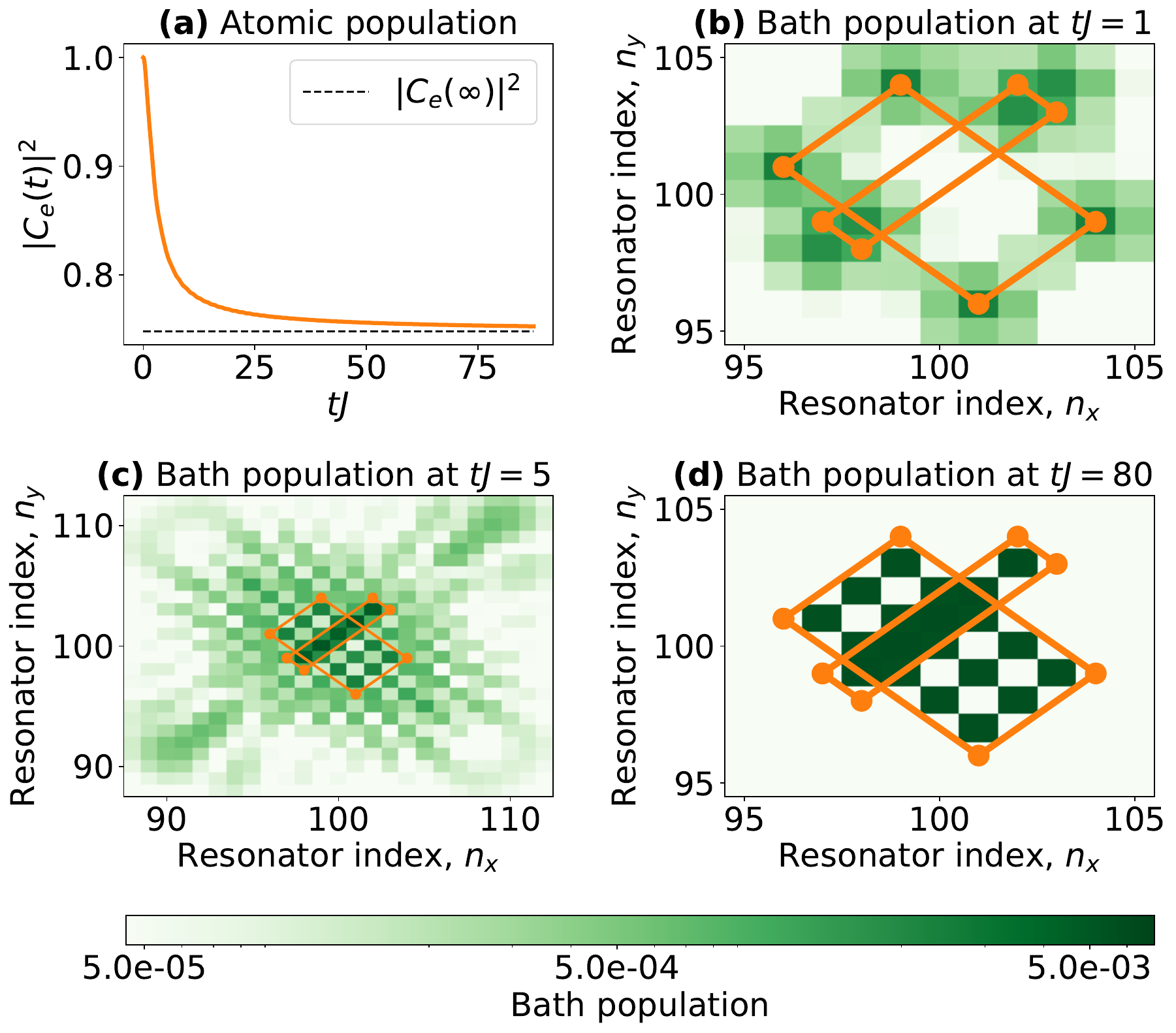}
    \caption{Subradiant dynamics of a giant atom with 8 coupling points (orange markers connected by lines), all with coupling strength $g/J=0.088$ ($G/J=0.25$). The atom is tuned to the middle of the band ($\detuning/J=0$) in a lattice of $200\times200$ coupled resonators. (a) Population of the atom, starting in the excited state and saturating at $\abs{C_e(\infty)}^2$. (b,c,d) Population of the bath in real space, at different times $tJ=1, 5, 80$. Panel (d) shows no interference between the photonic part of the bound states in the continuum of the two subradiant sets of coupling points.}
    \label{fig:8cPts_no-interference}
\end{figure}

\subsubsection{Perfect subradiance with multiple sets of four coupling points}

Beyond the minimal configurations with four coupling points discussed above, perfect subradiance can also be achieved by a GA with $\ncpts = 4p\; (p\in \N)$, where each set of four coupling points has the same coupling strength and is subradiant in itself.
In such a case, the photonic BIC peaks generated by each set may interfere with the photonic BIC peaks from the other sets, either constructively (\figref{fig:8cPts_constructive}), destructively (\figref{fig:8cPts_destructive}), or not at all (\figref{fig:8cPts_no-interference}).
The interference pattern depends on the distribution of the coupling points, and arises from the alternating phase shown in \eqref{eq:B_alt_phase}.

For an analytical derivation of the interference patterns in the examples shown in Figs.~\ref{fig:8cPts_constructive}--\ref{fig:8cPts_no-interference}, where $P=8$, we consider the self-energy of the atom.
It can be calculated as that of the two subradiant sets plus the interference between them:
\begin{equation}
    \Sigma_e(z) = \Sigma_1(z) + \Sigma_2(z)+\Sigma_\text{int}(z).
\end{equation}
Then, $\eval{\partial_z\Sigma_i(z)}_{z=0}$ is calculated as in \eqref{eq:partial_Sigma_e}, giving
\begin{equation}
    -\eval{\partial_z\Sigma_\text{int}(z)}_{z=0} = \xi \frac{2g^2}{J^2}\overlap^{+}\overlap^{-},
\end{equation}
where $\overlap^{\pm}$ is the overlap between the photonic BIC peaks generated by the two subsets along the $\smqty[\;\;\,1\\ \pm 1]$ diagonals and
\begin{equation}
    \xi = \begin{cases}
        +1 & \text{constructive interference}\\
        \;\;\:0 & \text{no interference}\\
        -1 & \text{destructive interference}.
    \end{cases}
\end{equation}
In the particular case where both subsets of coupling points are centered around the same point in the bath (e.g., \figref{fig:8cPts_constructive} and \figref{fig:8cPts_destructive}),
\begin{equation}
\begin{aligned}
    \overlap^{+} &= 2 \min(n^+_1, n^+_2) + 1 , \\
    \overlap^{-} &= 2 \min(n^-_1, n^-_2) + 1 , \\
    \xi &= (-1)^{n^+_1+n^+_2+n^-_1+n^-_2}.
\end{aligned}
\end{equation}

We highlight that, to the best of our knowledge, this interference between the photonic components of different BICs has not been reported before.

\subsubsection{Other perfectly subradiant configurations of coupling points}

\begin{figure}
    \centering
    \includegraphics[width=0.7\columnwidth]{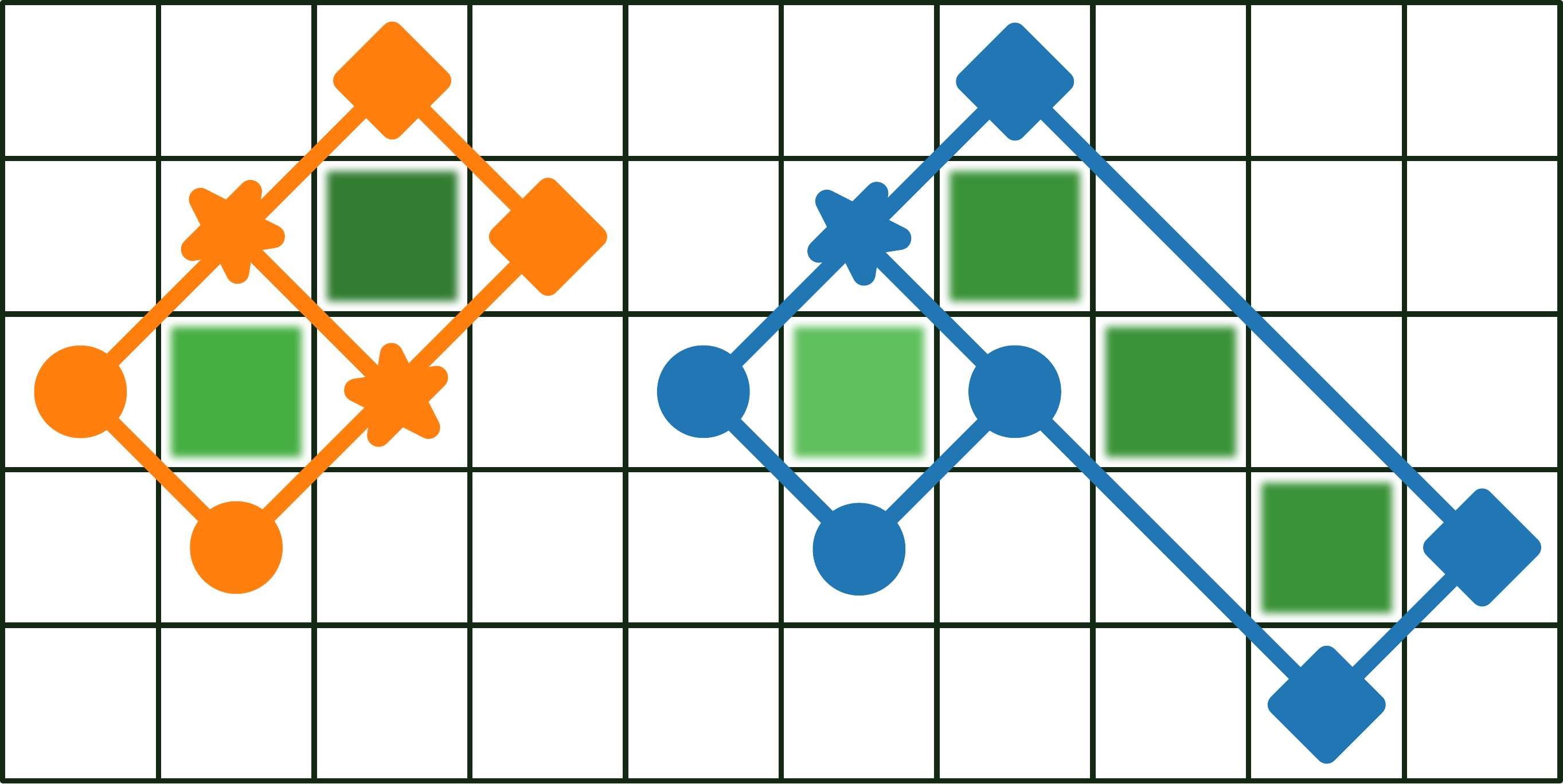}
    \caption{Two perfectly subradiant giant atoms with a number of coupling points that is not a multiple of four. The orange and blue marks linked by a solid line denote the coupling points of each atom, respectively. The circles indicate points with coupling strength $g_1$, the squares indicate points with strength $g_2 > g_1$, and the stars couple with strength $g_1+g_2$. The green lattice sites show the photonic excitation of the bound state in the continuum, with the intensity of the color referring to the population on each site (darker color indicates higher population).}
    \label{fig:6cPts_7cPts}
\end{figure}

One can also achieve perfectly subradiant configurations with $P \neq 4p$ by superimposing subradiant setups with four coupling points in such a way that some coupling points end up on top of one another.
Each set of overlapping coupling points can then be replaced with a single coupling point, whose coupling strength equals the sum of the coupling strengths of the overlapping points.
For example, two superimposed coupling points with coupling strengths $g_1$ and $g_2$ would be replaced by a single coupling point with coupling strength $g = g_1 + g_2$, as shown in \figref{fig:6cPts_7cPts}.
We note that this additive property for coupling points also holds if we allow negative (or complex) coupling strengths.

\subsection{Multiple giant atoms --- decoherence-free interaction}

Unlike small atoms, GAs can interact without decohering when tuned to the continuum.
This is one of the most intriguing, and potentially useful, properties of GAs; it has been reported in 1D continuous waveguides~\cite{FriskKockum2018a, Kannan2020, Carollo2020, Du2021a, Soro2022, Du2023b} and 1D structured waveguides~\cite{Longhi2021, Soro2023}.
Here, we show how such decoherence-free interaction (DFI) between GAs is also possible in 2D, and how it differs from the interaction outside the continuum (i.e., in the band gap), which is also possible for small atoms.
We first consider two GAs and then also explore how many GAs can have pairwise DFI in various configurations.

Outside the continuum, it is well known that atom-photon bound states are formed, with photons becoming exponentially localized in the vicinity of the atoms (small or giant), thus inhibiting their decay~\cite{Calajo2016, Gonzalez-Tudela2017, Soro2023}.
Furthermore, multiple atoms coupled to the same reservoir can interact through the overlap of their bound-state photonic wave functions~\cite{Bay1997, Lambropoulos2000, Shahmoon2013}, and since the atoms are decoupled from the propagating modes, this interaction is inherently decoherence-free.

\begin{figure}[t]
    \centering
    \begin{tikzpicture}
\tikzset{every node}=[font=\Large\sffamily]

\node[anchor = north west] at (-3.5, 2.45) {(a) Braided};

\draw[ultra thick] (-3, 0) -- (3, 0);

\draw[line width = 2pt, python_blue, inner color=white, outer color=python_blue!25] (0, 1) circle (15pt);
\draw[line width = 2pt, python_blue] (-0.3, 1.2) -- (0.3, 1.2);
\draw[line width = 2pt, python_blue] (-0.3, 0.8) -- (0.3, 0.8);
\fill[python_blue] (-2.5, 0) circle (5pt);
\draw[line width = 2pt, python_blue] (-2.5, 0) -- (-2.5, 1.02);
\draw[line width = 2pt, python_blue] (-2.5, 1) -- (-0.53, 1);
\fill[python_blue] (1, 0) circle (5pt);
\draw[line width = 2pt, python_blue] (1, 0) -- (1, 1.02);
\draw[line width = 2pt, python_blue] (0.99, 1) -- (0.52, 1);

\draw[line width = 2pt, python_orange, inner color=white, outer color=python_orange!25] (0, -1) circle (15pt);
\draw[line width = 2pt, python_orange] (-0.3, -1.2) -- (0.3, -1.2);
\draw[line width = 2pt, python_orange] (-0.3, -0.8) -- (0.3, -0.8);
\fill[python_orange] (-1, 0) circle (5pt);
\draw[line width = 2pt, python_orange] (-1, 0) -- (-1, -1.02);
\draw[line width = 2pt, python_orange] (-1.01, -1) -- (-0.52, -1);
\fill[python_orange] (2.5, 0) circle (5pt);
\draw[line width = 2pt, python_orange] (2.5, 0) -- (2.5, -1.02);
\draw[line width = 2pt, python_orange] (2.51, -1) -- (0.52, -1.0);

\end{tikzpicture}
    \raisebox{11pt}{\includegraphics[width=0.49\columnwidth]{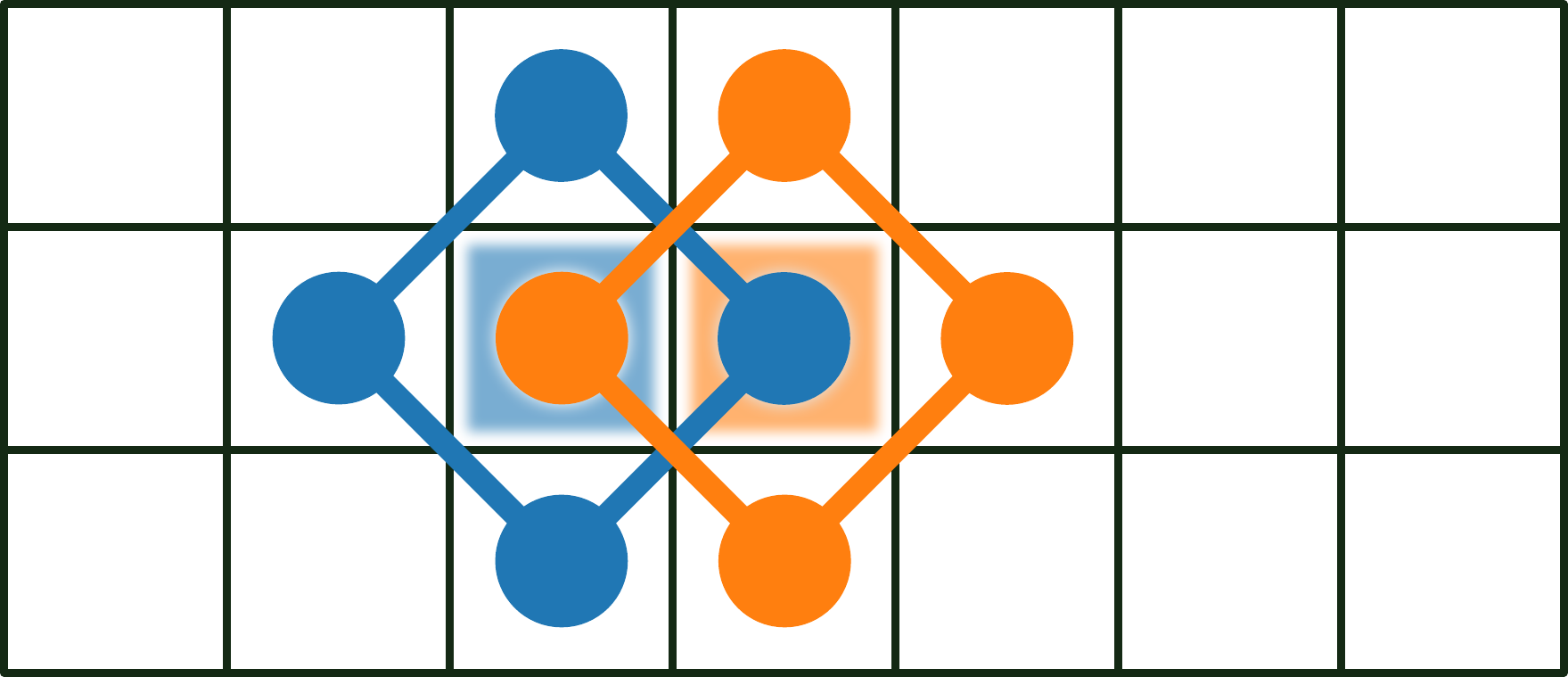}}\\[6pt]
    \begin{tikzpicture}
\tikzset{every node}=[font=\Large\sffamily]

\node[anchor = north west] at (-3, 2.45) {(b) Separate};

\draw[ultra thick] (-2.5, 0) -- (3.5, 0);

\draw[line width = 2pt, python_blue, inner color=white, outer color=python_blue!25] (-1, 1) circle (15pt);
\draw[line width = 2pt, python_blue] (-1.3, 1.2) -- (-0.7, 1.2);
\draw[line width = 2pt, python_blue] (-1.3, 0.8) -- (-0.7, 0.8);
\fill[python_blue] (-2, 0) circle (5pt);
\draw[line width = 2pt, python_blue] (-2, 0) -- (-2, 1.02);
\draw[line width = 2pt, python_blue] (-2.01, 1) -- (-1.53, 1);

\fill[python_blue] (0, 0) circle (5pt);
\draw[line width = 2pt, python_blue] (0, 0) -- (0, 1.02);
\draw[line width = 2pt, python_blue] (0.01, 1) -- (-0.48, 1);

\draw[line width = 2pt, python_orange, inner color=white, outer color=python_orange!25] (2,-1) circle (15pt);
\draw[line width = 2pt, python_orange] (2.3, -1.2) -- (1.7, -1.2);
\draw[line width = 2pt, python_orange] (2.3, -0.8) -- (1.7, -0.8);
\fill[python_orange] (1, 0) circle (5pt);
\draw[line width = 2pt, python_orange] (1, 0) -- (1, -1.02);
\draw[line width = 2pt, python_orange] (0.99, -1) -- (1.47, -1);

\fill[python_orange] (3, 0) circle (5pt);
\draw[line width = 2pt, python_orange] (3, 0) -- (3, -1.02);
\draw[line width = 2pt, python_orange] (3.01, -1) -- (2.52, -1.0);

\end{tikzpicture}
   \includegraphics[width=0.49\columnwidth]{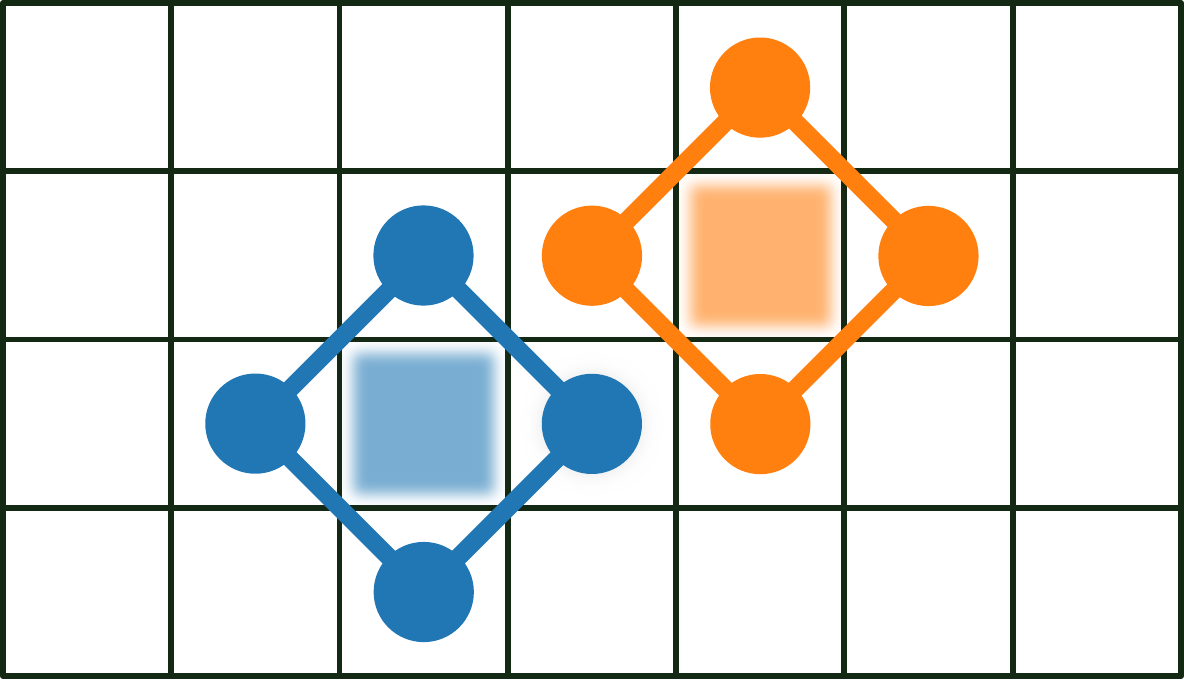}\\[20pt]
    \raisebox{30pt}{\begin{tikzpicture}
\tikzset{every node}=[font=\Large\sffamily]

\node[anchor = north west] at (-3.5, 2.45) {(c) Nested};

\draw[ultra thick] (-3, 0) -- (3, 0);

\draw[line width = 2pt, python_blue, inner color=white, outer color=python_blue!25] (0, 1) circle (15pt);
\draw[line width = 2pt, python_blue] (-0.3, 1.2) -- (0.3, 1.2);
\draw[line width = 2pt, python_blue] (-0.3, 0.8) -- (0.3, 0.8);
\fill[python_blue] (-2.5, 0) circle (5pt);
\draw[line width = 2pt, python_blue] (-2.5, 0) -- (-2.5, 1.02);
\draw[line width = 2pt, python_blue] (-2.5, 1) -- (-0.53, 1);
\fill[python_blue] (2.5, 0) circle (5pt);
\draw[line width = 2pt, python_blue] (2.5, 0) -- (2.5, 1.02);
\draw[line width = 2pt, python_blue] (2.51, 1) -- (0.52, 1.0);

\draw[line width = 2pt, python_orange, inner color=white, outer color=python_orange!25] (0, -1) circle (15pt);
\draw[line width = 2pt, python_orange] (-0.3, -1.2) -- (0.3, -1.2);
\draw[line width = 2pt, python_orange] (-0.3, -0.8) -- (0.3, -0.8);
\fill[python_orange] (-1, 0) circle (5pt);
\draw[line width = 2pt, python_orange] (-1, 0) -- (-1, -1.02);
\draw[line width = 2pt, python_orange] (-1.01, -1) -- (-0.52, -1);

\fill[python_orange] (1, 0) circle (5pt);
\draw[line width = 2pt, python_orange] (1, 0) -- (1, -1.02);
\draw[line width = 2pt, python_orange] (0.99, -1) -- (0.52, -1);

\end{tikzpicture}}
    \includegraphics[width=0.49\columnwidth]{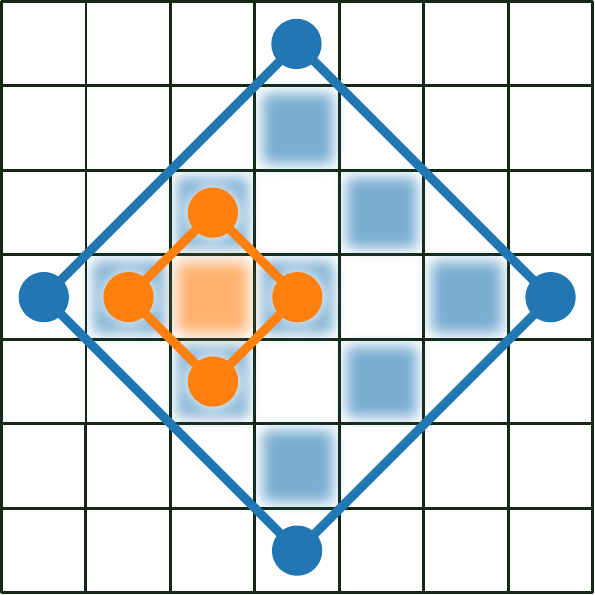}
    \caption{Analogy between the different configurations of giant atoms coupled to a 1D waveguide (left) and to a 2D square lattice (right). Left, 1D: the black line represents the waveguide. The two atoms (blue and orange) are coupled to it at the points marked by the blue and orange dots, respectively. Right, 2D: The grid denotes the lattice of coupled cavities, with each square corresponding to one lattice site. The blue (orange) dots united by a continuous line denote the four coupling points of the first (second) giant atom. The blue (orange) lattice sites show the photonic part of the bound states in the continuum of the blue (orange) atom. These parts mediate the decoherence-free interaction if the orange (blue) atom has a coupling point on a blue (orange) lattice site and vice versa.
    (a) Braided configuration --- the only one that allows decoherence-free interaction. (b) Separate configuration. (c) Nested configuration. 
    }
    \label{fig:DFI-sketch}
\end{figure}
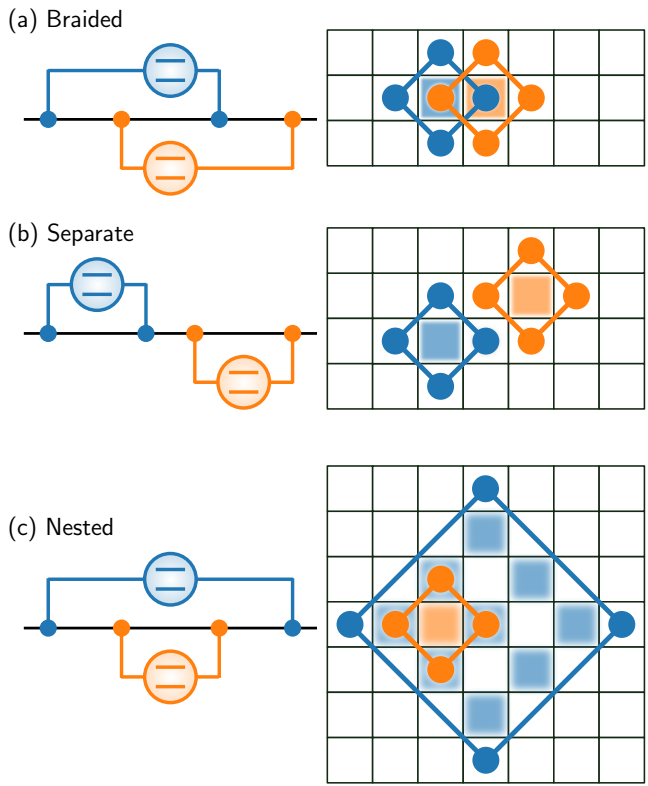

We find that the interaction mechanism in the continuum is not quite the same: for DFI to take place, each of the GAs need to be perfectly subradiant and have at least one of their coupling points in a cavity populated by a BIC associated with the other atom (see \figref{fig:DFI-sketch}).
Note that this mechanism also applies to 1D structured waveguides, although that has not been reported before with such clarity.
In fact, for two GAs, this configuration is analogous to the so-called \textit{braided} one in 1D~\cite{FriskKockum2018a, Soro2023}, which is typically the only one allowing DFI.
Similarly to 1D, 2D configurations analogous to the \textit{nested} and \textit{separate} ones (see \figref{fig:DFI-sketch}) do not allow for DFI either.

\subsubsection{Two giant atoms}
\label{sec:2GAs}

\begin{figure}[t]
    \centering
    \includegraphics[width=\columnwidth]{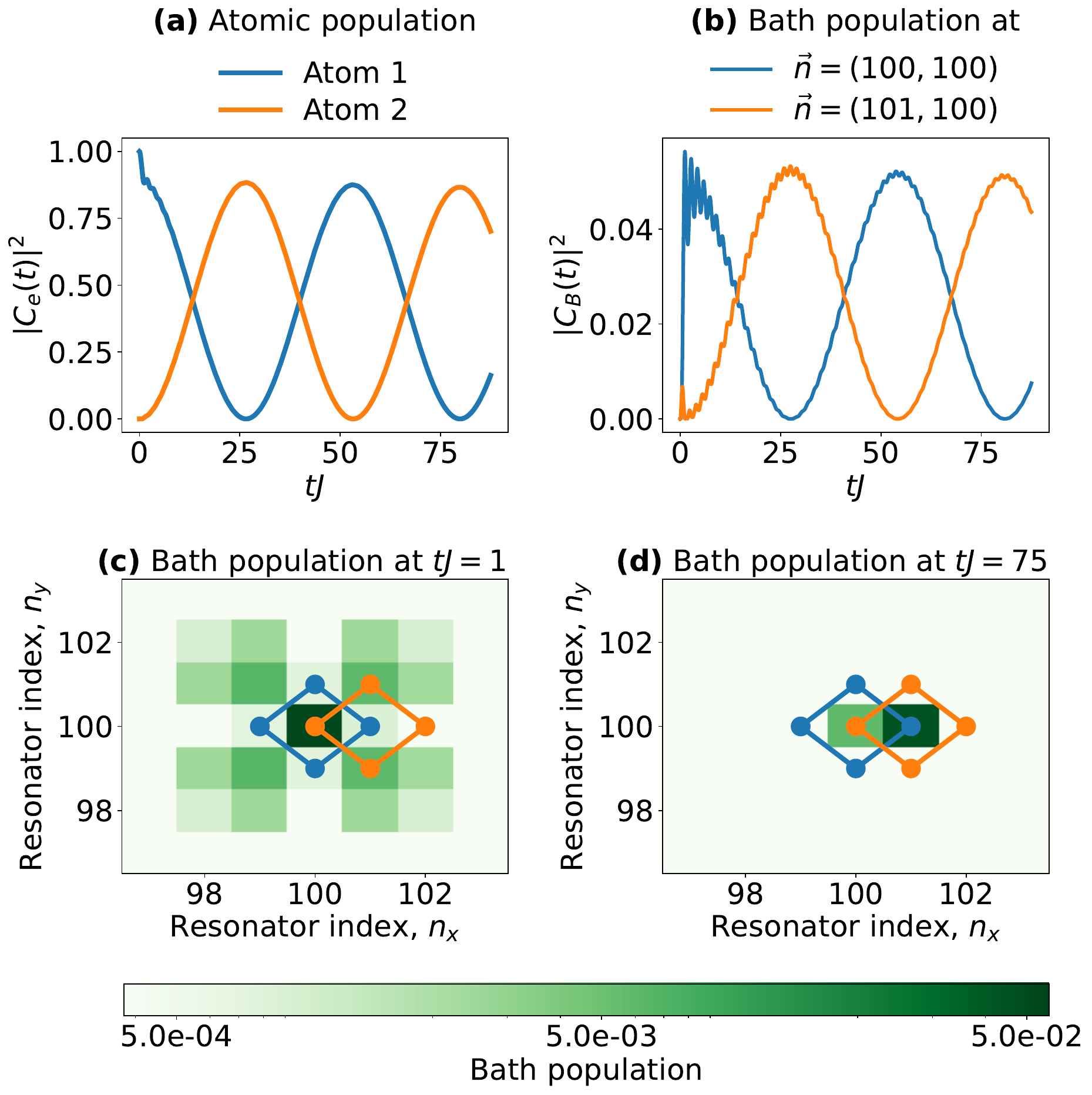}
    \caption{Decoherence-free dynamics of two giant atoms with 4 coupling points each (blue and orange dots linked by solid lines), all with coupling strength $g/J=0.25$ ($G/J=0.5$). The atoms are tuned to the middle of the band ($\detuning/J=0$), in a lattice of $200\times200$ coupled resonators. (a) Population of the atoms, starting in the bare excited state of atom 1. (b) Population of the photonic parts of bound states 1 (blue) and 2 (orange), i.e., of the cavities enclosed by atoms 1 and 2, respectively. (c,d) Population of the bath in real space, at different times $tJ=1, 75$.}
    \label{fig:DFI-dynamics}
\end{figure}

In \figref{fig:DFI-dynamics}, we depict the canonical signature of DFI, i.e., the time evolution of two braided GAs exchanging an excitation back and forth. We note that this evolution is nearly identical to that in 1D~\cite{Soro2023}: starting with the first atom in its bare excited state, there is an initial exponential decay corresponding to the buildup of the interference, for a time 
\begin{equation}
\tau = \frac{d}{v_g} = \frac{2\max(n^+, n^-)+1}{2J},
\end{equation}
which is then followed by weakly damped population exchanges [$\cos[2](z_R t)\e^{-2z_It}$].
As shown in the 1D case~\cite{Soro2023}, the interaction rate $z_R$ is given by the real part of the energy of the two BICs ($z_R=\abs{\Re{z_1}-\Re{z_2}}/2$), whereas the damping rate is given by the imaginary part ($z_I=\abs{\Im{z_1}+\Im{z_2}}/2$).
Note that the damping, which occurs due to a combination of the retardation effects and the exchange interaction being nonzero, makes these states \textit{quasi-bound}.

\begin{figure}
    \centering
    \includegraphics[width=0.7\columnwidth]{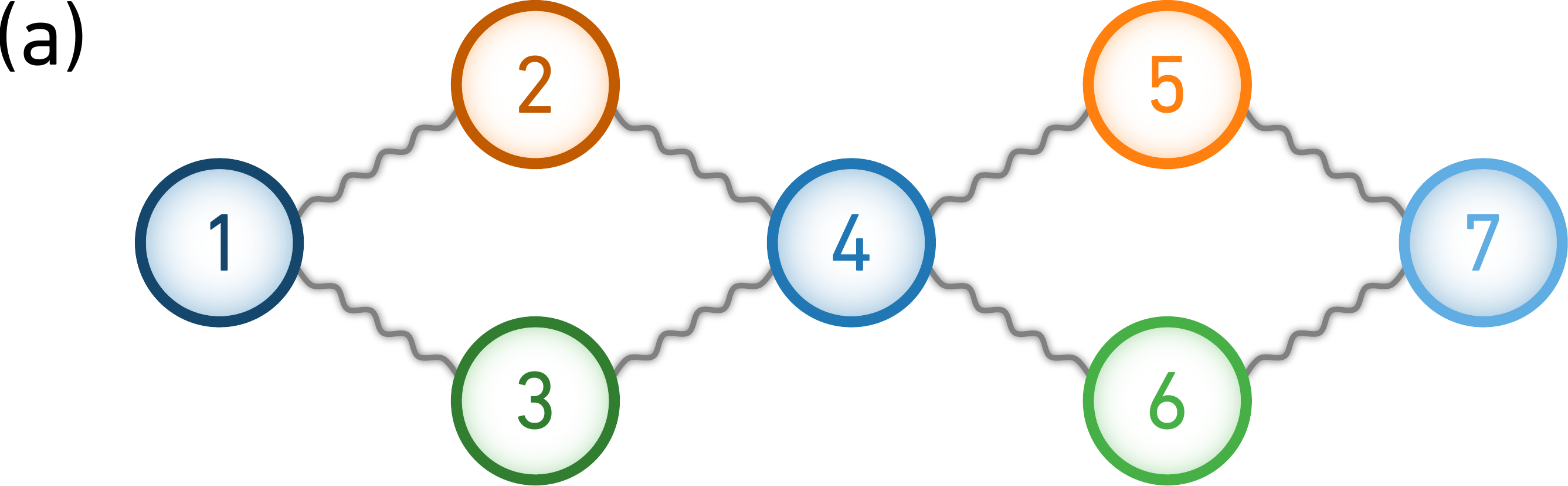}\\[4pt]
    \includegraphics[width=0.45\columnwidth]{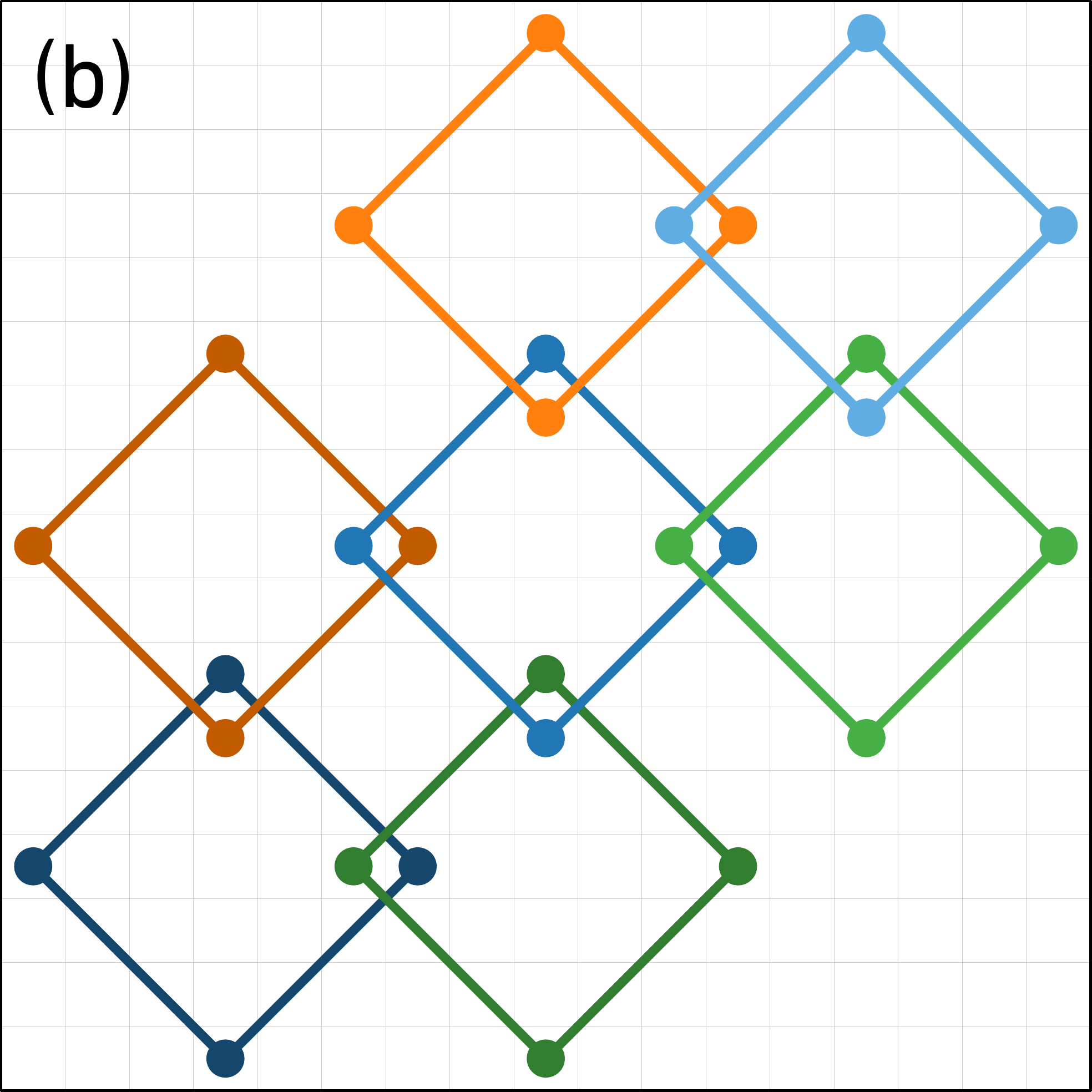}
    \includegraphics[width=0.45\columnwidth]{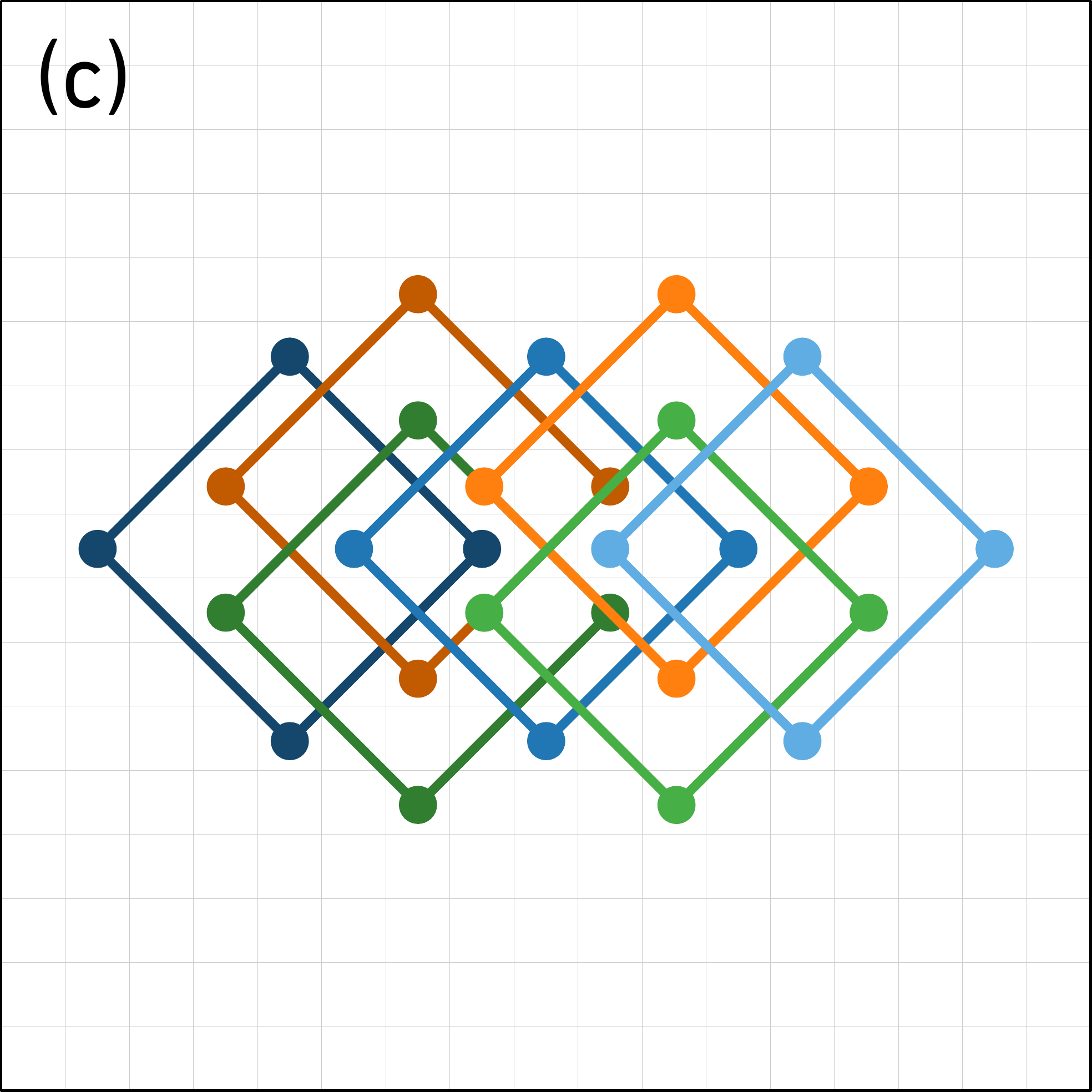}
    \caption{Decoherence-free braided chains of seven giant atoms. The subchain formed by atoms 1-2-4-5-7 (i.e., without the green atoms) is also a DFI braided chain.  (a) Sketch of the connections between the seven atoms. Gray squiggly lines denote atoms that interact with each other without decohering. (b,c) Two possible implementations of the chain shown in (a) with seven atoms having four coupling points each, tuned to the middle of the band ($\detuning/J=0$). The dots linked by a solid line denote the four coupling points of each giant atom. (b) Loosely braided. (c) Tightly braided. Note that even if atoms 2-5, 1-4-7 and 3-6 also seem to be braided, they do not directly interact with each other, since none of their coupling points connect to a cavity containing some photonic part of the bound state in the continuum of the other atom(s).}
    \label{fig:DFI_chain}
\end{figure}

Finally, we identify the Rabi swaps between the photonic parts of the bound states (\figref{fig:DFI-dynamics}, top right panel) as the two-atom analogue of the so-called \textit{oscillating BICs}~\cite{Guo2020, Guo2020a, Longhi2021, Terradas-Brianso2022, Noachtar2022, Lim2023}.
The latter have been shown to appear in GAs with three or more coupling points to the waveguide (continuous or structured), where multiple BIC solutions coexist and give rise to dynamic oscillations. 
In the 2D setups we study here, the multiplicity of BICs does not come from the multiplicity of coupling points, but instead, from the multiplicity of atoms.

\subsubsection{Many giant atoms}
\label{sec:manyGAs}

As one may suspect, DFI is also possible beyond two GAs in decoherence-free chains with pairwise interaction (reported before in 1D for continuous waveguides~\cite{FriskKockum2018a, Soro2022}).
Many configurations of such chains, both 1D and 2D, are feasible in order to transport excitations, as shown in \figref{fig:DFI_chain}.

\begin{figure}[t]
    \centering
    \includegraphics[width=\columnwidth]{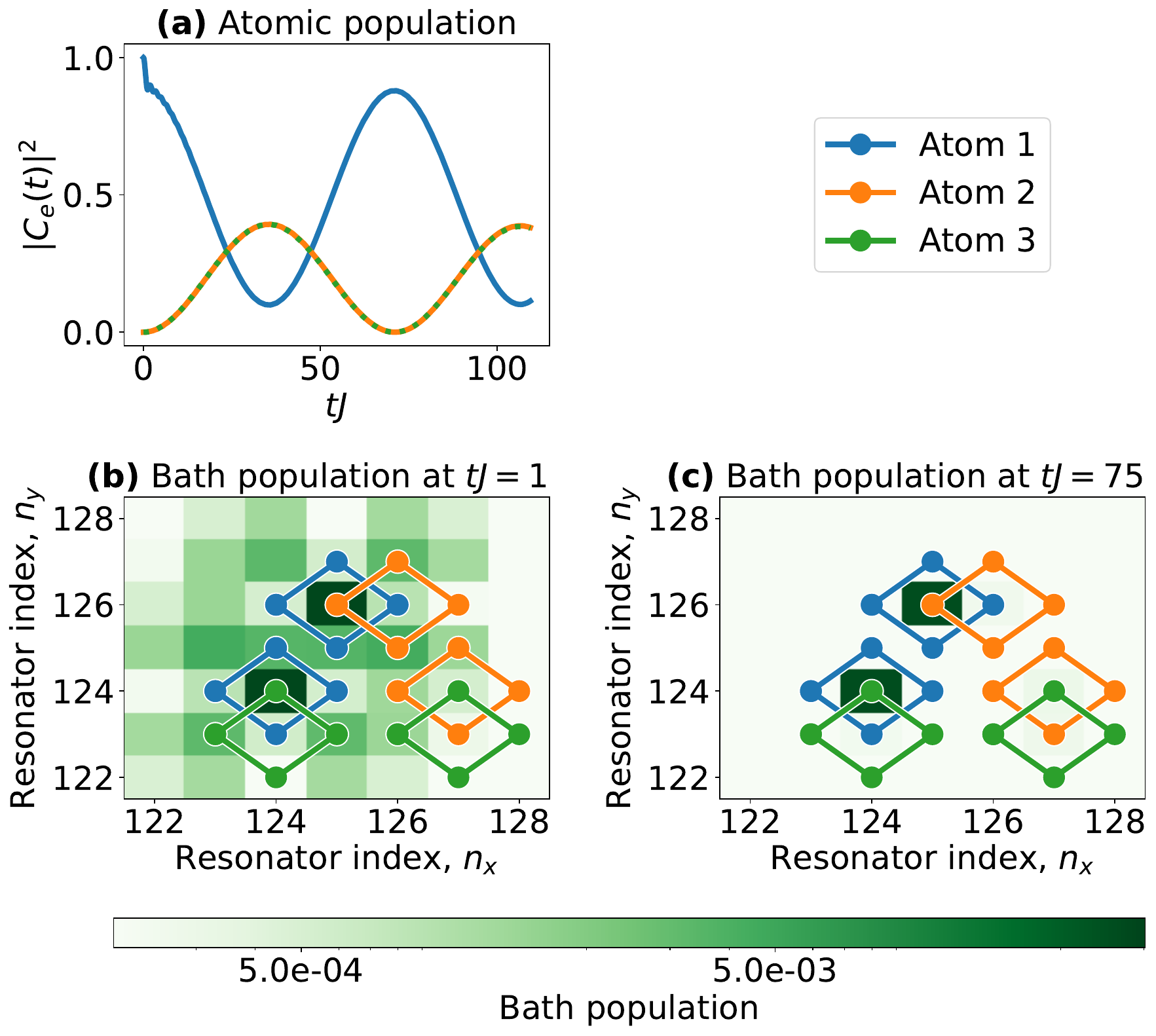}
    \caption{All-to-all interaction between three giant atoms with 8 coupling points each, all with coupling strength $g/J=0.177$ ($G/J=0.5$). The atoms are tuned to the band ($\detuning/J=0$) in a lattice of $250\times250$ coupled resonators. (a) Population of the atoms, starting in the bare excited state of atom 1. (b,c) Population of the bath in real space, at different times $tJ=1, 75$. The blue (atom 1), orange (atom 2), and green (atom 3) markers linked by solid lines of the same color indicate the sets of coupling points that are in a perfectly subradiant configuration.}
\label{fig:all-to-all}
\end{figure}

Moreover, these GAs allow \emph{all-to-all interaction}, which is possible in 1D continuous waveguides in the minimal configuration (i.e., three GAs having two coupling points each)~\cite{FriskKockum2018a}.
In the 2D square lattice, however, the minimal configuration of three GAs with four coupling points each does not suffice, and achieving all-to-all interaction requires additional coupling points.
For example, in \figref{fig:all-to-all}, we show a layout of three GAs with eight coupling points each, in which all atoms are perfectly subradiant, and each atom has at least one coupling point fully surrounded by each of the other atoms.

\begin{figure}[t]
    \centering
    \raisebox{10pt}{\includegraphics[width=0.4\columnwidth]{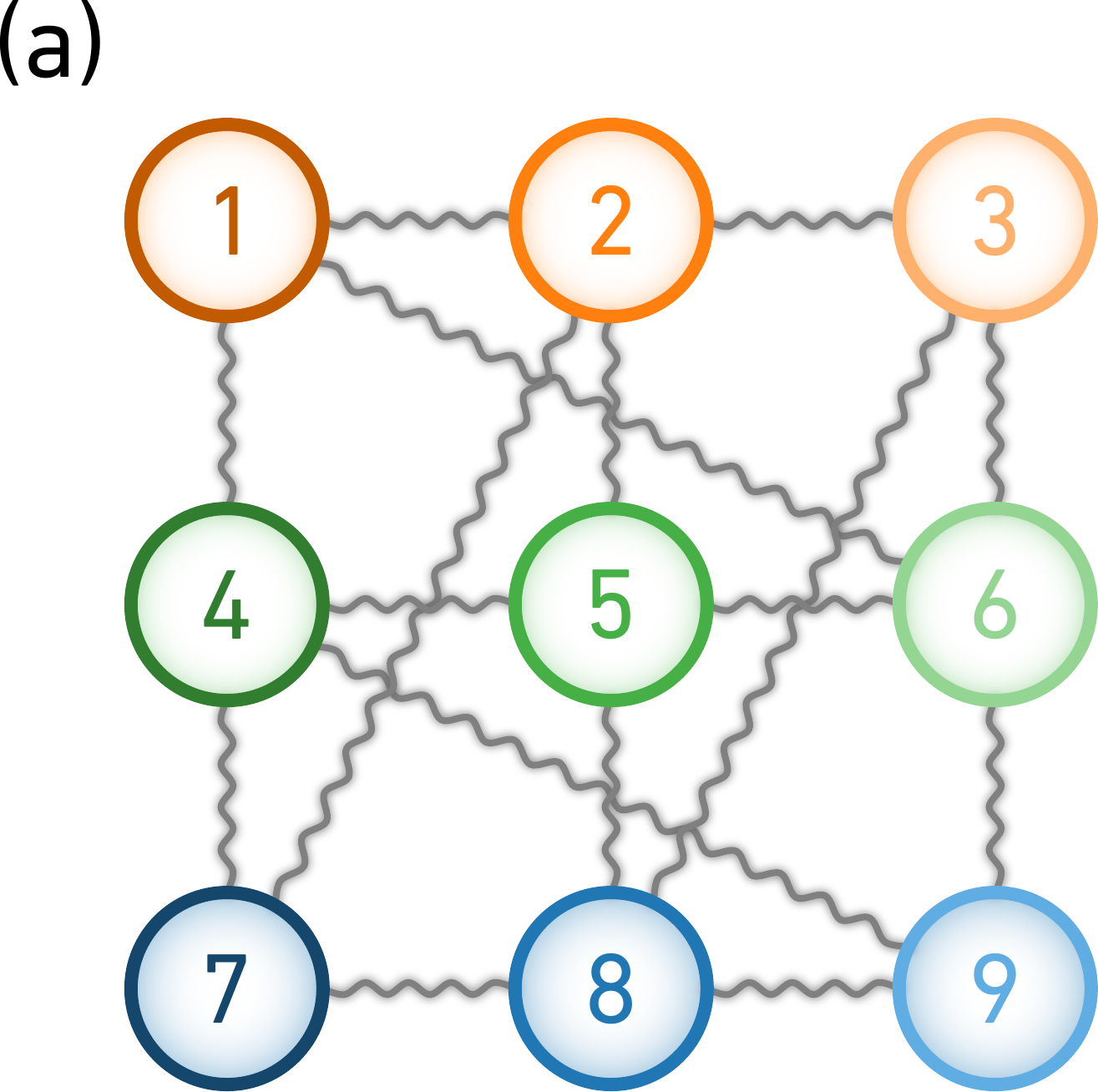}}\qquad
    \includegraphics[width=0.45\columnwidth]{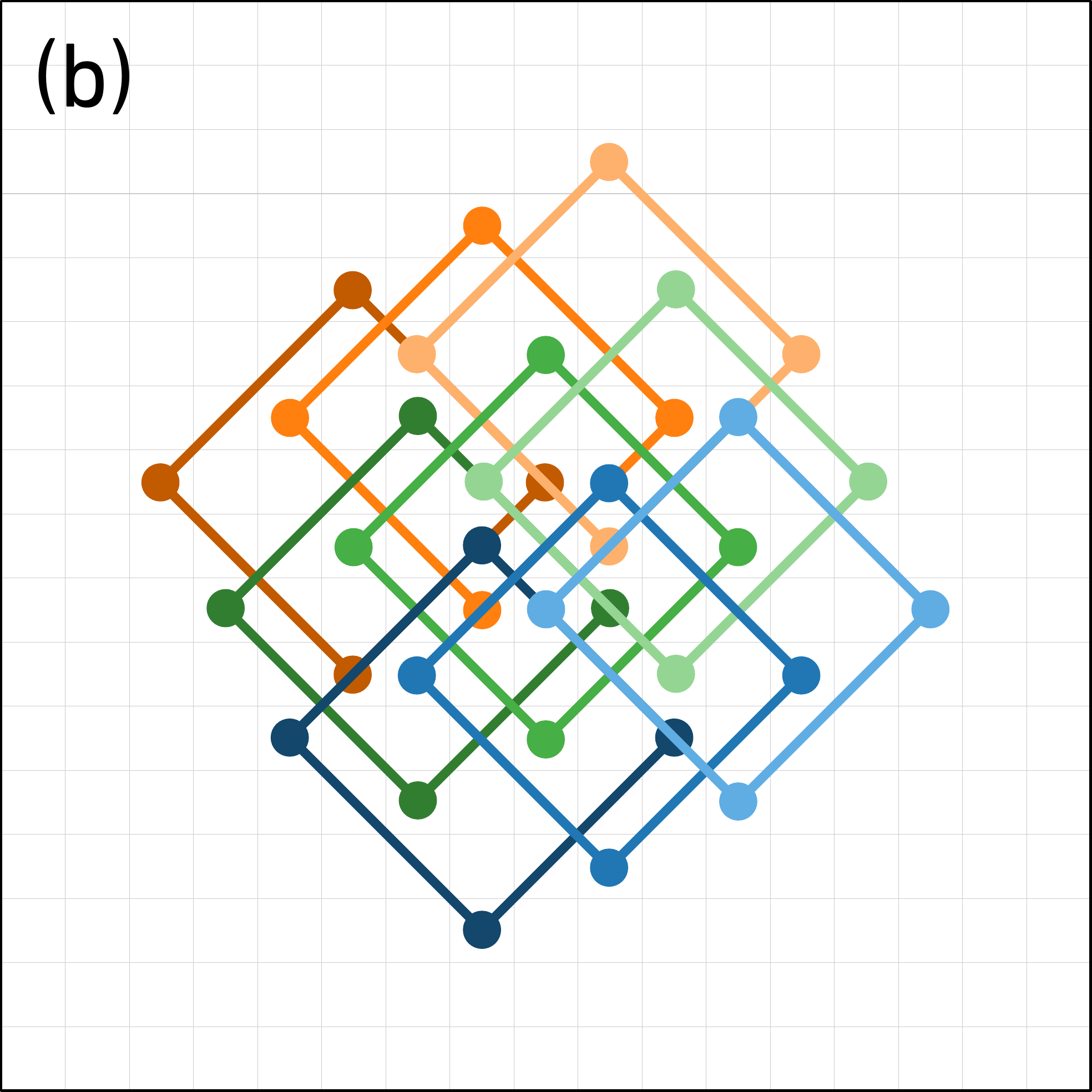}
    \caption{High-connectivity grid of nine identical giant atoms (the different colors are for readability). (a) Sketch of the connections between the nine atoms. Gray squiggly lines denote atoms that interact with each other without decohering. (b) A possible implementation of the grid shown in (a) with nine giant atoms having four coupling points each, tuned to the middle of the band ($\detuning/J = 0$). The dots linked by a solid line denote the four coupling points of each atom. 
}
    \label{fig:high-connectivity}
\end{figure}

Finally, without having all-to-all interaction, we can still build high-connectivity grids with GAs that just have four coupling points each, like the arrangement shown in \figref{fig:high-connectivity}.
In this example, we establish long-range DFI between distant atoms that are mediated by a bath containing only nearest-neighbor couplings. 
This kind of setup could prove useful in quantum simulation of systems beyond nearest-neighbor interaction~\cite{Korenblit2012, Zhang2023, Qiao2024, Fauseweh2024}, as well as in the implementation of nonlocal quantum gates.

\section{Conclusion}
\label{sec:conclusion}

We conducted a detailed theoretical study of giant atoms (GAs) in 2D structured environments.
We focused on the case when this environment is a square lattice of cavities, which leads to band gaps and an energy band that comes with an anisotropic energy dispersion at its center.
In this setting, we showed how GAs can avoid relaxing into the environment.

For a single GA with a transition frequency in the center of the band, we found that it can display perfect subradiance, i.e., completely suppressed emission of energy into the environment, if the atom couples to the bath through at least four cavities.
This suppression is because the propagation of excitations in the environment is restricted to be diagonal at the band center; with four coupling points it becomes possible to cancel emission along these four propagation directions through destructive interference between the emissions through pairs of coupling points.
We also showed that such perfect subradiance can be achieved with more coupling points, if their number is a multiple of four and the coupling strength is equal at all coupling points.
Allowing for coupling strengths to vary between coupling points (including becoming negative), we further showed how subradiance can be achieved in setups with other numbers of coupling points, both even and odd.

For multiple GAs, we showed that the concept of decoherence-free interaction (DFI; two GAs interacting through an environment without losing energy into that environment), previously demonstrated in a 1D setting, also can be realized in 2D. 
More specifically, we showed that this DFI can take place if the individual atoms have their coupling points arranged to be subradiant and each atom has some (but not all) of its coupling points enclosed by coupling points from the other atom.
This setup constitutes a generalization of the so-called braided setup of coupling points required for GAs in 1D~\cite{FriskKockum2018a, Soro2023}.
We further showed how this DFI for GAs in 2D can be extended to more atoms, forming effective high-connectivity (even all-to-all coupling) lattices of atoms connected through DFI.

The results we found for both single and multiple GAs can be understood as manifestations of bound states in the continuum (BICs).
For a single GA in a subradiant configuration, an initial excitation in the atom mostly remains there.
A small amount of energy leaks into the environment outside the atom before the destructive interference between emissions from the different coupling points kicks in, but about half that energy remains in the photonic part of a BIC formed in-between the coupling points of the atom.
For multiple GAs, DFI between pairs of them is only possible when some coupling point(s) of each atom in the pair are placed in locations that contain a part of the BIC associated with the other atom.

We note that these results were enabled by nontrivial extensions of analytical and numerical methods previously employed to study small atoms in 2D structured environments.
Through these extensions, we were able to obtain analytical expressions for the steady-state populations of GAs and their BICs, and perform numerical simulations of how the atomic and photonic populations evolve in time for both single and multiple GAs.
For DFI between a pair of multiple GAs, we observed that both the atomic populations and the photonic parts of the BIC populations undergo Rabi oscillations when one of the atoms is initialized in its excited state and the rest of the system is initialized in its ground state.

As we discuss at the end of \secref{sec:theory}, the effects revealed in this study should be possible to observe with existing technology in several experimental platforms.
We believe that a setup with superconducting qubits coupled to a lattice of microwave resonators seems most promising for such experimental demonstrations. 
We note that arranging for the multiple coupling points in such a setup may be aided by flip-chip technology~\cite{Rosenberg2017, Rahamim2017, Kosen2022}.

The properties of GAs in 2D structured environments that we have discussed in this paper may find applications in quantum computing and quantum simulation.
In these fields, the ability to protect qubits from decoherence while at the same time enabling them to interact, preferably with a high connectivity, is crucial, and we have shown that the DFI between GAs in 2D provides such capability.
In particular, the DFI between GAs may be especially interesting for simulation of open quantum systems, since the interaction of the atoms with the environment can be turned on and off by tuning the atomic frequency, which is possible, e.g., with superconducting qubits~\cite{Kannan2020, Chen2024}.

There are several possible directions for future work.
Beyond more detailed studies of the possible applications discussed in the preceding paragraph, one could study other 2D lattices than the square grid or introduce varying hopping rates between the cavities in such a lattice.
Such changes could lead to other band structures, including ones with topological properties.
Given the importance of BICs and subradiance, it would also be interesting to extend the model to more than one excitation to study superradiance and multiphoton BICs.
It may furthermore be valuable to consider more than two energy levels in each GA, since additional atomic levels are crucial in several quantum-optics effects.


\begin{acknowledgments}

We thank Alberto del \'{A}ngel, Simon Pettersson Fors, Th\'{e}o S\'{e}pulcre, and Alejandro Gonz\'{a}lez-Tudela for useful discussions.
We acknowledge support from the Swedish Research Council (grant number 2019-03696), from the Knut and Alice Wallenberg Foundation through the Wallenberg Centre for Quantum Technology (WACQT), and from the Swedish Foundation for Strategic Research.
\end{acknowledgments}


\section*{Note added}

While finishing writing this manuscript, we became aware of the related Ref.~\cite{Leonforte2024}, which was made available as a preprint at the same time as our paper.


\appendix

\section{Resolvent formalism}
\label{app:resolvent}

The formalism used here is based on Chapter 3 of Ref.~\cite{Cohen-Tannoudji} and adapted to the particular case of giant atoms (GAs) coupled to a 2D square lattice through $\ncpts$ coupling points.
For the 1D structured waveguide case, see the derivation in Ref.~\cite{Soro2023}.

\subsection{A single giant atom --- derivation of the self-energy and probability amplitude}
We consider the total Hamiltonian of a single GA coupled to a structured lattice from \eqref{eq:H}, $H=H_0+H_\text{int}$, where
\begin{align}
    H_0 &= H_A + H_B = \detuning \sigma^+\sigma^- + \sum_k \omega(k) a_k^\dagger a_k,\\
    \Hint &= \sum_{p=1}^{P} \frac{g_{p}}{N} \sum_{\vec{k}} \mleft(e^{-i\vec{k}\cdot\vec{n}_{p}} a_{\vec{k}}\sigma^+ + \Hc \mright),
\end{align}
In the single-excitation subspace, the eigenstates of the bare Hamiltonian $H_0$ are $\ket{e}:=\ket{e,0}$ and $\ket{\vec{k}} := \ket{g,\vec{k}}$, for $\vec{k}=(k_x, k_y)$ and $k_x,k_y \in \{-\pi, \dots, \pi-\frac{2\pi}{N}\}$.
The interaction term $\Hint$ couples these atomic and photonic subspaces $\qty{\ket{e}}$ and $\qty{\ket{\vec{k}}}$ to one another.

The \emph{resolvent} of the Hamiltonian is defined by 
\begin{equation}
    G(z) = \frac{1}{z-H}.
\end{equation}
In general, for a projection $\mathcal{P}$ onto a subset spanned by a set of eigenvectors of $H_0$ and its complement $\mathcal{Q} = 1 -\mathcal{P}$, the resolvent obeys
\begin{equation}
    \mathcal{P}G(z)\mathcal{P} = \frac{\mathcal{P}}{z - \mathcal{P}H_0\mathcal{P} - \mathcal{P}\Sigma(z)\mathcal{P}},
\end{equation}
where
\begin{align}
    \Sigma (z) =& H_\text{int} + H_\text{int} \frac{\mathcal{Q}}{z-\mathcal{Q}H_0\mathcal{Q} - \mathcal{Q}H_\text{int}\mathcal{Q}}H_\text{int}\nonumber\\
                \approx& H_\text{int} + H_\text{int} \frac{\mathcal{Q}}{z-H_0}H_\text{int}
    \label{eq:app_sigma_z}
\end{align}
is the \emph{level-shift operator}.
Note that the approximation symbol above denotes the second-order perturbative expansion in powers of $H_\text{int}$, a truncation that is justified since $H_\text{int}$ is small compared to $H_0$.
In particular, when $\mathcal{P}$ is the projector onto a single state $\ket{\alpha}$ with energy $E_\alpha$,
\begin{equation}
    G_\alpha (z) = \frac{1}{z-E_\alpha - \Sigma_\alpha (z)},
    \label{eq:G_alpha}
\end{equation}
with $G_\alpha(z) = \mel{\alpha}{G(z)}{\alpha}$ and $\Sigma_\alpha(z) = \mel{\alpha}{\Sigma(z)}{\alpha}$.

In our case, we define $\mathcal{P}=\ketbra{e}$ and its complement $\mathcal{Q}=\sum_{\vec{k}}\ketbra{\vec{k}}$.
Then we can write the self-energy of the atom $\Sigma_e(z) = \mel{e}{\Sigma(z)}{e}$ as follows:
\begin{align}
    \Sigma_e(z) = &\cancelto{0}{\mel{e}{H_\text{int}}{e}} \quad+ \sum_{\vec{k}} \bra{e}H_\text{int}\frac{\ketbra{\vec{k}}}{z-H_0}H_\text{int}\ket{e} \nonumber\\
    =& \frac{1}{N^2}\sum_{\vec{k}} \frac{\mleft(\sum_p^\ncpts g_p e^{-i\vec{k}\cdot\vec{n_p}}\mright)\mleft(\sum_q^\ncpts g_q^* e^{i\vec{k}\cdot\vec{n_q}}\mright)}{z-\omega(\vec{k})} .
\end{align}
Henceforth, we use the dispersion relation $\omega(\vec{k})=-2J[\cos(k_x)+\cos(k_y)]$ and that the distance vector between two different coupling points is $\Delta\vec{n}$.
For simplicity, let us also assume that $g_p = g_q = g \in \R$.
Then,
\begin{equation}
    \Sigma_e(z) = \frac{g^2}{N^2}\sum_{\vec{k}} \frac{\ncpts + 2\sum_{\Delta\vec{n}} \cos(\vec{k}\cdot\Delta\vec{n})}{z+2J[\cos(k_x)+\cos(k_y)]}.
\end{equation}
In the continuum limit, i.e., when $N\to \infty$, the sum over $\vec{k}$ becomes a double integral: $\sum_{\vec{k}}\left(\frac{2\pi}{N}\right)^2\to \iint_{\vec{k}} \dd[2]{\vec{k}}.$
Therefore, we can write the self-energy as
\begin{align}
    \Sigma_e(z) \to& \frac{g^2}{(2\pi)^2}\iint_{-\pi}^\pi \dd[2]{\vec{k}} \frac{\ncpts + 2\sum_{\Delta\vec{n}} \cos(\vec{k}\cdot\Delta\vec{n})}{z+2J[\cos(k_x)+\cos(k_y)]}
    \label{eq:Sigma_e_iint}\\
    =&  \ncpts \underbrace{\frac{g^2}{(2\pi)^2}\iint_{-\pi}^\pi \dd[2]{\vec{k}} \frac{1}{z+2J[\cos(k_x)+\cos(k_y)]}}_{\Sigma_{\text{SA}}(z)} \nonumber\\
    +&2\sum_{\Delta\vec{n}} \underbrace{
    \frac{g^2}{(2\pi)^2}\iint_{-\pi}^\pi \dd[2]{\vec{k}} \frac{\cos(\vec{k}\cdot\Delta\vec{n})}{z+2J[\cos(k_x)+\cos(k_y)]}}_{\Sigma_{\Delta\vec{n}}(z)},
    \label{eq:Sigma_e_SA_interf}
\end{align}
where the self-energy of a small atom $\Sigma_{\text{SA}}(z)$ is calculated from the particular case of a GA [\eqref{eq:Sigma_e_iint}] with $\ncpts=1$ (and therefore $\Delta\vec{n} = 0$, since there is only one coupling point); and $\Sigma_{\Delta\vec{n}}(z)$ denotes the contribution to the self-energy from the interference between coupling points that are spaced by $\Delta\vec{n}$.
For example, consider the diamond configuration shown in \figref{fig:subradiant_dynamics}. The self-energy of the atom in such a case, according to \eqref{eq:Sigma_e_SA_interf}, is
\begin{equation}
    \Sigma_\Diamond (z) = 4 \Sigma_\text{SA}(z) + 2 \mleft[4 \Sigma_{\tiny\mqty[1\\1]}(z) + 2 \Sigma_{\tiny\mqty[2\\0]}(z)\mright].
    \label{eq:Sigma_diamond}
\end{equation}

As shown in Ref.~\cite{Morita1971}, it is convenient to rewrite $\Sigma_\text{SA}$ and $\Sigma_{\Delta\vec{n}}$ in a different basis, such that instead of integrating in the $k_{x,y}$ horizontal and vertical directions, we integrate in the $q_\pm$ diagonal directions.
For that, we apply the following change of variables:
\begin{align}
q_\pm &= \frac{1}{2}(k_x \pm k_y),\\
\Delta n_\pm &= \Delta n_x \pm \Delta n_y.
\end{align}
Note that this implies $\dd[2]\vec{q} = \frac{1}{2}\dd[2]\vec{k}$ and that the integration area in the $k_{x,y}$ direction is twice the area of that in the $q_\pm$ direction.
Using the trigonometric expressions $\cos(\alpha\pm\beta) = \cos(\alpha)\cos(\beta) \mp \sin(\alpha)\sin(\beta)$ and $\cos(2\alpha) = 2\cos^2(\alpha)-1 = 1-2\sin^2(\alpha)$, we can show that
\begin{align}
\Sigma_{\text{SA}} (z) &= \frac{g^2}{(2\pi)^2}\iint_{-\pi}^\pi \dd[2]{\vec{k}} \frac{1}{z+2J[\cos(k_x)+\cos(k_y)]} \nonumber\\
&= \frac{g^2}{(2\pi)^2}\iint_{-\pi}^\pi \dd[2]{\vec{q}} \frac{1}{z+4J\cos(q_+)\cos(q_-)},\\
\Sigma_{\Delta\vec{n}} (z)&=  \frac{g^2}{(2\pi)^2}\iint_{-\pi}^\pi \dd[2]{\vec{k}}\; \frac{\cos(\vec{k}\Delta\vec{n})}{z+2J[\cos(k_x)+\cos(k_y)]} \nonumber\\
&= \frac{g^2}{(2\pi)^2} \iint_{-\pi}^{\pi} \dd[2]\vec{q}\;\frac{\cos(q_+\Delta n_+)\cos(q_-\Delta n_-)}{z+4J\cos(q_+)\cos(q_-)}.
\end{align}
In this basis, the expressions above can be integrated by parts and rewritten in a compact way in terms of elliptic integrals~\cite{Morita1971}.
For instance, from Ref.~\cite{Gonzalez-Tudela2017}:
\begin{align}
    \Sigma_{\text{SA}}(z) &= \frac{2g^2}{\pi z} K[m(z)],
    \label{eq:Sigma_SA} \\
    \Sigma_{\tiny\mqty[1\\1]}(z) &= \frac{2g^2}{\pi z} \mleft\{\mleft[\frac{2}{m(z)}-1 \mright]K[m(z)] - \frac{2}{m(z)} E[m(z)] \mright\}, \label{eq:Sigma_11}\\
     \Sigma_{\tiny\mqty[1\\0]}(z) &= \frac{g^2}{4J} - \frac{g^2}{2\pi J} K[m(z)], \label{eq:Sigma_10}
\end{align}
where 
\begin{equation}
    m(z) = \mleft(\frac{4J}{z}\mright)^2,
\end{equation}
and $K$ and $E$ are the complete elliptical integrals of the first and second kind, respectively:
\begin{align}
    K(m) &= \int_0^{\pi/2} \frac{\dd\phi}{\sqrt{1-m\sin^2(\phi)}}, \\
    E(m) &= \int_0^{\pi/2} \dd\phi \sqrt{1-m\sin^2(\phi)}.
\end{align}
Finally, using the recursive formulas in Ref.~\cite{Morita1971}, we can obtain the self-energy at any arbitrary site.
For example, using 
\begin{align}
    \Sigma_{\tiny\mqty[m+1\\0]}(z) = 
    -\frac{1}{2J} 
    \Big[2z \:\Sigma_{\tiny\mqty[m\\0]}(z) 
    &+ 2J\: \Sigma_{\tiny\mqty[m-1\\0]}(z) \nonumber\\
    &+ 4J\: \Sigma_{\tiny\mqty[m\\1]}(z)\Big],
\end{align}
together with \eqsref{eq:Sigma_SA}{eq:Sigma_11}, we can show that
\begin{align}
    \Sigma_{\tiny\mqty[2\\0]}(z) &= 
    -\frac{z}{J} \Sigma_{\tiny\mqty[1\\0]}(z) 
    - \underbrace{\Sigma_{\tiny\mqty[0\\0]}(z)}_{\Sigma_{\text{SA}}(z)} 
    -2 \Sigma_{\tiny\mqty[1\\1]}(z) \nonumber\\
    &= \frac{2g^2}{\pi z} \mleft(K[m(z)] + \frac{4}{m(z)} \mleft\{E[m(z)] - \frac{\pi}{2}\mright\}\mright). \label{eq:Sigma_20}
\end{align}

Going back to the diamond configuration [\eqref{eq:Sigma_diamond}], we have now derived an expression for the self-energy that is integrable in the first Riemann sheet, i.e., $\abs{z}>4J$.
Then, according to \eqref{eq:G_alpha}, the resolvent-operator element corresponding to the excited state of the atom is
\begin{equation}
    G_e(z) = \frac{1}{z-\detuning-\Sigma_e(z)},
    \label{eq:Green}
\end{equation}
with $\detuning$ the atom-cavity detuning.
Lastly, we can express the probability amplitude of an initially excited GA, for $t>0$, as follows:
\begin{equation}
    C_e(t) = -\frac{1}{2\pi i}\int_{-\infty}^{\infty} G_e(E + i 0^+) e^{-iEt} \,dE,
    \label{eq:C_e}
\end{equation}
i.e., as the Fourier transform of the retarded Green's function $G_e$.

\subsection{Integration contour of the probability amplitude}

\begin{figure}[t]
    \centering
    \includegraphics[width=\columnwidth]{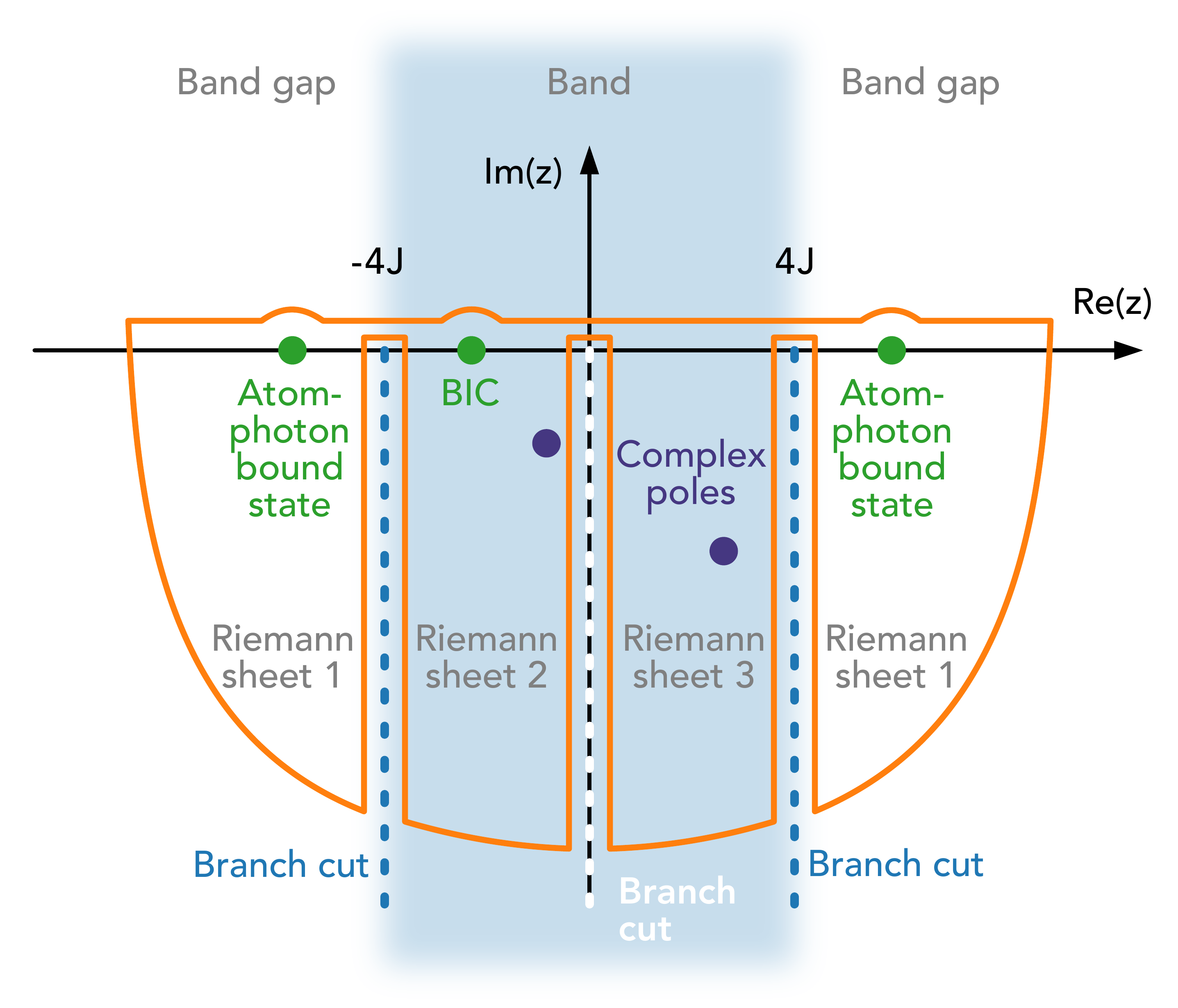}
    \caption{Contour of the integral in \eqref{eq:C_e}, with contributions from the poles of the Green's function [\eqref{eq:Green}], as well as the branch cuts at the band edges and at the middle of the band.}
    \label{fig:ContourIntegral}
\end{figure}

Similarly to the 1D coupled-cavity array, the energy dispersion of the 2D square lattice introduces branch cuts at the band edges, making the integral in \eqref{eq:C_e} contour around them (see \figref{fig:ContourIntegral}).
Moreover, an additional branch point that is not present in the 1D case arises in the middle of the band due to the additional van Hove singularity.
Therefore, the branch cuts divide the surface enclosed by the contour into three Riemann sheets:
the first Riemann sheet corresponds to the energy values outside the band ($\abs{z} > 4J$) and contains real poles of the Green's function [\eqref{eq:Green}], which are associated to the atom-photon bound states; while the second ($-4J<z<0$) and third ($0<z<4J$) Riemann sheets extend over the band and contain the complex poles of the Green's function, which are responsible for the spontaneous emission into the bath. 
The poles in the second and third Riemann sheets that are real are responsible for the existence of the bound states in the continuum (BICs). 

Essentially, this means that the atomic population $\abs{C_e(t)}^2$ is affected by two elements: detours around the branch cuts and poles of the Green's function.
In fact, as explained in Refs.~\cite{Gonzalez-Tudela2017, Soro2023}, the probability amplitude $C_e(t)$ can be calculated as a sum of the different contributions:
\begin{equation}
    C_e(t) = \sum_{\alpha\in\substack{\text{branch}\\[-0.5pt]\text{cuts}}} C_{\alpha} (t) + \sum_{\beta \in \text{poles}} R_{\beta} e^{-iz_\beta t},
    \label{eq:contributions}
\end{equation}
where $C_\alpha$ has the form of \eqref{eq:C_e}, and $R_\beta$ is the residue of the poles (real and complex) that we obtain through the residue theorem, which is the overlap of the initial wave function with the poles, i.e.,
\begin{equation}
    R_\beta = \eval{\frac{1}{1-\partial_z \Sigma_{e}(z)}}_{z=z_\beta}.
\end{equation}

The singularities that give rise to the branch cuts not only affect the contour of the probability-amplitude integral, but also limit the domain of definition of the self-energy.
In fact, the expressions given in \eqsref{eq:Sigma_SA}{eq:Sigma_10} and \eqref{eq:Sigma_20} are only valid for the first Riemann sheet.
The analytical continuation into the second and third Riemann sheets can be obtained by transforming the elliptic integrals in the following way~\cite{Gonzalez-Tudela2017}:
\begin{align}
K^{\rm II [III]} (m) &= K(m) \pm 2i K(1-m), \\
E^{\rm II [III]} (m) &= E(m) \pm 2i [K(1-m) -E(1-m)].
\end{align}

\section{Efficient computation of the time-evolution operator $U_A$ for giant atoms}
\label{app:numerics}

As outlined in \secref{sec:numerics}, the time evolution generated by the atom-interaction Hamiltonian
\begin{equation}
    H_A + \Hint = \begin{bmatrix}
        D & \Ga \\
        \GT & \mbb{0}
    \end{bmatrix}
\end{equation}
can be modelled efficiently by using $M$ different effective Hamiltonians $H_i$, each modeling the interaction between one of the atoms and the bath. In the GA case, the $D$ matrix still looks the same as in the small-atom case: $D_{ij} = \De_i \de_{ij}$. However, $\Ga$ looks slightly different:
\begin{equation}
    \qty[\Ga]_{in} = \begin{cases}
        g_{ip_n} & \text{if atom $i$ couples to cavity $n$},\\
        0 & \text{otherwise},
    \end{cases}
\end{equation}
where $p_n$ is the point index corresponding to the coupling point at cavity $n$.

In this appendix, we show the details of how this efficient modeling can be done by deriving \eqref{eq:UA_els} from how $H_A + \Hint$ behaves when exponentiated. For simplicity, we henceforth refer to $H_A + \Hint$ as $H$ in this appendix, since we will not need to refer to the full Hamiltonian $H_B + H_A + \Hint$ much. The reason that understanding the structure of $H^k$ is a good place to start is that by definition
\begin{equation}
    U_A(\De t) = e^{-i H \De t} = \sum_{k \geq 0} \frac{(-i \De t)^k}{k!} H^k,
\end{equation}
so knowing how $H^k$ looks should (and indeed does) enable us to find explicit expressions for the matrix elements of $U_A(\De t)$---namely, those of \eqref{eq:UA_els}.

Computing $H^k$ explicitly for $k = 0$ and $k=1$ is trivial: $H^0 = \mathds{1}$ and $H^1 = H$. For the first nontrivial case, $k=2$, we find that
\begin{equation}
    H^2 = \begin{bmatrix}
        D^2 + \Ga \GT & D \Ga \\
        \GT D & \GT \Ga
    \end{bmatrix}.
\end{equation}
Since $D$ is diagonal, so is $D^2$---specifically,
\begin{equation}
    \qty[D^2]_{ij} = \Delta_i^2 \de_{ij}.
\end{equation}
Examining the definition of $\Ga$ leads us to the conclusion that $\Ga \GT$ is also diagonal:
\begin{equation}
\begin{aligned}
    \qty[\Ga \GT]_{ij} &= \sum_{n=1}^{N^2} \Ga_{in} \Ga_{jn} = \\
    &= \qty{\text{no cavity couples to $> 1$ atom}} = \\
    &= \sum_{p=1}^{P_i} g_{ip}^2 \de_{ij},
\end{aligned}
\end{equation}
Introducing the notation $\Lm_k$ for the top-left (“atomic”) block of $H^k$, so that $\Lm_1 = D$ and $\Lm_0 = \mathds{1}$, we see that $\Lm_2$, the top-left block of $H^2$, is still diagonal, with elements
\begin{equation}
    \qty[\Lm_2]_{ij} = (\Delta_i^2 + G_i^2) \de_{ij},
\end{equation}
if we define the effective coupling strength
\begin{equation}
    G_i = \sqrt{ \sum_{p=1}^{P_i} g_{ip}^2 }.
\end{equation}
If we instead take a look at the top-right block, we see that since $D$ is diagonal,
\begin{equation}
    \qty[D \Ga]_{in} = \begin{cases}
        \Delta_i g_{ip_n} & \text{if atom $i$ couples to cavity $n$},\\
        0 & \text{otherwise}.
    \end{cases}
\end{equation}
Thus, the structure of this block also remains the same---squaring $H$ simply results in the non-zero matrix elements contained therein changing from $g_{ip_n}$ to $\Delta_i g_{ip_n}$. Since $H$ is symmetric (recalling the assumption we made for calculational convenience that all coupling strengths $g_{ip}$ are real), the bottom-left block is simply the transpose of the top-right. 

Finally, the bottom-right block of $H^2$ has matrix elements
\begin{equation}
\begin{aligned}
    \qty[\GT \Ga]_{mn} &= \sum_{i=1}^M \Ga_{im} \Ga_{in} = \\
    &= \qty{\text{no cavity couples to $> 1$ atom}} = \\
    &= \begin{cases}
        g_{ip_m} g_{ip_n} & \text{if atom $i$ couples cavities $m$ and $n$},\\
        0 & \text{otherwise}.
    \end{cases}
\end{aligned}
\end{equation}
Note that all of the elements in the four blocks involve only a single atom each, which means that the time evolution caused by $U_A$ to one atom can be treated independently from all the others, just as we noted in \secref{sec:numerics}.

Examining the case $k=3$, we find
\begin{equation}
\begin{aligned}
    H^3 &= \begin{bmatrix}
        D \Lm_2 + \Ga \GT \Lm_1 & \Lm_2 \Ga \\
        \GT \Lm_2 & \GT \Lm_1 \Ga
    \end{bmatrix} \\
    &= \begin{bmatrix}
        D^3 + 2 \Ga \GT D & \qty(D^2 + \Ga \GT) \Ga \\
        \GT \qty(D^2 + \Ga \GT) & \GT D \Ga
    \end{bmatrix}.
\end{aligned}    
\end{equation}
A pattern starts to emerge. The top-left block $\Lm_3 = D \Lm_2 + \Ga \GT \Lm_1$ is clearly still diagonal, since it is a linear combination of products of the diagonal matrices $D$ and $\Ga \GT$. Similarly, $\Lm_2 \Ga$ must have the same structure as $\Ga$, since it is the product of a diagonal matrix and $\Ga$, and the bottom-left block is still the transpose of the top-right one. Lastly, the structure of the bottom-right block also remains unchanged, since $D \Ga$ has the same structure as $\Ga$. Specifically,
it must be that $\GT D \Ga$ has the same structure as $\GT \Ga$, but with $g_{ip_m} g_{ip_n} \to \Delta_i g_{ip_m} g_{ip_n}$.

In fact, as can be shown via induction, we can write $H^k$ for any $k \geq 1$ as
\begin{equation}
    H^k = \begin{bmatrix}
        \Lambda_k & \Lambda_{k-1} \Ga \\
        \GT \Lambda_{k-1} & \GT \Lambda_{k-2} \Ga
    \end{bmatrix},
    \label{eq:Htothek}
\end{equation}
where $\qty{\Lambda_k}_{k=-1}^\infty$ is the sequence of diagonal matrices defined by the recursive formula
\begin{equation}
    \Lambda_{k+1} = D \Lambda_k + \Ga \GT \Lambda_{k-1}
    \label{eq:RecLambda}
\end{equation}
and the seed values
\begin{equation}
    \Lambda_{-1} = \mbb{0}, \quad \Lambda_0 = \mathds{1},
\end{equation}
where $\mbb{0}$ and $\mathds{1}$ are the $M \times M$ zero and identity matrices, respectively.

Thus, defining the diagonal matrices $D_{0,1,2}$ via
\begin{equation}
    D_l = \sum_{k \geq 1} \frac{\qty(-i \De t)^k}{k!} \Lm_{k-l},
\end{equation}
we find that
\begin{equation}
\begin{aligned}
    U_A(\De t) &= \mathds{1} + \begin{bmatrix}
        D_0 & D_1 \Ga \\
        \GT D_1 & \GT D_2 \Ga
    \end{bmatrix} \\
    &= \begin{bmatrix}
        \mathds{1}_M + D_0 & D_1 \Ga \\
        \GT D_1 & \mathds{1}_{N^2} + \GT D_2 \Ga
    \end{bmatrix},
    \label{eq:UA_and_Dl_appendix}
\end{aligned}
\end{equation}
which is where \eqref{eq:UA_and_Dl} comes from.

To derive \eqref{eq:UA_els} from this, we need to examine the elements of the $D_l$ matrices. Since both $D_l$ and $\Lm_k$ are diagonal, let us introduce the notation $d_{l,i} = \qty[D_l]_{ii}$ and $\lm_{k,i} = \qty[\Lm_k]_{ii}$ (no sum) for the $i$th element along their main diagonal, so that
\begin{equation}
    d_{l,i} =  \sum_{k \geq 1} \frac{\qty(-i \De t)^k}{k!} \lm_{k-l,i},
\end{equation}
where $\qty{\lm_{k,i}}_{k=-1}^{\infty}$ satisfies the recursive formula
\begin{equation}
    \lm_{k+1,i} = \De_i \lm_{k,i} + G_i^2 \lm_{k-1,i}
\end{equation}
with the seed values $\lm_{0,i} = 1$ and $\lm_{-1,i} = 0$.

Defining the effective Hamiltonian
\begin{equation}
    H_i = \begin{bmatrix}
        \De_i & G_i \\
        G_i & 0
    \end{bmatrix},
\end{equation}
we see that its associated evolution operator $U_i(\De t) = \exp( -i H_i \De t )$ must be equal to
\begin{equation}
\begin{aligned}
    U_i(\De t) &= \mathds{1} + \begin{bmatrix}
        d_{0,i} & G_i d_{1,i} \\
        G_i d_{1,i} & G_i^2 d_{2,i}
    \end{bmatrix} \\
    &= \begin{bmatrix}
        1 + d_{0,i} & G_i d_{1,i} \\
        G_i d_{1,i} & 1 + G_i^2 d_{2,i}
    \end{bmatrix}.
\end{aligned}
\end{equation}
This follows from the fact that $H_i$ can be seen as a special case of $H$ with $M,N = 1$, $D = \qty[ \De_i ]$ and $\Ga = \qty[ G_i ]$.

Comparing this to the elements of $U_A$, i.e.,
\begin{widetext}
\begin{equation}
\begin{dcases}
    \qty[\mathds{1}_M + D_0]_{ij} = \qty(1 + d_{0,i}) \de_{ij} \\
    \qty[D_1 \Ga]_{in} = \qty[\GT D_1]_{ni} = \begin{dcases}
        g_{i p_n} d_{1,i} & \text{if cavity $n$ couples to atom $i$ at point $p_n$} \\
        0 & \text{otherwise},
    \end{dcases} \\
    \qty[\mathds{1}_{N^d} + \GT D_2 \Ga]_{mn} = \begin{dcases}
        \de_{mn} + g_{i p_m} g_{i p_n} d_{2,i} & \text{if cavities $m,n$ couple to atom $i$ at $p_m, p_n$} \\
        \de_{mn} & \text{otherwise},
    \end{dcases}
\end{dcases}
\end{equation}
we finally arrive at equation \eqref{eq:UA_els}:
\begin{equation}
\begin{dcases}
    \qty[\mathds{1}_M + D_0]_{ij} = \de_{ij} \qty[{U}_i]_{11}, \\
    \qty[D_1 \Ga]_{in} = \qty[\GT D_1]_{ni} = \begin{dcases}
        \frac{g_{i p_n}}{G_i} \qty[{U}_i]_{12} & \text{if cavity $n$ couples to atom $i$ at point $p_n$} \\
        0 & \text{otherwise},
    \end{dcases} \\
    \qty[\mathds{1}_{N^d} + \GT D_2 \Ga]_{mn} = \begin{dcases}
        \de_{mn} + \frac{g_{i p_m} g_{i p_n}}{G_i^2} \qty( \qty[{U}_i]_{22} - 1 ) & \text{if cavities $m,n$ couple to atom $i$ at $p_m, p_n$} \\
        \de_{mn} & \text{otherwise}.
    \end{dcases} 
\end{dcases}
\label{eq:UA_els_appendix}
\end{equation}
\end{widetext}

It is possible to save a bit more computation time by precomputing $U_i$ analytically. Diagonalizing the $H_i$ matrix, we find that its eigenvalues are
\begin{equation}
    \kappa_{\pm,i} = \frac{\Delta_i}{2} \pm \sqrt{\frac{\Delta_i^2}{4} + G_i^2}.
\end{equation}
By examining the expression for $\qty[H_i^k]_{11} = \lm_{k,i}$ obtained from diagonalization one can then show that
\begin{equation}
    \lm_{k,i} = \frac{\ka_{+,i}^{k+1} - \ka_{-,i}^{k+1}}{\ka_{+,i} - \ka_{-,i}}.
\end{equation}
This in turn gives us an analytic expression for the elements of $D_l$, namely
\begin{equation}
\begin{aligned}
    d_{l,i} &= \frac{1}{\ka_{+} - \ka_{-}} \sum_{k=1}^\infty \frac{(-i \Delta t)^k}{k!}  \qty(\ka_{+}^{k-l+1} - \ka_{-}^{k-l+1}) \\
    &= \frac{\ka_{+}^{1-l} \qty( e^{-i \De t \ka_+} - 1 ) - \ka_{-}^{1-l} \qty( e^{-i \De t \ka_-} - 1 )}{\ka_{+} - \ka_{-}},
\end{aligned}
\end{equation}
where we have suppressed the $i$ index on $\ka_{\pm,i}$.
Inserting these findings into our expressions for the elements of $U_i$ then gives us
\begin{equation}
\begin{aligned}
    \qty[U_i]_{11} &= \frac{\ka_{+} e^{-i \De t \ka_+} - \ka_{-} e^{-i \De t \ka_-}}{\ka_{+} - \ka_{-}}, \\
    \qty[U_i]_{12} = \qty[U_i]_{21}  &= G_i \frac{ e^{-i \Delta t \ka_{+}} - e^{-i \Delta t \ka_{-}} }{\ka_{+} - \ka_{-}}, \\
    \qty[U_i]_{22} &= \frac{\ka_{+} e^{-i \Delta t \ka_{-}} - \ka_{-} e^{-i \Delta t \ka_{+}}}{\ka_{+} - \ka_{-}}.
    \label{eq:Ui_precomp}
\end{aligned}
\end{equation}

It should be noted that while it is slightly faster to calculate $d_{l,i}$ and $U_i$ by inserting all the relevant quantities into these expressions rather than by numerically computing the $2 \times 2$ matrix exponential $\exp(-i H_i \Delta t)$ [at least with SciPy's \texttt{linalg.expm()}], the effect of doing so on the speed of the overall time-evolution algorithm is negligible. This is because, as discussed in \secref{sec:numerics}, the FFT and iFFT performed every time step are where the overwhelming majority of the computational cost of the algorithm comes from.
However, calculating the elements of $U_i$ using \eqref{eq:Ui_precomp} tends to give slightly better unitarity than that achieved by naively exponentiating $-i H_i \Delta t$ using SciPy's \texttt{linalg.expm()}.


\bibliography{References.bib}

\begin{thebibliography}{124}%
\makeatletter
\providecommand \@ifxundefined [1]{%
 \@ifx{#1\undefined}
}%
\providecommand \@ifnum [1]{%
 \ifnum #1\expandafter \@firstoftwo
 \else \expandafter \@secondoftwo
 \fi
}%
\providecommand \@ifx [1]{%
 \ifx #1\expandafter \@firstoftwo
 \else \expandafter \@secondoftwo
 \fi
}%
\providecommand \natexlab [1]{#1}%
\providecommand \enquote  [1]{``#1''}%
\providecommand \bibnamefont  [1]{#1}%
\providecommand \bibfnamefont [1]{#1}%
\providecommand \citenamefont [1]{#1}%
\providecommand \href@noop [0]{\@secondoftwo}%
\providecommand \href [0]{\begingroup \@sanitize@url \@href}%
\providecommand \@href[1]{\@@startlink{#1}\@@href}%
\providecommand \@@href[1]{\endgroup#1\@@endlink}%
\providecommand \@sanitize@url [0]{\catcode `\\12\catcode `\$12\catcode
  `\&12\catcode `\#12\catcode `\^12\catcode `\_12\catcode `\%12\relax}%
\providecommand \@@startlink[1]{}%
\providecommand \@@endlink[0]{}%
\providecommand \url  [0]{\begingroup\@sanitize@url \@url }%
\providecommand \@url [1]{\endgroup\@href {#1}{\urlprefix }}%
\providecommand \urlprefix  [0]{URL }%
\providecommand \Eprint [0]{\href }%
\providecommand \doibase [0]{http://dx.doi.org/}%
\providecommand \selectlanguage [0]{\@gobble}%
\providecommand \bibinfo  [0]{\@secondoftwo}%
\providecommand \bibfield  [0]{\@secondoftwo}%
\providecommand \translation [1]{[#1]}%
\providecommand \BibitemOpen [0]{}%
\providecommand \bibitemStop [0]{}%
\providecommand \bibitemNoStop [0]{.\EOS\space}%
\providecommand \EOS [0]{\spacefactor3000\relax}%
\providecommand \BibitemShut  [1]{\csname bibitem#1\endcsname}%
\let\auto@bib@innerbib\@empty
\bibitem [{\citenamefont {{Frisk Kockum}}(2021)}]{FriskKockum2021}%
  \BibitemOpen
  \bibfield  {author} {\bibinfo {author} {\bibfnamefont {A.}~\bibnamefont
  {{Frisk Kockum}}},\ }\bibfield  {title} {\enquote {\bibinfo {title} {{Quantum
  Optics with Giant Atoms---the First Five Years}},}\ }in\ \href {\doibase
  10.1007/978-981-15-5191-8_12} {\emph {\bibinfo {booktitle} {International
  Symposium on Mathematics, Quantum Theory, and Cryptography}}},\ \bibinfo
  {editor} {edited by\ \bibinfo {editor} {\bibfnamefont {T.}~\bibnamefont
  {Takagi}}, \bibinfo {editor} {\bibfnamefont {M.}~\bibnamefont {Wakayama}},
  \bibinfo {editor} {\bibfnamefont {K.}~\bibnamefont {Tanaka}}, \bibinfo
  {editor} {\bibfnamefont {N.}~\bibnamefont {Kunihiro}}, \bibinfo {editor}
  {\bibfnamefont {K.}~\bibnamefont {Kimoto}}, \ and\ \bibinfo {editor}
  {\bibfnamefont {Y.}~\bibnamefont {Ikematsu}}}\ (\bibinfo  {publisher}
  {Springer Singapore},\ \bibinfo {address} {Singapore},\ \bibinfo {year}
  {2021})\ pp.\ \bibinfo {pages} {125--146}\BibitemShut {NoStop}%
\bibitem [{\citenamefont {{Frisk Kockum}}\ \emph {et~al.}(2014)\citenamefont
  {{Frisk Kockum}}, \citenamefont {Delsing},\ and\ \citenamefont
  {Johansson}}]{Kockum2014}%
  \BibitemOpen
  \bibfield  {author} {\bibinfo {author} {\bibfnamefont {A.}~\bibnamefont
  {{Frisk Kockum}}}, \bibinfo {author} {\bibfnamefont {P.}~\bibnamefont
  {Delsing}}, \ and\ \bibinfo {author} {\bibfnamefont {G.}~\bibnamefont
  {Johansson}},\ }\bibfield  {title} {\enquote {\bibinfo {title} {{Designing
  frequency-dependent relaxation rates and Lamb shifts for a giant artificial
  atom}},}\ }\href {\doibase 10.1103/PhysRevA.90.013837} {\bibfield  {journal}
  {\bibinfo  {journal} {Physical Review A}\ }\textbf {\bibinfo {volume} {90}},\
  \bibinfo {pages} {013837} (\bibinfo {year} {2014})}\BibitemShut {NoStop}%
\bibitem [{\citenamefont {Vadiraj}\ \emph {et~al.}(2021)\citenamefont
  {Vadiraj}, \citenamefont {Ask}, \citenamefont {McConkey}, \citenamefont
  {Nsanzineza}, \citenamefont {Chang}, \citenamefont {Kockum},\ and\
  \citenamefont {Wilson}}]{Vadiraj2021}%
  \BibitemOpen
  \bibfield  {author} {\bibinfo {author} {\bibfnamefont {A.~M.}\ \bibnamefont
  {Vadiraj}}, \bibinfo {author} {\bibfnamefont {A.}~\bibnamefont {Ask}},
  \bibinfo {author} {\bibfnamefont {T.~G.}\ \bibnamefont {McConkey}}, \bibinfo
  {author} {\bibfnamefont {I.}~\bibnamefont {Nsanzineza}}, \bibinfo {author}
  {\bibfnamefont {C.~W.~S.}\ \bibnamefont {Chang}}, \bibinfo {author}
  {\bibfnamefont {A.~F.}\ \bibnamefont {Kockum}}, \ and\ \bibinfo {author}
  {\bibfnamefont {C.~M.}\ \bibnamefont {Wilson}},\ }\bibfield  {title}
  {\enquote {\bibinfo {title} {{Engineering the level structure of a giant
  artificial atom in waveguide quantum electrodynamics}},}\ }\href {\doibase
  10.1103/PhysRevA.103.023710} {\bibfield  {journal} {\bibinfo  {journal}
  {Physical Review A}\ }\textbf {\bibinfo {volume} {103}},\ \bibinfo {pages}
  {023710} (\bibinfo {year} {2021})}\BibitemShut {NoStop}%
\bibitem [{\citenamefont {{Frisk Kockum}}\ \emph {et~al.}(2018)\citenamefont
  {{Frisk Kockum}}, \citenamefont {Johansson},\ and\ \citenamefont
  {Nori}}]{FriskKockum2018a}%
  \BibitemOpen
  \bibfield  {author} {\bibinfo {author} {\bibfnamefont {A.}~\bibnamefont
  {{Frisk Kockum}}}, \bibinfo {author} {\bibfnamefont {G.}~\bibnamefont
  {Johansson}}, \ and\ \bibinfo {author} {\bibfnamefont {F.}~\bibnamefont
  {Nori}},\ }\bibfield  {title} {\enquote {\bibinfo {title} {{Decoherence-Free
  Interaction between Giant Atoms in Waveguide Quantum Electrodynamics}},}\
  }\href {\doibase 10.1103/PhysRevLett.120.140404} {\bibfield  {journal}
  {\bibinfo  {journal} {Physical Review Letters}\ }\textbf {\bibinfo {volume}
  {120}},\ \bibinfo {pages} {140404} (\bibinfo {year} {2018})}\BibitemShut
  {NoStop}%
\bibitem [{\citenamefont {Kannan}\ \emph {et~al.}(2020)\citenamefont {Kannan},
  \citenamefont {Ruckriegel}, \citenamefont {Campbell}, \citenamefont {{Frisk
  Kockum}}, \citenamefont {Braum{\"{u}}ller}, \citenamefont {Kim},
  \citenamefont {Kjaergaard}, \citenamefont {Krantz}, \citenamefont {Melville},
  \citenamefont {Niedzielski}, \citenamefont {Veps{\"{a}}l{\"{a}}inen},
  \citenamefont {Winik}, \citenamefont {Yoder}, \citenamefont {Nori},
  \citenamefont {Orlando}, \citenamefont {Gustavsson},\ and\ \citenamefont
  {Oliver}}]{Kannan2020}%
  \BibitemOpen
  \bibfield  {author} {\bibinfo {author} {\bibfnamefont {B.}~\bibnamefont
  {Kannan}}, \bibinfo {author} {\bibfnamefont {M.~J.}\ \bibnamefont
  {Ruckriegel}}, \bibinfo {author} {\bibfnamefont {D.~L.}\ \bibnamefont
  {Campbell}}, \bibinfo {author} {\bibfnamefont {A.}~\bibnamefont {{Frisk
  Kockum}}}, \bibinfo {author} {\bibfnamefont {J.}~\bibnamefont
  {Braum{\"{u}}ller}}, \bibinfo {author} {\bibfnamefont {D.~K.}\ \bibnamefont
  {Kim}}, \bibinfo {author} {\bibfnamefont {M.}~\bibnamefont {Kjaergaard}},
  \bibinfo {author} {\bibfnamefont {P.}~\bibnamefont {Krantz}}, \bibinfo
  {author} {\bibfnamefont {A.}~\bibnamefont {Melville}}, \bibinfo {author}
  {\bibfnamefont {B.~M.}\ \bibnamefont {Niedzielski}}, \bibinfo {author}
  {\bibfnamefont {A.}~\bibnamefont {Veps{\"{a}}l{\"{a}}inen}}, \bibinfo
  {author} {\bibfnamefont {R.}~\bibnamefont {Winik}}, \bibinfo {author}
  {\bibfnamefont {J.~L.}\ \bibnamefont {Yoder}}, \bibinfo {author}
  {\bibfnamefont {F.}~\bibnamefont {Nori}}, \bibinfo {author} {\bibfnamefont
  {T.~P.}\ \bibnamefont {Orlando}}, \bibinfo {author} {\bibfnamefont
  {S.}~\bibnamefont {Gustavsson}}, \ and\ \bibinfo {author} {\bibfnamefont
  {W.~D.}\ \bibnamefont {Oliver}},\ }\bibfield  {title} {\enquote {\bibinfo
  {title} {{Waveguide quantum electrodynamics with superconducting artificial
  giant atoms}},}\ }\href {\doibase 10.1038/s41586-020-2529-9} {\bibfield
  {journal} {\bibinfo  {journal} {Nature}\ }\textbf {\bibinfo {volume} {583}},\
  \bibinfo {pages} {775} (\bibinfo {year} {2020})}\BibitemShut {NoStop}%
\bibitem [{\citenamefont {Carollo}\ \emph {et~al.}(2020)\citenamefont
  {Carollo}, \citenamefont {Cilluffo},\ and\ \citenamefont
  {Ciccarello}}]{Carollo2020}%
  \BibitemOpen
  \bibfield  {author} {\bibinfo {author} {\bibfnamefont {A.}~\bibnamefont
  {Carollo}}, \bibinfo {author} {\bibfnamefont {D.}~\bibnamefont {Cilluffo}}, \
  and\ \bibinfo {author} {\bibfnamefont {F.}~\bibnamefont {Ciccarello}},\
  }\bibfield  {title} {\enquote {\bibinfo {title} {{Mechanism of
  decoherence-free coupling between giant atoms}},}\ }\href {\doibase
  10.1103/PhysRevResearch.2.043184} {\bibfield  {journal} {\bibinfo  {journal}
  {Physical Review Research}\ }\textbf {\bibinfo {volume} {2}},\ \bibinfo
  {pages} {043184} (\bibinfo {year} {2020})}\BibitemShut {NoStop}%
\bibitem [{\citenamefont {Soro}\ and\ \citenamefont {Kockum}(2022)}]{Soro2022}%
  \BibitemOpen
  \bibfield  {author} {\bibinfo {author} {\bibfnamefont {A.}~\bibnamefont
  {Soro}}\ and\ \bibinfo {author} {\bibfnamefont {A.~F.}\ \bibnamefont
  {Kockum}},\ }\bibfield  {title} {\enquote {\bibinfo {title} {{Chiral quantum
  optics with giant atoms}},}\ }\href {\doibase 10.1103/PhysRevA.105.023712}
  {\bibfield  {journal} {\bibinfo  {journal} {Physical Review A}\ }\textbf
  {\bibinfo {volume} {105}},\ \bibinfo {pages} {023712} (\bibinfo {year}
  {2022})}\BibitemShut {NoStop}%
\bibitem [{\citenamefont {Soro}\ \emph {et~al.}(2023)\citenamefont {Soro},
  \citenamefont {Mu{\~{n}}oz},\ and\ \citenamefont {Kockum}}]{Soro2023}%
  \BibitemOpen
  \bibfield  {author} {\bibinfo {author} {\bibfnamefont {A.}~\bibnamefont
  {Soro}}, \bibinfo {author} {\bibfnamefont {C.~S.}\ \bibnamefont
  {Mu{\~{n}}oz}}, \ and\ \bibinfo {author} {\bibfnamefont {A.~F.}\ \bibnamefont
  {Kockum}},\ }\bibfield  {title} {\enquote {\bibinfo {title} {{Interaction
  between giant atoms in a one-dimensional structured environment}},}\ }\href
  {\doibase 10.1103/PhysRevA.107.013710} {\bibfield  {journal} {\bibinfo
  {journal} {Physical Review A}\ }\textbf {\bibinfo {volume} {107}},\ \bibinfo
  {pages} {013710} (\bibinfo {year} {2023})}\BibitemShut {NoStop}%
\bibitem [{\citenamefont {Du}\ \emph {et~al.}(2023{\natexlab{a}})\citenamefont
  {Du}, \citenamefont {Guo},\ and\ \citenamefont {Li}}]{Du2023b}%
  \BibitemOpen
  \bibfield  {author} {\bibinfo {author} {\bibfnamefont {L.}~\bibnamefont
  {Du}}, \bibinfo {author} {\bibfnamefont {L.}~\bibnamefont {Guo}}, \ and\
  \bibinfo {author} {\bibfnamefont {Y.}~\bibnamefont {Li}},\ }\bibfield
  {title} {\enquote {\bibinfo {title} {{Complex decoherence-free interactions
  between giant atoms}},}\ }\href {\doibase 10.1103/PhysRevA.107.023705}
  {\bibfield  {journal} {\bibinfo  {journal} {Physical Review A}\ }\textbf
  {\bibinfo {volume} {107}},\ \bibinfo {pages} {023705} (\bibinfo {year}
  {2023}{\natexlab{a}})}\BibitemShut {NoStop}%
\bibitem [{\citenamefont {Guo}\ \emph {et~al.}(2020{\natexlab{a}})\citenamefont
  {Guo}, \citenamefont {Kockum}, \citenamefont {Marquardt},\ and\ \citenamefont
  {Johansson}}]{Guo2020}%
  \BibitemOpen
  \bibfield  {author} {\bibinfo {author} {\bibfnamefont {L.}~\bibnamefont
  {Guo}}, \bibinfo {author} {\bibfnamefont {A.~F.}\ \bibnamefont {Kockum}},
  \bibinfo {author} {\bibfnamefont {F.}~\bibnamefont {Marquardt}}, \ and\
  \bibinfo {author} {\bibfnamefont {G.}~\bibnamefont {Johansson}},\ }\bibfield
  {title} {\enquote {\bibinfo {title} {{Oscillating bound states for a giant
  atom}},}\ }\href {\doibase 10.1103/PhysRevResearch.2.043014} {\bibfield
  {journal} {\bibinfo  {journal} {Physical Review Research}\ }\textbf {\bibinfo
  {volume} {2}},\ \bibinfo {pages} {043014} (\bibinfo {year}
  {2020}{\natexlab{a}})}\BibitemShut {NoStop}%
\bibitem [{\citenamefont {Guo}\ \emph {et~al.}(2020{\natexlab{b}})\citenamefont
  {Guo}, \citenamefont {Wang}, \citenamefont {Purdy},\ and\ \citenamefont
  {Taylor}}]{Guo2020a}%
  \BibitemOpen
  \bibfield  {author} {\bibinfo {author} {\bibfnamefont {S.}~\bibnamefont
  {Guo}}, \bibinfo {author} {\bibfnamefont {Y.}~\bibnamefont {Wang}}, \bibinfo
  {author} {\bibfnamefont {T.}~\bibnamefont {Purdy}}, \ and\ \bibinfo {author}
  {\bibfnamefont {J.}~\bibnamefont {Taylor}},\ }\bibfield  {title} {\enquote
  {\bibinfo {title} {{Beyond spontaneous emission: Giant atom bounded in the
  continuum}},}\ }\href {\doibase 10.1103/PhysRevA.102.033706} {\bibfield
  {journal} {\bibinfo  {journal} {Physical Review A}\ }\textbf {\bibinfo
  {volume} {102}},\ \bibinfo {pages} {033706} (\bibinfo {year}
  {2020}{\natexlab{b}})}\BibitemShut {NoStop}%
\bibitem [{\citenamefont {Terradas-Brians{\'{o}}}\ \emph
  {et~al.}(2022)\citenamefont {Terradas-Brians{\'{o}}}, \citenamefont
  {Gonz{\'{a}}lez-Guti{\'{e}}rrez}, \citenamefont {Nori}, \citenamefont
  {Mart{\'{i}}n-Moreno},\ and\ \citenamefont {Zueco}}]{Terradas-Brianso2022}%
  \BibitemOpen
  \bibfield  {author} {\bibinfo {author} {\bibfnamefont {S.}~\bibnamefont
  {Terradas-Brians{\'{o}}}}, \bibinfo {author} {\bibfnamefont {C.~A.}\
  \bibnamefont {Gonz{\'{a}}lez-Guti{\'{e}}rrez}}, \bibinfo {author}
  {\bibfnamefont {F.}~\bibnamefont {Nori}}, \bibinfo {author} {\bibfnamefont
  {L.}~\bibnamefont {Mart{\'{i}}n-Moreno}}, \ and\ \bibinfo {author}
  {\bibfnamefont {D.}~\bibnamefont {Zueco}},\ }\bibfield  {title} {\enquote
  {\bibinfo {title} {{Ultrastrong waveguide QED with giant atoms}},}\ }\href
  {\doibase 10.1103/PhysRevA.106.063717} {\bibfield  {journal} {\bibinfo
  {journal} {Physical Review A}\ }\textbf {\bibinfo {volume} {106}},\ \bibinfo
  {pages} {063717} (\bibinfo {year} {2022})}\BibitemShut {NoStop}%
\bibitem [{\citenamefont {Noachtar}\ \emph {et~al.}(2022)\citenamefont
  {Noachtar}, \citenamefont {Kn{\"{o}}rzer},\ and\ \citenamefont
  {Jonsson}}]{Noachtar2022}%
  \BibitemOpen
  \bibfield  {author} {\bibinfo {author} {\bibfnamefont {D.~D.}\ \bibnamefont
  {Noachtar}}, \bibinfo {author} {\bibfnamefont {J.}~\bibnamefont
  {Kn{\"{o}}rzer}}, \ and\ \bibinfo {author} {\bibfnamefont {R.~H.}\
  \bibnamefont {Jonsson}},\ }\bibfield  {title} {\enquote {\bibinfo {title}
  {{Nonperturbative treatment of giant atoms using chain transformations}},}\
  }\href {\doibase 10.1103/PhysRevA.106.013702} {\bibfield  {journal} {\bibinfo
   {journal} {Physical Review A}\ }\textbf {\bibinfo {volume} {106}},\ \bibinfo
  {pages} {013702} (\bibinfo {year} {2022})}\BibitemShut {NoStop}%
\bibitem [{\citenamefont {Lim}\ \emph {et~al.}(2023)\citenamefont {Lim},
  \citenamefont {Mok},\ and\ \citenamefont {Kwek}}]{Lim2023}%
  \BibitemOpen
  \bibfield  {author} {\bibinfo {author} {\bibfnamefont {K.~H.}\ \bibnamefont
  {Lim}}, \bibinfo {author} {\bibfnamefont {W.~K.}\ \bibnamefont {Mok}}, \ and\
  \bibinfo {author} {\bibfnamefont {L.~C.}\ \bibnamefont {Kwek}},\ }\bibfield
  {title} {\enquote {\bibinfo {title} {{Oscillating bound states in
  non-Markovian photonic lattices}},}\ }\href {\doibase
  10.1103/PhysRevA.107.023716} {\bibfield  {journal} {\bibinfo  {journal}
  {Physical Review A}\ }\textbf {\bibinfo {volume} {107}},\ \bibinfo {pages}
  {023716} (\bibinfo {year} {2023})}\BibitemShut {NoStop}%
\bibitem [{\citenamefont {Gustafsson}\ \emph {et~al.}(2014)\citenamefont
  {Gustafsson}, \citenamefont {Aref}, \citenamefont {Kockum}, \citenamefont
  {Ekstr{\"{o}}m}, \citenamefont {Johansson},\ and\ \citenamefont
  {Delsing}}]{Gustafsson14}%
  \BibitemOpen
  \bibfield  {author} {\bibinfo {author} {\bibfnamefont {M.~V.}\ \bibnamefont
  {Gustafsson}}, \bibinfo {author} {\bibfnamefont {T.}~\bibnamefont {Aref}},
  \bibinfo {author} {\bibfnamefont {A.~F.}\ \bibnamefont {Kockum}}, \bibinfo
  {author} {\bibfnamefont {M.~K.}\ \bibnamefont {Ekstr{\"{o}}m}}, \bibinfo
  {author} {\bibfnamefont {G.}~\bibnamefont {Johansson}}, \ and\ \bibinfo
  {author} {\bibfnamefont {P.}~\bibnamefont {Delsing}},\ }\bibfield  {title}
  {\enquote {\bibinfo {title} {{Propagating phonons coupled to an artificial
  atom}},}\ }\href {\doibase 10.1126/science.1257219} {\bibfield  {journal}
  {\bibinfo  {journal} {Science}\ }\textbf {\bibinfo {volume} {346}},\ \bibinfo
  {pages} {207} (\bibinfo {year} {2014})}\BibitemShut {NoStop}%
\bibitem [{\citenamefont {Aref}\ \emph {et~al.}(2016)\citenamefont {Aref},
  \citenamefont {Delsing}, \citenamefont {Ekstr{\"{o}}m}, \citenamefont
  {Kockum}, \citenamefont {Gustafsson}, \citenamefont {Johansson},
  \citenamefont {Leek}, \citenamefont {Magnusson},\ and\ \citenamefont
  {Manenti}}]{Aref2016}%
  \BibitemOpen
  \bibfield  {author} {\bibinfo {author} {\bibfnamefont {T.}~\bibnamefont
  {Aref}}, \bibinfo {author} {\bibfnamefont {P.}~\bibnamefont {Delsing}},
  \bibinfo {author} {\bibfnamefont {M.~K.}\ \bibnamefont {Ekstr{\"{o}}m}},
  \bibinfo {author} {\bibfnamefont {A.~F.}\ \bibnamefont {Kockum}}, \bibinfo
  {author} {\bibfnamefont {M.~V.}\ \bibnamefont {Gustafsson}}, \bibinfo
  {author} {\bibfnamefont {G.}~\bibnamefont {Johansson}}, \bibinfo {author}
  {\bibfnamefont {P.~J.}\ \bibnamefont {Leek}}, \bibinfo {author}
  {\bibfnamefont {E.}~\bibnamefont {Magnusson}}, \ and\ \bibinfo {author}
  {\bibfnamefont {R.}~\bibnamefont {Manenti}},\ }\bibfield  {title} {\enquote
  {\bibinfo {title} {{Quantum Acoustics with Surface Acoustic Waves}},}\ }in\
  \href {\doibase 10.1007/978-3-319-24091-6_9} {\emph {\bibinfo {booktitle}
  {Superconducting Devices in Quantum Optics}}},\ \bibinfo {editor} {edited by\
  \bibinfo {editor} {\bibfnamefont {R.~H.}\ \bibnamefont {Hadfield}}\ and\
  \bibinfo {editor} {\bibfnamefont {G.}~\bibnamefont {Johansson}}}\ (\bibinfo
  {publisher} {Springer},\ \bibinfo {year} {2016})\BibitemShut {NoStop}%
\bibitem [{\citenamefont {Manenti}\ \emph {et~al.}(2017)\citenamefont
  {Manenti}, \citenamefont {Kockum}, \citenamefont {Patterson}, \citenamefont
  {Behrle}, \citenamefont {Rahamim}, \citenamefont {Tancredi}, \citenamefont
  {Nori},\ and\ \citenamefont {Leek}}]{Manenti2017}%
  \BibitemOpen
  \bibfield  {author} {\bibinfo {author} {\bibfnamefont {R.}~\bibnamefont
  {Manenti}}, \bibinfo {author} {\bibfnamefont {A.~F.}\ \bibnamefont {Kockum}},
  \bibinfo {author} {\bibfnamefont {A.}~\bibnamefont {Patterson}}, \bibinfo
  {author} {\bibfnamefont {T.}~\bibnamefont {Behrle}}, \bibinfo {author}
  {\bibfnamefont {J.}~\bibnamefont {Rahamim}}, \bibinfo {author} {\bibfnamefont
  {G.}~\bibnamefont {Tancredi}}, \bibinfo {author} {\bibfnamefont
  {F.}~\bibnamefont {Nori}}, \ and\ \bibinfo {author} {\bibfnamefont {P.~J.}\
  \bibnamefont {Leek}},\ }\bibfield  {title} {\enquote {\bibinfo {title}
  {{Circuit quantum acoustodynamics with surface acoustic waves}},}\ }\href
  {\doibase 10.1038/s41467-017-01063-9} {\bibfield  {journal} {\bibinfo
  {journal} {Nature Communications}\ }\textbf {\bibinfo {volume} {8}},\
  \bibinfo {pages} {975} (\bibinfo {year} {2017})}\BibitemShut {NoStop}%
\bibitem [{\citenamefont {Noguchi}\ \emph {et~al.}(2017)\citenamefont
  {Noguchi}, \citenamefont {Yamazaki}, \citenamefont {Tabuchi},\ and\
  \citenamefont {Nakamura}}]{Noguchi2017}%
  \BibitemOpen
  \bibfield  {author} {\bibinfo {author} {\bibfnamefont {A.}~\bibnamefont
  {Noguchi}}, \bibinfo {author} {\bibfnamefont {R.}~\bibnamefont {Yamazaki}},
  \bibinfo {author} {\bibfnamefont {Y.}~\bibnamefont {Tabuchi}}, \ and\
  \bibinfo {author} {\bibfnamefont {Y.}~\bibnamefont {Nakamura}},\ }\bibfield
  {title} {\enquote {\bibinfo {title} {{Qubit-Assisted Transduction for a
  Detection of Surface Acoustic Waves near the Quantum Limit}},}\ }\href
  {\doibase 10.1103/PhysRevLett.119.180505} {\bibfield  {journal} {\bibinfo
  {journal} {Physical Review Letters}\ }\textbf {\bibinfo {volume} {119}},\
  \bibinfo {pages} {180505} (\bibinfo {year} {2017})}\BibitemShut {NoStop}%
\bibitem [{\citenamefont {Satzinger}\ \emph {et~al.}(2018)\citenamefont
  {Satzinger}, \citenamefont {Zhong}, \citenamefont {Chang}, \citenamefont
  {Peairs}, \citenamefont {Bienfait}, \citenamefont {Chou}, \citenamefont
  {Cleland}, \citenamefont {Conner}, \citenamefont {Dumur}, \citenamefont
  {Grebel}, \citenamefont {Gutierrez}, \citenamefont {November}, \citenamefont
  {Povey}, \citenamefont {Whiteley}, \citenamefont {Awschalom}, \citenamefont
  {Schuster},\ and\ \citenamefont {Cleland}}]{Satzinger2018}%
  \BibitemOpen
  \bibfield  {author} {\bibinfo {author} {\bibfnamefont {K.~J.}\ \bibnamefont
  {Satzinger}}, \bibinfo {author} {\bibfnamefont {Y.~P.}\ \bibnamefont
  {Zhong}}, \bibinfo {author} {\bibfnamefont {H.-S.}\ \bibnamefont {Chang}},
  \bibinfo {author} {\bibfnamefont {G.~A.}\ \bibnamefont {Peairs}}, \bibinfo
  {author} {\bibfnamefont {A.}~\bibnamefont {Bienfait}}, \bibinfo {author}
  {\bibfnamefont {M.-H.}\ \bibnamefont {Chou}}, \bibinfo {author}
  {\bibfnamefont {A.~Y.}\ \bibnamefont {Cleland}}, \bibinfo {author}
  {\bibfnamefont {C.~R.}\ \bibnamefont {Conner}}, \bibinfo {author}
  {\bibfnamefont {{\'{E}}.}~\bibnamefont {Dumur}}, \bibinfo {author}
  {\bibfnamefont {J.}~\bibnamefont {Grebel}}, \bibinfo {author} {\bibfnamefont
  {I.}~\bibnamefont {Gutierrez}}, \bibinfo {author} {\bibfnamefont {B.~H.}\
  \bibnamefont {November}}, \bibinfo {author} {\bibfnamefont {R.~G.}\
  \bibnamefont {Povey}}, \bibinfo {author} {\bibfnamefont {S.~J.}\ \bibnamefont
  {Whiteley}}, \bibinfo {author} {\bibfnamefont {D.~D.}\ \bibnamefont
  {Awschalom}}, \bibinfo {author} {\bibfnamefont {D.~I.}\ \bibnamefont
  {Schuster}}, \ and\ \bibinfo {author} {\bibfnamefont {A.~N.}\ \bibnamefont
  {Cleland}},\ }\bibfield  {title} {\enquote {\bibinfo {title} {{Quantum
  control of surface acoustic-wave phonons}},}\ }\href {\doibase
  10.1038/s41586-018-0719-5} {\bibfield  {journal} {\bibinfo  {journal}
  {Nature}\ }\textbf {\bibinfo {volume} {563}},\ \bibinfo {pages} {661}
  (\bibinfo {year} {2018})}\BibitemShut {NoStop}%
\bibitem [{\citenamefont {Moores}\ \emph {et~al.}(2018)\citenamefont {Moores},
  \citenamefont {Sletten}, \citenamefont {Viennot},\ and\ \citenamefont
  {Lehnert}}]{Moores2018}%
  \BibitemOpen
  \bibfield  {author} {\bibinfo {author} {\bibfnamefont {B.~A.}\ \bibnamefont
  {Moores}}, \bibinfo {author} {\bibfnamefont {L.~R.}\ \bibnamefont {Sletten}},
  \bibinfo {author} {\bibfnamefont {J.~J.}\ \bibnamefont {Viennot}}, \ and\
  \bibinfo {author} {\bibfnamefont {K.~W.}\ \bibnamefont {Lehnert}},\
  }\bibfield  {title} {\enquote {\bibinfo {title} {{Cavity Quantum Acoustic
  Device in the Multimode Strong Coupling Regime}},}\ }\href {\doibase
  10.1103/PhysRevLett.120.227701} {\bibfield  {journal} {\bibinfo  {journal}
  {Physical Review Letters}\ }\textbf {\bibinfo {volume} {120}},\ \bibinfo
  {pages} {227701} (\bibinfo {year} {2018})}\BibitemShut {NoStop}%
\bibitem [{\citenamefont {Bolgar}\ \emph {et~al.}(2018)\citenamefont {Bolgar},
  \citenamefont {Zotova}, \citenamefont {Kirichenko}, \citenamefont {Besedin},
  \citenamefont {Semenov}, \citenamefont {Shaikhaidarov},\ and\ \citenamefont
  {Astafiev}}]{Bolgar2018}%
  \BibitemOpen
  \bibfield  {author} {\bibinfo {author} {\bibfnamefont {A.~N.}\ \bibnamefont
  {Bolgar}}, \bibinfo {author} {\bibfnamefont {J.~I.}\ \bibnamefont {Zotova}},
  \bibinfo {author} {\bibfnamefont {D.~D.}\ \bibnamefont {Kirichenko}},
  \bibinfo {author} {\bibfnamefont {I.~S.}\ \bibnamefont {Besedin}}, \bibinfo
  {author} {\bibfnamefont {A.~V.}\ \bibnamefont {Semenov}}, \bibinfo {author}
  {\bibfnamefont {R.~S.}\ \bibnamefont {Shaikhaidarov}}, \ and\ \bibinfo
  {author} {\bibfnamefont {O.~V.}\ \bibnamefont {Astafiev}},\ }\bibfield
  {title} {\enquote {\bibinfo {title} {{Quantum Regime of a Two-Dimensional
  Phonon Cavity}},}\ }\href {\doibase 10.1103/PhysRevLett.120.223603}
  {\bibfield  {journal} {\bibinfo  {journal} {Physical Review Letters}\
  }\textbf {\bibinfo {volume} {120}},\ \bibinfo {pages} {223603} (\bibinfo
  {year} {2018})}\BibitemShut {NoStop}%
\bibitem [{\citenamefont {Sletten}\ \emph {et~al.}(2019)\citenamefont
  {Sletten}, \citenamefont {Moores}, \citenamefont {Viennot},\ and\
  \citenamefont {Lehnert}}]{Sletten2019}%
  \BibitemOpen
  \bibfield  {author} {\bibinfo {author} {\bibfnamefont {L.~R.}\ \bibnamefont
  {Sletten}}, \bibinfo {author} {\bibfnamefont {B.~A.}\ \bibnamefont {Moores}},
  \bibinfo {author} {\bibfnamefont {J.~J.}\ \bibnamefont {Viennot}}, \ and\
  \bibinfo {author} {\bibfnamefont {K.~W.}\ \bibnamefont {Lehnert}},\
  }\bibfield  {title} {\enquote {\bibinfo {title} {{Resolving Phonon Fock
  States in a Multimode Cavity with a Double-Slit Qubit}},}\ }\href {\doibase
  10.1103/PhysRevX.9.021056} {\bibfield  {journal} {\bibinfo  {journal}
  {Physical Review X}\ }\textbf {\bibinfo {volume} {9}},\ \bibinfo {pages}
  {021056} (\bibinfo {year} {2019})}\BibitemShut {NoStop}%
\bibitem [{\citenamefont {Bienfait}\ \emph {et~al.}(2019)\citenamefont
  {Bienfait}, \citenamefont {Satzinger}, \citenamefont {Zhong}, \citenamefont
  {Chang}, \citenamefont {Chou}, \citenamefont {Conner}, \citenamefont {Dumur},
  \citenamefont {Grebel}, \citenamefont {Peairs}, \citenamefont {Povey},\ and\
  \citenamefont {Cleland}}]{Bienfait2019}%
  \BibitemOpen
  \bibfield  {author} {\bibinfo {author} {\bibfnamefont {A.}~\bibnamefont
  {Bienfait}}, \bibinfo {author} {\bibfnamefont {K.~J.}\ \bibnamefont
  {Satzinger}}, \bibinfo {author} {\bibfnamefont {Y.~P.}\ \bibnamefont
  {Zhong}}, \bibinfo {author} {\bibfnamefont {H.-S.}\ \bibnamefont {Chang}},
  \bibinfo {author} {\bibfnamefont {M.-H.}\ \bibnamefont {Chou}}, \bibinfo
  {author} {\bibfnamefont {C.~R.}\ \bibnamefont {Conner}}, \bibinfo {author}
  {\bibfnamefont {{\'{E}}.}~\bibnamefont {Dumur}}, \bibinfo {author}
  {\bibfnamefont {J.}~\bibnamefont {Grebel}}, \bibinfo {author} {\bibfnamefont
  {G.~A.}\ \bibnamefont {Peairs}}, \bibinfo {author} {\bibfnamefont {R.~G.}\
  \bibnamefont {Povey}}, \ and\ \bibinfo {author} {\bibfnamefont {A.~N.}\
  \bibnamefont {Cleland}},\ }\bibfield  {title} {\enquote {\bibinfo {title}
  {{Phonon-mediated quantum state transfer and remote qubit entanglement}},}\
  }\href {\doibase 10.1126/science.aaw8415} {\bibfield  {journal} {\bibinfo
  {journal} {Science}\ }\textbf {\bibinfo {volume} {364}},\ \bibinfo {pages}
  {368} (\bibinfo {year} {2019})}\BibitemShut {NoStop}%
\bibitem [{\citenamefont {Andersson}\ \emph {et~al.}(2019)\citenamefont
  {Andersson}, \citenamefont {Suri}, \citenamefont {Guo}, \citenamefont
  {Aref},\ and\ \citenamefont {Delsing}}]{Andersson2019}%
  \BibitemOpen
  \bibfield  {author} {\bibinfo {author} {\bibfnamefont {G.}~\bibnamefont
  {Andersson}}, \bibinfo {author} {\bibfnamefont {B.}~\bibnamefont {Suri}},
  \bibinfo {author} {\bibfnamefont {L.}~\bibnamefont {Guo}}, \bibinfo {author}
  {\bibfnamefont {T.}~\bibnamefont {Aref}}, \ and\ \bibinfo {author}
  {\bibfnamefont {P.}~\bibnamefont {Delsing}},\ }\bibfield  {title} {\enquote
  {\bibinfo {title} {{Non-exponential decay of a giant artificial atom}},}\
  }\href {\doibase 10.1038/s41567-019-0605-6} {\bibfield  {journal} {\bibinfo
  {journal} {Nature Physics}\ }\textbf {\bibinfo {volume} {15}},\ \bibinfo
  {pages} {1123} (\bibinfo {year} {2019})}\BibitemShut {NoStop}%
\bibitem [{\citenamefont {Bienfait}\ \emph {et~al.}(2020)\citenamefont
  {Bienfait}, \citenamefont {Zhong}, \citenamefont {Chang}, \citenamefont
  {Chou}, \citenamefont {Conner}, \citenamefont {Dumur}, \citenamefont
  {Grebel}, \citenamefont {Peairs}, \citenamefont {Povey}, \citenamefont
  {Satzinger},\ and\ \citenamefont {Cleland}}]{Bienfait2020}%
  \BibitemOpen
  \bibfield  {author} {\bibinfo {author} {\bibfnamefont {A.}~\bibnamefont
  {Bienfait}}, \bibinfo {author} {\bibfnamefont {Y.~P.}\ \bibnamefont {Zhong}},
  \bibinfo {author} {\bibfnamefont {H.-S.}\ \bibnamefont {Chang}}, \bibinfo
  {author} {\bibfnamefont {M.-H.}\ \bibnamefont {Chou}}, \bibinfo {author}
  {\bibfnamefont {C.~R.}\ \bibnamefont {Conner}}, \bibinfo {author}
  {\bibfnamefont {{\'{E}}.}~\bibnamefont {Dumur}}, \bibinfo {author}
  {\bibfnamefont {J.}~\bibnamefont {Grebel}}, \bibinfo {author} {\bibfnamefont
  {G.~A.}\ \bibnamefont {Peairs}}, \bibinfo {author} {\bibfnamefont {R.~G.}\
  \bibnamefont {Povey}}, \bibinfo {author} {\bibfnamefont {K.~J.}\ \bibnamefont
  {Satzinger}}, \ and\ \bibinfo {author} {\bibfnamefont {A.~N.}\ \bibnamefont
  {Cleland}},\ }\bibfield  {title} {\enquote {\bibinfo {title} {{Quantum
  Erasure Using Entangled Surface Acoustic Phonons}},}\ }\href {\doibase
  10.1103/PhysRevX.10.021055} {\bibfield  {journal} {\bibinfo  {journal}
  {Physical Review X}\ }\textbf {\bibinfo {volume} {10}},\ \bibinfo {pages}
  {021055} (\bibinfo {year} {2020})}\BibitemShut {NoStop}%
\bibitem [{\citenamefont {Andersson}\ \emph {et~al.}(2020)\citenamefont
  {Andersson}, \citenamefont {Ekstr{\"{o}}m},\ and\ \citenamefont
  {Delsing}}]{Andersson2020}%
  \BibitemOpen
  \bibfield  {author} {\bibinfo {author} {\bibfnamefont {G.}~\bibnamefont
  {Andersson}}, \bibinfo {author} {\bibfnamefont {M.~K.}\ \bibnamefont
  {Ekstr{\"{o}}m}}, \ and\ \bibinfo {author} {\bibfnamefont {P.}~\bibnamefont
  {Delsing}},\ }\bibfield  {title} {\enquote {\bibinfo {title}
  {{Electromagnetically Induced Acoustic Transparency with a Superconducting
  Circuit}},}\ }\href {\doibase 10.1103/PhysRevLett.124.240402} {\bibfield
  {journal} {\bibinfo  {journal} {Physical Review Letters}\ }\textbf {\bibinfo
  {volume} {124}},\ \bibinfo {pages} {240402} (\bibinfo {year}
  {2020})}\BibitemShut {NoStop}%
\bibitem [{\citenamefont {Joshi}\ \emph {et~al.}(2023)\citenamefont {Joshi},
  \citenamefont {Yang},\ and\ \citenamefont {Mirhosseini}}]{Joshi2023}%
  \BibitemOpen
  \bibfield  {author} {\bibinfo {author} {\bibfnamefont {C.}~\bibnamefont
  {Joshi}}, \bibinfo {author} {\bibfnamefont {F.}~\bibnamefont {Yang}}, \ and\
  \bibinfo {author} {\bibfnamefont {M.}~\bibnamefont {Mirhosseini}},\
  }\bibfield  {title} {\enquote {\bibinfo {title} {{Resonance Fluorescence of a
  Chiral Artificial Atom}},}\ }\href {\doibase 10.1103/PhysRevX.13.021039}
  {\bibfield  {journal} {\bibinfo  {journal} {Physical Review X}\ }\textbf
  {\bibinfo {volume} {13}},\ \bibinfo {pages} {021039} (\bibinfo {year}
  {2023})}\BibitemShut {NoStop}%
\bibitem [{\citenamefont {Gonz{\'{a}}lez-Tudela}\ \emph
  {et~al.}(2019)\citenamefont {Gonz{\'{a}}lez-Tudela}, \citenamefont
  {Mu{\~{n}}oz},\ and\ \citenamefont {Cirac}}]{Gonzalez-Tudela2019}%
  \BibitemOpen
  \bibfield  {author} {\bibinfo {author} {\bibfnamefont {A.}~\bibnamefont
  {Gonz{\'{a}}lez-Tudela}}, \bibinfo {author} {\bibfnamefont {C.~S.}\
  \bibnamefont {Mu{\~{n}}oz}}, \ and\ \bibinfo {author} {\bibfnamefont {J.~I.}\
  \bibnamefont {Cirac}},\ }\bibfield  {title} {\enquote {\bibinfo {title}
  {{Engineering and Harnessing Giant Atoms in High-Dimensional Baths: A
  Proposal for Implementation with Cold Atoms}},}\ }\href {\doibase
  10.1103/PhysRevLett.122.203603} {\bibfield  {journal} {\bibinfo  {journal}
  {Physical Review Letters}\ }\textbf {\bibinfo {volume} {122}},\ \bibinfo
  {pages} {203603} (\bibinfo {year} {2019})}\BibitemShut {NoStop}%
\bibitem [{\citenamefont {Du}\ \emph {et~al.}(2022{\natexlab{a}})\citenamefont
  {Du}, \citenamefont {Zhang}, \citenamefont {Wu}, \citenamefont {Kockum},\
  and\ \citenamefont {Li}}]{Du2022}%
  \BibitemOpen
  \bibfield  {author} {\bibinfo {author} {\bibfnamefont {L.}~\bibnamefont
  {Du}}, \bibinfo {author} {\bibfnamefont {Y.}~\bibnamefont {Zhang}}, \bibinfo
  {author} {\bibfnamefont {J.~H.}\ \bibnamefont {Wu}}, \bibinfo {author}
  {\bibfnamefont {A.~F.}\ \bibnamefont {Kockum}}, \ and\ \bibinfo {author}
  {\bibfnamefont {Y.}~\bibnamefont {Li}},\ }\bibfield  {title} {\enquote
  {\bibinfo {title} {{Giant Atoms in a Synthetic Frequency Dimension}},}\
  }\href {\doibase 10.1103/PhysRevLett.128.223602} {\bibfield  {journal}
  {\bibinfo  {journal} {Physical Review Letters}\ }\textbf {\bibinfo {volume}
  {128}},\ \bibinfo {pages} {223602} (\bibinfo {year}
  {2022}{\natexlab{a}})}\BibitemShut {NoStop}%
\bibitem [{\citenamefont {Guimond}\ \emph {et~al.}(2020)\citenamefont
  {Guimond}, \citenamefont {Vermersch}, \citenamefont {Juan}, \citenamefont
  {Sharafiev}, \citenamefont {Kirchmair},\ and\ \citenamefont
  {Zoller}}]{Guimond2020}%
  \BibitemOpen
  \bibfield  {author} {\bibinfo {author} {\bibfnamefont {P.~O.}\ \bibnamefont
  {Guimond}}, \bibinfo {author} {\bibfnamefont {B.}~\bibnamefont {Vermersch}},
  \bibinfo {author} {\bibfnamefont {M.~L.}\ \bibnamefont {Juan}}, \bibinfo
  {author} {\bibfnamefont {A.}~\bibnamefont {Sharafiev}}, \bibinfo {author}
  {\bibfnamefont {G.}~\bibnamefont {Kirchmair}}, \ and\ \bibinfo {author}
  {\bibfnamefont {P.}~\bibnamefont {Zoller}},\ }\bibfield  {title} {\enquote
  {\bibinfo {title} {{A unidirectional on-chip photonic interface for
  superconducting circuits}},}\ }\href {\doibase 10.1038/s41534-020-0261-9}
  {\bibfield  {journal} {\bibinfo  {journal} {npj Quantum Information}\
  }\textbf {\bibinfo {volume} {6}},\ \bibinfo {pages} {32} (\bibinfo {year}
  {2020})}\BibitemShut {NoStop}%
\bibitem [{\citenamefont {Gheeraert}\ \emph {et~al.}(2020)\citenamefont
  {Gheeraert}, \citenamefont {Kono},\ and\ \citenamefont
  {Nakamura}}]{Gheeraert2020}%
  \BibitemOpen
  \bibfield  {author} {\bibinfo {author} {\bibfnamefont {N.}~\bibnamefont
  {Gheeraert}}, \bibinfo {author} {\bibfnamefont {S.}~\bibnamefont {Kono}}, \
  and\ \bibinfo {author} {\bibfnamefont {Y.}~\bibnamefont {Nakamura}},\
  }\bibfield  {title} {\enquote {\bibinfo {title} {{Programmable directional
  emitter and receiver of itinerant microwave photons in a waveguide}},}\
  }\href {\doibase 10.1103/PhysRevA.102.053720} {\bibfield  {journal} {\bibinfo
   {journal} {Physical Review A}\ }\textbf {\bibinfo {volume} {102}},\ \bibinfo
  {pages} {053720} (\bibinfo {year} {2020})}\BibitemShut {NoStop}%
\bibitem [{\citenamefont {Zhang}\ \emph {et~al.}(2021)\citenamefont {Zhang},
  \citenamefont {{I Carceller}}, \citenamefont {Kjaergaard},\ and\
  \citenamefont {S{\o}rensen}}]{Zhang2021}%
  \BibitemOpen
  \bibfield  {author} {\bibinfo {author} {\bibfnamefont {Y.~X.}\ \bibnamefont
  {Zhang}}, \bibinfo {author} {\bibfnamefont {C.~R.}\ \bibnamefont {{I
  Carceller}}}, \bibinfo {author} {\bibfnamefont {M.}~\bibnamefont
  {Kjaergaard}}, \ and\ \bibinfo {author} {\bibfnamefont {A.~S.}\ \bibnamefont
  {S{\o}rensen}},\ }\bibfield  {title} {\enquote {\bibinfo {title}
  {{Charge-Noise Insensitive Chiral Photonic Interface for Waveguide Circuit
  QED}},}\ }\href {\doibase 10.1103/PhysRevLett.127.233601} {\bibfield
  {journal} {\bibinfo  {journal} {Physical Review Letters}\ }\textbf {\bibinfo
  {volume} {127}},\ \bibinfo {pages} {233601} (\bibinfo {year}
  {2021})}\BibitemShut {NoStop}%
\bibitem [{\citenamefont {Yin}\ \emph {et~al.}(2022{\natexlab{a}})\citenamefont
  {Yin}, \citenamefont {Liu}, \citenamefont {Huang},\ and\ \citenamefont
  {Liao}}]{Yin2022a}%
  \BibitemOpen
  \bibfield  {author} {\bibinfo {author} {\bibfnamefont {X.-L.}\ \bibnamefont
  {Yin}}, \bibinfo {author} {\bibfnamefont {Y.-H.}\ \bibnamefont {Liu}},
  \bibinfo {author} {\bibfnamefont {J.-F.}\ \bibnamefont {Huang}}, \ and\
  \bibinfo {author} {\bibfnamefont {J.-Q.}\ \bibnamefont {Liao}},\ }\bibfield
  {title} {\enquote {\bibinfo {title} {{Single-photon scattering in a
  giant-molecule waveguide-QED system}},}\ }\href {\doibase
  10.1103/PhysRevA.106.013715} {\bibfield  {journal} {\bibinfo  {journal}
  {Physical Review A}\ }\textbf {\bibinfo {volume} {106}},\ \bibinfo {pages}
  {013715} (\bibinfo {year} {2022}{\natexlab{a}})}\BibitemShut {NoStop}%
\bibitem [{\citenamefont {Kannan}\ \emph {et~al.}(2023)\citenamefont {Kannan},
  \citenamefont {Almanakly}, \citenamefont {Sung}, \citenamefont {{Di Paolo}},
  \citenamefont {Rower}, \citenamefont {Braum{\"{u}}ller}, \citenamefont
  {Melville}, \citenamefont {Niedzielski}, \citenamefont {Karamlou},
  \citenamefont {Serniak}, \citenamefont {Veps{\"{a}}l{\"{a}}inen},
  \citenamefont {Schwartz}, \citenamefont {Yoder}, \citenamefont {Winik},
  \citenamefont {Wang}, \citenamefont {Orlando}, \citenamefont {Gustavsson},
  \citenamefont {Grover},\ and\ \citenamefont {Oliver}}]{Kannan2023}%
  \BibitemOpen
  \bibfield  {author} {\bibinfo {author} {\bibfnamefont {B.}~\bibnamefont
  {Kannan}}, \bibinfo {author} {\bibfnamefont {A.}~\bibnamefont {Almanakly}},
  \bibinfo {author} {\bibfnamefont {Y.}~\bibnamefont {Sung}}, \bibinfo {author}
  {\bibfnamefont {A.}~\bibnamefont {{Di Paolo}}}, \bibinfo {author}
  {\bibfnamefont {D.~A.}\ \bibnamefont {Rower}}, \bibinfo {author}
  {\bibfnamefont {J.}~\bibnamefont {Braum{\"{u}}ller}}, \bibinfo {author}
  {\bibfnamefont {A.}~\bibnamefont {Melville}}, \bibinfo {author}
  {\bibfnamefont {B.~M.}\ \bibnamefont {Niedzielski}}, \bibinfo {author}
  {\bibfnamefont {A.}~\bibnamefont {Karamlou}}, \bibinfo {author}
  {\bibfnamefont {K.}~\bibnamefont {Serniak}}, \bibinfo {author} {\bibfnamefont
  {A.}~\bibnamefont {Veps{\"{a}}l{\"{a}}inen}}, \bibinfo {author}
  {\bibfnamefont {M.~E.}\ \bibnamefont {Schwartz}}, \bibinfo {author}
  {\bibfnamefont {J.~L.}\ \bibnamefont {Yoder}}, \bibinfo {author}
  {\bibfnamefont {R.}~\bibnamefont {Winik}}, \bibinfo {author} {\bibfnamefont
  {J.~I.}\ \bibnamefont {Wang}}, \bibinfo {author} {\bibfnamefont {T.~P.}\
  \bibnamefont {Orlando}}, \bibinfo {author} {\bibfnamefont {S.}~\bibnamefont
  {Gustavsson}}, \bibinfo {author} {\bibfnamefont {J.~A.}\ \bibnamefont
  {Grover}}, \ and\ \bibinfo {author} {\bibfnamefont {W.~D.}\ \bibnamefont
  {Oliver}},\ }\bibfield  {title} {\enquote {\bibinfo {title} {{On-demand
  directional microwave photon emission using waveguide quantum
  electrodynamics}},}\ }\href {\doibase 10.1038/s41567-022-01869-5} {\bibfield
  {journal} {\bibinfo  {journal} {Nature Physics}\ }\textbf {\bibinfo {volume}
  {19}},\ \bibinfo {pages} {394} (\bibinfo {year} {2023})}\BibitemShut
  {NoStop}%
\bibitem [{\citenamefont {Wang}\ \emph {et~al.}(2022)\citenamefont {Wang},
  \citenamefont {Wang}, \citenamefont {Yao}, \citenamefont {Shen},
  \citenamefont {Wu}, \citenamefont {Qian}, \citenamefont {Li}, \citenamefont
  {Zhu},\ and\ \citenamefont {You}}]{Wang2022}%
  \BibitemOpen
  \bibfield  {author} {\bibinfo {author} {\bibfnamefont {Z.~Q.}\ \bibnamefont
  {Wang}}, \bibinfo {author} {\bibfnamefont {Y.~P.}\ \bibnamefont {Wang}},
  \bibinfo {author} {\bibfnamefont {J.}~\bibnamefont {Yao}}, \bibinfo {author}
  {\bibfnamefont {R.~C.}\ \bibnamefont {Shen}}, \bibinfo {author}
  {\bibfnamefont {W.~J.}\ \bibnamefont {Wu}}, \bibinfo {author} {\bibfnamefont
  {J.}~\bibnamefont {Qian}}, \bibinfo {author} {\bibfnamefont {J.}~\bibnamefont
  {Li}}, \bibinfo {author} {\bibfnamefont {S.~Y.}\ \bibnamefont {Zhu}}, \ and\
  \bibinfo {author} {\bibfnamefont {J.~Q.}\ \bibnamefont {You}},\ }\bibfield
  {title} {\enquote {\bibinfo {title} {{Giant spin ensembles in waveguide
  magnonics}},}\ }\href {\doibase 10.1038/s41467-022-35174-9} {\bibfield
  {journal} {\bibinfo  {journal} {Nature Communications}\ }\textbf {\bibinfo
  {volume} {13}},\ \bibinfo {pages} {7580} (\bibinfo {year}
  {2022})}\BibitemShut {NoStop}%
\bibitem [{\citenamefont {Guo}\ \emph {et~al.}(2017)\citenamefont {Guo},
  \citenamefont {Grimsmo}, \citenamefont {Kockum}, \citenamefont {Pletyukhov},\
  and\ \citenamefont {Johansson}}]{Guo2017}%
  \BibitemOpen
  \bibfield  {author} {\bibinfo {author} {\bibfnamefont {L.}~\bibnamefont
  {Guo}}, \bibinfo {author} {\bibfnamefont {A.~L.}\ \bibnamefont {Grimsmo}},
  \bibinfo {author} {\bibfnamefont {A.~F.}\ \bibnamefont {Kockum}}, \bibinfo
  {author} {\bibfnamefont {M.}~\bibnamefont {Pletyukhov}}, \ and\ \bibinfo
  {author} {\bibfnamefont {G.}~\bibnamefont {Johansson}},\ }\bibfield  {title}
  {\enquote {\bibinfo {title} {{Giant acoustic atom: A single quantum system
  with a deterministic time delay}},}\ }\href {\doibase
  10.1103/PhysRevA.95.053821} {\bibfield  {journal} {\bibinfo  {journal}
  {Physical Review A}\ }\textbf {\bibinfo {volume} {95}},\ \bibinfo {pages}
  {053821} (\bibinfo {year} {2017})}\BibitemShut {NoStop}%
\bibitem [{\citenamefont {Karg}\ \emph {et~al.}(2019)\citenamefont {Karg},
  \citenamefont {Gouraud}, \citenamefont {Treutlein},\ and\ \citenamefont
  {Hammerer}}]{Karg2019}%
  \BibitemOpen
  \bibfield  {author} {\bibinfo {author} {\bibfnamefont {T.~M.}\ \bibnamefont
  {Karg}}, \bibinfo {author} {\bibfnamefont {B.}~\bibnamefont {Gouraud}},
  \bibinfo {author} {\bibfnamefont {P.}~\bibnamefont {Treutlein}}, \ and\
  \bibinfo {author} {\bibfnamefont {K.}~\bibnamefont {Hammerer}},\ }\bibfield
  {title} {\enquote {\bibinfo {title} {{Remote Hamiltonian interactions
  mediated by light}},}\ }\href {\doibase 10.1103/PhysRevA.99.063829}
  {\bibfield  {journal} {\bibinfo  {journal} {Physical Review A}\ }\textbf
  {\bibinfo {volume} {99}},\ \bibinfo {pages} {063829} (\bibinfo {year}
  {2019})}\BibitemShut {NoStop}%
\bibitem [{\citenamefont {Ask}\ \emph {et~al.}(2020)\citenamefont {Ask},
  \citenamefont {Fang},\ and\ \citenamefont {Kockum}}]{Ask2020}%
  \BibitemOpen
  \bibfield  {author} {\bibinfo {author} {\bibfnamefont {A.}~\bibnamefont
  {Ask}}, \bibinfo {author} {\bibfnamefont {Y.-L.~L.}\ \bibnamefont {Fang}}, \
  and\ \bibinfo {author} {\bibfnamefont {A.~F.}\ \bibnamefont {Kockum}},\
  }\href@noop {} {\enquote {\bibinfo {title} {{Synthesizing electromagnetically
  induced transparency without a control field in waveguide QED using small and
  giant atoms}},}\ } (\bibinfo {year} {2020}),\ \Eprint
  {http://arxiv.org/abs/2011.15077} {arXiv:2011.15077} \BibitemShut {NoStop}%
\bibitem [{\citenamefont {Du}\ and\ \citenamefont {Li}(2021)}]{Du2021}%
  \BibitemOpen
  \bibfield  {author} {\bibinfo {author} {\bibfnamefont {L.}~\bibnamefont
  {Du}}\ and\ \bibinfo {author} {\bibfnamefont {Y.}~\bibnamefont {Li}},\
  }\bibfield  {title} {\enquote {\bibinfo {title} {{Single-photon frequency
  conversion via a giant $\Lambda$-type atom}},}\ }\href {\doibase
  10.1103/PhysRevA.104.023712} {\bibfield  {journal} {\bibinfo  {journal}
  {Physical Review A}\ }\textbf {\bibinfo {volume} {104}},\ \bibinfo {pages}
  {023712} (\bibinfo {year} {2021})}\BibitemShut {NoStop}%
\bibitem [{\citenamefont {Feng}\ and\ \citenamefont {Jia}(2021)}]{Feng2021}%
  \BibitemOpen
  \bibfield  {author} {\bibinfo {author} {\bibfnamefont {S.~L.}\ \bibnamefont
  {Feng}}\ and\ \bibinfo {author} {\bibfnamefont {W.~Z.}\ \bibnamefont {Jia}},\
  }\bibfield  {title} {\enquote {\bibinfo {title} {{Manipulating single-photon
  transport in a waveguide-QED structure containing two giant atoms}},}\ }\href
  {\doibase 10.1103/PhysRevA.104.063712} {\bibfield  {journal} {\bibinfo
  {journal} {Physical Review A}\ }\textbf {\bibinfo {volume} {104}},\ \bibinfo
  {pages} {063712} (\bibinfo {year} {2021})}\BibitemShut {NoStop}%
\bibitem [{\citenamefont {Cai}\ and\ \citenamefont {Jia}(2021)}]{Cai2021}%
  \BibitemOpen
  \bibfield  {author} {\bibinfo {author} {\bibfnamefont {Q.~Y.}\ \bibnamefont
  {Cai}}\ and\ \bibinfo {author} {\bibfnamefont {W.~Z.}\ \bibnamefont {Jia}},\
  }\bibfield  {title} {\enquote {\bibinfo {title} {{Coherent single-photon
  scattering spectra for a giant-atom waveguide-QED system beyond the dipole
  approximation}},}\ }\href {\doibase 10.1103/PhysRevA.104.033710} {\bibfield
  {journal} {\bibinfo  {journal} {Physical Review A}\ }\textbf {\bibinfo
  {volume} {104}},\ \bibinfo {pages} {033710} (\bibinfo {year}
  {2021})}\BibitemShut {NoStop}%
\bibitem [{\citenamefont {Yin}\ \emph {et~al.}(2022{\natexlab{b}})\citenamefont
  {Yin}, \citenamefont {Luo},\ and\ \citenamefont {Liao}}]{Yin2022}%
  \BibitemOpen
  \bibfield  {author} {\bibinfo {author} {\bibfnamefont {X.-L.}\ \bibnamefont
  {Yin}}, \bibinfo {author} {\bibfnamefont {W.-B.}\ \bibnamefont {Luo}}, \ and\
  \bibinfo {author} {\bibfnamefont {J.-Q.}\ \bibnamefont {Liao}},\ }\bibfield
  {title} {\enquote {\bibinfo {title} {{Non-Markovian disentanglement dynamics
  in double-giant-atom waveguide-QED systems}},}\ }\href {\doibase
  10.1103/PhysRevA.106.063703} {\bibfield  {journal} {\bibinfo  {journal}
  {Physical Review A}\ }\textbf {\bibinfo {volume} {106}},\ \bibinfo {pages}
  {063703} (\bibinfo {year} {2022}{\natexlab{b}})}\BibitemShut {NoStop}%
\bibitem [{\citenamefont {Chen}\ \emph {et~al.}(2022)\citenamefont {Chen},
  \citenamefont {Du}, \citenamefont {Guo}, \citenamefont {Wang}, \citenamefont
  {Zhang}, \citenamefont {Li},\ and\ \citenamefont {Wu}}]{Chen2022}%
  \BibitemOpen
  \bibfield  {author} {\bibinfo {author} {\bibfnamefont {Y.-T.}\ \bibnamefont
  {Chen}}, \bibinfo {author} {\bibfnamefont {L.}~\bibnamefont {Du}}, \bibinfo
  {author} {\bibfnamefont {L.}~\bibnamefont {Guo}}, \bibinfo {author}
  {\bibfnamefont {Z.}~\bibnamefont {Wang}}, \bibinfo {author} {\bibfnamefont
  {Y.}~\bibnamefont {Zhang}}, \bibinfo {author} {\bibfnamefont
  {Y.}~\bibnamefont {Li}}, \ and\ \bibinfo {author} {\bibfnamefont {J.-H.}\
  \bibnamefont {Wu}},\ }\bibfield  {title} {\enquote {\bibinfo {title}
  {{Nonreciprocal and chiral single-photon scattering for giant atoms}},}\
  }\href {\doibase 10.1038/s42005-022-00991-3} {\bibfield  {journal} {\bibinfo
  {journal} {Communications Physics}\ }\textbf {\bibinfo {volume} {5}},\
  \bibinfo {pages} {215} (\bibinfo {year} {2022})}\BibitemShut {NoStop}%
\bibitem [{\citenamefont {Du}\ \emph {et~al.}(2022{\natexlab{b}})\citenamefont
  {Du}, \citenamefont {Chen}, \citenamefont {Zhang},\ and\ \citenamefont
  {Li}}]{Du2022a}%
  \BibitemOpen
  \bibfield  {author} {\bibinfo {author} {\bibfnamefont {L.}~\bibnamefont
  {Du}}, \bibinfo {author} {\bibfnamefont {Y.-T.}\ \bibnamefont {Chen}},
  \bibinfo {author} {\bibfnamefont {Y.}~\bibnamefont {Zhang}}, \ and\ \bibinfo
  {author} {\bibfnamefont {Y.}~\bibnamefont {Li}},\ }\bibfield  {title}
  {\enquote {\bibinfo {title} {{Giant atoms with time-dependent couplings}},}\
  }\href {\doibase 10.1103/PhysRevResearch.4.023198} {\bibfield  {journal}
  {\bibinfo  {journal} {Physical Review Research}\ }\textbf {\bibinfo {volume}
  {4}},\ \bibinfo {pages} {023198} (\bibinfo {year}
  {2022}{\natexlab{b}})}\BibitemShut {NoStop}%
\bibitem [{\citenamefont {Du}\ \emph {et~al.}(2023{\natexlab{b}})\citenamefont
  {Du}, \citenamefont {Zhang},\ and\ \citenamefont {Li}}]{Du2023a}%
  \BibitemOpen
  \bibfield  {author} {\bibinfo {author} {\bibfnamefont {L.}~\bibnamefont
  {Du}}, \bibinfo {author} {\bibfnamefont {Y.}~\bibnamefont {Zhang}}, \ and\
  \bibinfo {author} {\bibfnamefont {Y.}~\bibnamefont {Li}},\ }\bibfield
  {title} {\enquote {\bibinfo {title} {{A giant atom with modulated transition
  frequency}},}\ }\href {\doibase 10.1007/s11467-022-1215-9} {\bibfield
  {journal} {\bibinfo  {journal} {Frontiers of Physics}\ }\textbf {\bibinfo
  {volume} {18}},\ \bibinfo {pages} {12301} (\bibinfo {year}
  {2023}{\natexlab{b}})}\BibitemShut {NoStop}%
\bibitem [{\citenamefont {Santos}\ and\ \citenamefont
  {Bachelard}(2023)}]{Santos2023}%
  \BibitemOpen
  \bibfield  {author} {\bibinfo {author} {\bibfnamefont {A.~C.}\ \bibnamefont
  {Santos}}\ and\ \bibinfo {author} {\bibfnamefont {R.}~\bibnamefont
  {Bachelard}},\ }\bibfield  {title} {\enquote {\bibinfo {title} {{Generation
  of Maximally Entangled Long-Lived States with Giant Atoms in a Waveguide}},}\
  }\href {\doibase 10.1103/PhysRevLett.130.053601} {\bibfield  {journal}
  {\bibinfo  {journal} {Physical Review Letters}\ }\textbf {\bibinfo {volume}
  {130}},\ \bibinfo {pages} {053601} (\bibinfo {year} {2023})}\BibitemShut
  {NoStop}%
\bibitem [{\citenamefont {Wang}\ \emph {et~al.}(2024)\citenamefont {Wang},
  \citenamefont {Zhu}, \citenamefont {Liu},\ and\ \citenamefont
  {Nori}}]{Wang2024}%
  \BibitemOpen
  \bibfield  {author} {\bibinfo {author} {\bibfnamefont {X.}~\bibnamefont
  {Wang}}, \bibinfo {author} {\bibfnamefont {H.-B.}\ \bibnamefont {Zhu}},
  \bibinfo {author} {\bibfnamefont {T.}~\bibnamefont {Liu}}, \ and\ \bibinfo
  {author} {\bibfnamefont {F.}~\bibnamefont {Nori}},\ }\bibfield  {title}
  {\enquote {\bibinfo {title} {Realizing quantum optics in structured
  environments with giant atoms},}\ }\href {\doibase
  10.1103/PhysRevResearch.6.013279} {\bibfield  {journal} {\bibinfo  {journal}
  {Physical Review Research}\ }\textbf {\bibinfo {volume} {6}},\ \bibinfo
  {pages} {013279} (\bibinfo {year} {2024})}\BibitemShut {NoStop}%
\bibitem [{\citenamefont {Zhou}\ \emph {et~al.}(2023)\citenamefont {Zhou},
  \citenamefont {Yin},\ and\ \citenamefont {Liao}}]{Zhou2023}%
  \BibitemOpen
  \bibfield  {author} {\bibinfo {author} {\bibfnamefont {J.}~\bibnamefont
  {Zhou}}, \bibinfo {author} {\bibfnamefont {X.-L.}\ \bibnamefont {Yin}}, \
  and\ \bibinfo {author} {\bibfnamefont {J.-Q.}\ \bibnamefont {Liao}},\
  }\bibfield  {title} {\enquote {\bibinfo {title} {{Chiral and nonreciprocal
  single-photon scattering in a chiral-giant-molecule waveguide-QED system}},}\
  }\href {\doibase 10.1103/PhysRevA.107.063703} {\bibfield  {journal} {\bibinfo
   {journal} {Physical Review A}\ }\textbf {\bibinfo {volume} {107}},\ \bibinfo
  {pages} {063703} (\bibinfo {year} {2023})}\BibitemShut {NoStop}%
\bibitem [{\citenamefont {Gu}\ \emph {et~al.}(2023)\citenamefont {Gu},
  \citenamefont {Huang}, \citenamefont {Yi}, \citenamefont {Chen},
  \citenamefont {Sun},\ and\ \citenamefont {Tan}}]{Gu2023}%
  \BibitemOpen
  \bibfield  {author} {\bibinfo {author} {\bibfnamefont {W.}~\bibnamefont
  {Gu}}, \bibinfo {author} {\bibfnamefont {H.}~\bibnamefont {Huang}}, \bibinfo
  {author} {\bibfnamefont {Z.}~\bibnamefont {Yi}}, \bibinfo {author}
  {\bibfnamefont {L.}~\bibnamefont {Chen}}, \bibinfo {author} {\bibfnamefont
  {L.}~\bibnamefont {Sun}}, \ and\ \bibinfo {author} {\bibfnamefont
  {H.}~\bibnamefont {Tan}},\ }\bibfield  {title} {\enquote {\bibinfo {title}
  {{Correlated two-photon scattering in a one-dimensional waveguide coupled to
  two- or three-level giant atoms}},}\ }\href {\doibase
  10.1103/PhysRevA.108.053718} {\bibfield  {journal} {\bibinfo  {journal}
  {Physical Review A}\ }\textbf {\bibinfo {volume} {108}},\ \bibinfo {pages}
  {053718} (\bibinfo {year} {2023})}\BibitemShut {NoStop}%
\bibitem [{\citenamefont {Xu}\ and\ \citenamefont {Guo}(2024)}]{Xu2024}%
  \BibitemOpen
  \bibfield  {author} {\bibinfo {author} {\bibfnamefont {L.}~\bibnamefont
  {Xu}}\ and\ \bibinfo {author} {\bibfnamefont {L.}~\bibnamefont {Guo}},\
  }\bibfield  {title} {\enquote {\bibinfo {title} {{Catch and release of
  propagating bosonic field with non-Markovian giant atom}},}\ }\href {\doibase
  10.1088/1367-2630/ad18ed} {\bibfield  {journal} {\bibinfo  {journal} {New
  Journal of Physics}\ }\textbf {\bibinfo {volume} {26}},\ \bibinfo {pages}
  {013025} (\bibinfo {year} {2024})}\BibitemShut {NoStop}%
\bibitem [{\citenamefont {Longhi}(2020)}]{Longhi2020}%
  \BibitemOpen
  \bibfield  {author} {\bibinfo {author} {\bibfnamefont {S.}~\bibnamefont
  {Longhi}},\ }\bibfield  {title} {\enquote {\bibinfo {title} {{Photonic
  simulation of giant atom decay}},}\ }\href {\doibase 10.1364/OL.393578}
  {\bibfield  {journal} {\bibinfo  {journal} {Optics Letters}\ }\textbf
  {\bibinfo {volume} {45}},\ \bibinfo {pages} {3017} (\bibinfo {year}
  {2020})}\BibitemShut {NoStop}%
\bibitem [{\citenamefont {Zhao}\ and\ \citenamefont {Wang}(2020)}]{Zhao2020}%
  \BibitemOpen
  \bibfield  {author} {\bibinfo {author} {\bibfnamefont {W.}~\bibnamefont
  {Zhao}}\ and\ \bibinfo {author} {\bibfnamefont {Z.}~\bibnamefont {Wang}},\
  }\bibfield  {title} {\enquote {\bibinfo {title} {{Single-photon scattering
  and bound states in an atom-waveguide system with two or multiple coupling
  points}},}\ }\href {\doibase 10.1103/PhysRevA.101.053855} {\bibfield
  {journal} {\bibinfo  {journal} {Physical Review A}\ }\textbf {\bibinfo
  {volume} {101}},\ \bibinfo {pages} {053855} (\bibinfo {year}
  {2020})}\BibitemShut {NoStop}%
\bibitem [{\citenamefont {Wang}\ \emph {et~al.}(2021)\citenamefont {Wang},
  \citenamefont {Liu}, \citenamefont {Kockum}, \citenamefont {Li},\ and\
  \citenamefont {Nori}}]{Wang21}%
  \BibitemOpen
  \bibfield  {author} {\bibinfo {author} {\bibfnamefont {X.}~\bibnamefont
  {Wang}}, \bibinfo {author} {\bibfnamefont {T.}~\bibnamefont {Liu}}, \bibinfo
  {author} {\bibfnamefont {A.~F.}\ \bibnamefont {Kockum}}, \bibinfo {author}
  {\bibfnamefont {H.-R.}\ \bibnamefont {Li}}, \ and\ \bibinfo {author}
  {\bibfnamefont {F.}~\bibnamefont {Nori}},\ }\bibfield  {title} {\enquote
  {\bibinfo {title} {{Tunable Chiral Bound States with Giant Atoms}},}\ }\href
  {\doibase 10.1103/PhysRevLett.126.043602} {\bibfield  {journal} {\bibinfo
  {journal} {Physical Review Letters}\ }\textbf {\bibinfo {volume} {126}},\
  \bibinfo {pages} {043602} (\bibinfo {year} {2021})}\BibitemShut {NoStop}%
\bibitem [{\citenamefont {Longhi}(2021)}]{Longhi2021}%
  \BibitemOpen
  \bibfield  {author} {\bibinfo {author} {\bibfnamefont {S.}~\bibnamefont
  {Longhi}},\ }\bibfield  {title} {\enquote {\bibinfo {title} {{Rabi
  oscillations of bound states in the continuum}},}\ }\href {\doibase
  10.1364/ol.424756} {\bibfield  {journal} {\bibinfo  {journal} {Optics
  Letters}\ }\textbf {\bibinfo {volume} {46}},\ \bibinfo {pages} {2091}
  (\bibinfo {year} {2021})}\BibitemShut {NoStop}%
\bibitem [{\citenamefont {Yu}\ \emph {et~al.}(2021)\citenamefont {Yu},
  \citenamefont {Wang},\ and\ \citenamefont {Wu}}]{Yu2021a}%
  \BibitemOpen
  \bibfield  {author} {\bibinfo {author} {\bibfnamefont {H.}~\bibnamefont
  {Yu}}, \bibinfo {author} {\bibfnamefont {Z.}~\bibnamefont {Wang}}, \ and\
  \bibinfo {author} {\bibfnamefont {J.-H.}\ \bibnamefont {Wu}},\ }\bibfield
  {title} {\enquote {\bibinfo {title} {{Entanglement preparation and
  nonreciprocal excitation evolution in giant atoms by controllable dissipation
  and coupling}},}\ }\href {\doibase 10.1103/PhysRevA.104.013720} {\bibfield
  {journal} {\bibinfo  {journal} {Physical Review A}\ }\textbf {\bibinfo
  {volume} {104}},\ \bibinfo {pages} {013720} (\bibinfo {year}
  {2021})}\BibitemShut {NoStop}%
\bibitem [{\citenamefont {Vega}\ \emph {et~al.}(2021)\citenamefont {Vega},
  \citenamefont {Bello}, \citenamefont {Porras},\ and\ \citenamefont
  {Gonz{\'{a}}lez-Tudela}}]{Vega2021}%
  \BibitemOpen
  \bibfield  {author} {\bibinfo {author} {\bibfnamefont {C.}~\bibnamefont
  {Vega}}, \bibinfo {author} {\bibfnamefont {M.}~\bibnamefont {Bello}},
  \bibinfo {author} {\bibfnamefont {D.}~\bibnamefont {Porras}}, \ and\ \bibinfo
  {author} {\bibfnamefont {A.}~\bibnamefont {Gonz{\'{a}}lez-Tudela}},\
  }\bibfield  {title} {\enquote {\bibinfo {title} {{Qubit-photon bound states
  in topological waveguides with long-range hoppings}},}\ }\href {\doibase
  10.1103/PhysRevA.104.053522} {\bibfield  {journal} {\bibinfo  {journal}
  {Physical Review A}\ }\textbf {\bibinfo {volume} {104}},\ \bibinfo {pages}
  {053522} (\bibinfo {year} {2021})}\BibitemShut {NoStop}%
\bibitem [{\citenamefont {Wang}\ and\ \citenamefont {Li}(2022)}]{Wang2021}%
  \BibitemOpen
  \bibfield  {author} {\bibinfo {author} {\bibfnamefont {X.}~\bibnamefont
  {Wang}}\ and\ \bibinfo {author} {\bibfnamefont {H.-r.}\ \bibnamefont {Li}},\
  }\bibfield  {title} {\enquote {\bibinfo {title} {{Chiral quantum network with
  giant atoms}},}\ }\href {\doibase 10.1088/2058-9565/ac6a04} {\bibfield
  {journal} {\bibinfo  {journal} {Quantum Science and Technology}\ }\textbf
  {\bibinfo {volume} {7}},\ \bibinfo {pages} {035007} (\bibinfo {year}
  {2022})}\BibitemShut {NoStop}%
\bibitem [{\citenamefont {Xiao}\ \emph {et~al.}(2022)\citenamefont {Xiao},
  \citenamefont {Wang}, \citenamefont {Li}, \citenamefont {Chen},\ and\
  \citenamefont {Yuan}}]{Xiao2022}%
  \BibitemOpen
  \bibfield  {author} {\bibinfo {author} {\bibfnamefont {H.}~\bibnamefont
  {Xiao}}, \bibinfo {author} {\bibfnamefont {L.}~\bibnamefont {Wang}}, \bibinfo
  {author} {\bibfnamefont {Z.}~\bibnamefont {Li}}, \bibinfo {author}
  {\bibfnamefont {X.}~\bibnamefont {Chen}}, \ and\ \bibinfo {author}
  {\bibfnamefont {L.}~\bibnamefont {Yuan}},\ }\bibfield  {title} {\enquote
  {\bibinfo {title} {{Bound state in a giant atom-modulated resonators
  system}},}\ }\href {\doibase 10.1038/s41534-022-00591-7} {\bibfield
  {journal} {\bibinfo  {journal} {npj Quantum Information}\ }\textbf {\bibinfo
  {volume} {8}},\ \bibinfo {pages} {80} (\bibinfo {year} {2022})}\BibitemShut
  {NoStop}%
\bibitem [{\citenamefont {Cheng}\ \emph {et~al.}(2022)\citenamefont {Cheng},
  \citenamefont {Wang},\ and\ \citenamefont {Liu}}]{Cheng2022}%
  \BibitemOpen
  \bibfield  {author} {\bibinfo {author} {\bibfnamefont {W.}~\bibnamefont
  {Cheng}}, \bibinfo {author} {\bibfnamefont {Z.}~\bibnamefont {Wang}}, \ and\
  \bibinfo {author} {\bibfnamefont {Y.-x.}\ \bibnamefont {Liu}},\ }\bibfield
  {title} {\enquote {\bibinfo {title} {{Topology and retardation effect of a
  giant atom in a topological waveguide}},}\ }\href {\doibase
  10.1103/PhysRevA.106.033522} {\bibfield  {journal} {\bibinfo  {journal}
  {Physical Review A}\ }\textbf {\bibinfo {volume} {106}},\ \bibinfo {pages}
  {033522} (\bibinfo {year} {2022})}\BibitemShut {NoStop}%
\bibitem [{\citenamefont {Zhang}\ \emph {et~al.}(2022)\citenamefont {Zhang},
  \citenamefont {Cheng}, \citenamefont {Gong}, \citenamefont {Zheng},\ and\
  \citenamefont {Wang}}]{Zhang2022}%
  \BibitemOpen
  \bibfield  {author} {\bibinfo {author} {\bibfnamefont {X.}~\bibnamefont
  {Zhang}}, \bibinfo {author} {\bibfnamefont {W.}~\bibnamefont {Cheng}},
  \bibinfo {author} {\bibfnamefont {Z.}~\bibnamefont {Gong}}, \bibinfo {author}
  {\bibfnamefont {T.}~\bibnamefont {Zheng}}, \ and\ \bibinfo {author}
  {\bibfnamefont {Z.}~\bibnamefont {Wang}},\ }\href@noop {} {\enquote {\bibinfo
  {title} {{Superconducting giant atom waveguide QED: Quantum Zeno and
  Anti-Zeno effects in ultrastrong coupling regime}},}\ } (\bibinfo {year}
  {2022}),\ \Eprint {http://arxiv.org/abs/2205.03674} {arXiv:2205.03674}
  \BibitemShut {NoStop}%
\bibitem [{\citenamefont {Du}\ \emph {et~al.}(2023{\natexlab{c}})\citenamefont
  {Du}, \citenamefont {Guo}, \citenamefont {Zhang},\ and\ \citenamefont
  {Kockum}}]{Du2023}%
  \BibitemOpen
  \bibfield  {author} {\bibinfo {author} {\bibfnamefont {L.}~\bibnamefont
  {Du}}, \bibinfo {author} {\bibfnamefont {L.}~\bibnamefont {Guo}}, \bibinfo
  {author} {\bibfnamefont {Y.}~\bibnamefont {Zhang}}, \ and\ \bibinfo {author}
  {\bibfnamefont {A.~F.}\ \bibnamefont {Kockum}},\ }\bibfield  {title}
  {\enquote {\bibinfo {title} {{Giant emitters in a structured bath with
  non-Hermitian skin effect}},}\ }\href {\doibase
  10.1103/PhysRevResearch.5.L042040} {\bibfield  {journal} {\bibinfo  {journal}
  {Physical Review Research}\ }\textbf {\bibinfo {volume} {5}},\ \bibinfo
  {pages} {L042040} (\bibinfo {year} {2023}{\natexlab{c}})}\BibitemShut
  {NoStop}%
\bibitem [{\citenamefont {Du}\ \emph {et~al.}(2023{\natexlab{d}})\citenamefont
  {Du}, \citenamefont {Chen}, \citenamefont {Zhang}, \citenamefont {Li},\ and\
  \citenamefont {Wu}}]{Du2023c}%
  \BibitemOpen
  \bibfield  {author} {\bibinfo {author} {\bibfnamefont {L.}~\bibnamefont
  {Du}}, \bibinfo {author} {\bibfnamefont {Y.-T.}\ \bibnamefont {Chen}},
  \bibinfo {author} {\bibfnamefont {Y.}~\bibnamefont {Zhang}}, \bibinfo
  {author} {\bibfnamefont {Y.}~\bibnamefont {Li}}, \ and\ \bibinfo {author}
  {\bibfnamefont {J.-H.}\ \bibnamefont {Wu}},\ }\bibfield  {title} {\enquote
  {\bibinfo {title} {{Decay dynamics of a giant atom in a structured bath with
  broken time-reversal symmetry}},}\ }\href {\doibase 10.1088/2058-9565/ace54c}
  {\bibfield  {journal} {\bibinfo  {journal} {Quantum Science and Technology}\
  }\textbf {\bibinfo {volume} {8}},\ \bibinfo {pages} {045010} (\bibinfo {year}
  {2023}{\natexlab{d}})}\BibitemShut {NoStop}%
\bibitem [{\citenamefont {Bag}\ and\ \citenamefont {Roy}(2023)}]{Bag2023}%
  \BibitemOpen
  \bibfield  {author} {\bibinfo {author} {\bibfnamefont {R.}~\bibnamefont
  {Bag}}\ and\ \bibinfo {author} {\bibfnamefont {D.}~\bibnamefont {Roy}},\
  }\bibfield  {title} {\enquote {\bibinfo {title} {{Quantum light-matter
  interactions in structured waveguides}},}\ }\href {\doibase
  10.1103/PhysRevA.108.053717} {\bibfield  {journal} {\bibinfo  {journal}
  {Physical Review A}\ }\textbf {\bibinfo {volume} {108}},\ \bibinfo {pages}
  {053717} (\bibinfo {year} {2023})}\BibitemShut {NoStop}%
\bibitem [{\citenamefont {Jia}\ and\ \citenamefont {Yu}(2024)}]{Jia2024}%
  \BibitemOpen
  \bibfield  {author} {\bibinfo {author} {\bibfnamefont {W.~Z.}\ \bibnamefont
  {Jia}}\ and\ \bibinfo {author} {\bibfnamefont {M.~T.}\ \bibnamefont {Yu}},\
  }\bibfield  {title} {\enquote {\bibinfo {title} {{Atom-photon dressed states
  in a waveguide-QED system with multiple giant atoms}},}\ }\href {\doibase
  10.1364/oe.518325} {\bibfield  {journal} {\bibinfo  {journal} {Optics
  Express}\ }\textbf {\bibinfo {volume} {32}},\ \bibinfo {pages} {9495}
  (\bibinfo {year} {2024})}\BibitemShut {NoStop}%
\bibitem [{\citenamefont {Gao}\ \emph {et~al.}(2023)\citenamefont {Gao},
  \citenamefont {Li}, \citenamefont {Li}, \citenamefont {Liu},\ and\
  \citenamefont {Wang}}]{Gao2023}%
  \BibitemOpen
  \bibfield  {author} {\bibinfo {author} {\bibfnamefont {Z.-M.}\ \bibnamefont
  {Gao}}, \bibinfo {author} {\bibfnamefont {J.-Q.}\ \bibnamefont {Li}},
  \bibinfo {author} {\bibfnamefont {Z.-W.}\ \bibnamefont {Li}}, \bibinfo
  {author} {\bibfnamefont {W.-X.}\ \bibnamefont {Liu}}, \ and\ \bibinfo
  {author} {\bibfnamefont {X.}~\bibnamefont {Wang}},\ }\href@noop {} {\enquote
  {\bibinfo {title} {{Circuit QED with a Giant Atom Coupling to Left-handed
  Superlattice Metamaterials}},}\ } (\bibinfo {year} {2023}),\ \Eprint
  {http://arxiv.org/abs/2309.06826} {arXiv:2309.06826} \BibitemShut {NoStop}%
\bibitem [{\citenamefont {Gonz{\'{a}}lez-Tudela}\ and\ \citenamefont
  {Cirac}(2017{\natexlab{a}})}]{Gonzalez-Tudela2017}%
  \BibitemOpen
  \bibfield  {author} {\bibinfo {author} {\bibfnamefont {A.}~\bibnamefont
  {Gonz{\'{a}}lez-Tudela}}\ and\ \bibinfo {author} {\bibfnamefont {J.~I.}\
  \bibnamefont {Cirac}},\ }\bibfield  {title} {\enquote {\bibinfo {title}
  {{Markovian and non-Markovian dynamics of quantum emitters coupled to
  two-dimensional structured reservoirs}},}\ }\href {\doibase
  10.1103/PhysRevA.96.043811} {\bibfield  {journal} {\bibinfo  {journal}
  {Physical Review A}\ }\textbf {\bibinfo {volume} {96}},\ \bibinfo {pages}
  {043811} (\bibinfo {year} {2017}{\natexlab{a}})}\BibitemShut {NoStop}%
\bibitem [{\citenamefont {Gonz{\'{a}}lez-Tudela}\ and\ \citenamefont
  {Cirac}(2017{\natexlab{b}})}]{Gonzalez-Tudela2017a}%
  \BibitemOpen
  \bibfield  {author} {\bibinfo {author} {\bibfnamefont {A.}~\bibnamefont
  {Gonz{\'{a}}lez-Tudela}}\ and\ \bibinfo {author} {\bibfnamefont {J.~I.}\
  \bibnamefont {Cirac}},\ }\bibfield  {title} {\enquote {\bibinfo {title}
  {{Quantum Emitters in Two-Dimensional Structured Reservoirs in the
  Nonperturbative Regime}},}\ }\href {\doibase 10.1103/PhysRevLett.119.143602}
  {\bibfield  {journal} {\bibinfo  {journal} {Physical Review Letters}\
  }\textbf {\bibinfo {volume} {119}},\ \bibinfo {pages} {143602} (\bibinfo
  {year} {2017}{\natexlab{b}})}\BibitemShut {NoStop}%
\bibitem [{\citenamefont {Galve}\ and\ \citenamefont
  {Zambrini}(2018)}]{Galve2018}%
  \BibitemOpen
  \bibfield  {author} {\bibinfo {author} {\bibfnamefont {F.}~\bibnamefont
  {Galve}}\ and\ \bibinfo {author} {\bibfnamefont {R.}~\bibnamefont
  {Zambrini}},\ }\bibfield  {title} {\enquote {\bibinfo {title} {{Completely
  Subradiant Multi‐Atom Architectures Through 2D Photonic Crystals}},}\
  }\href {\doibase 10.1002/andp.201800017} {\bibfield  {journal} {\bibinfo
  {journal} {Annalen der Physik}\ }\textbf {\bibinfo {volume} {530}},\ \bibinfo
  {pages} {1800017} (\bibinfo {year} {2018})}\BibitemShut {NoStop}%
\bibitem [{\citenamefont {Yu}\ \emph {et~al.}(2019)\citenamefont {Yu},
  \citenamefont {Muniz}, \citenamefont {Hung},\ and\ \citenamefont
  {Kimble}}]{Yu2019}%
  \BibitemOpen
  \bibfield  {author} {\bibinfo {author} {\bibfnamefont {S.-P.}\ \bibnamefont
  {Yu}}, \bibinfo {author} {\bibfnamefont {J.~A.}\ \bibnamefont {Muniz}},
  \bibinfo {author} {\bibfnamefont {C.-L.}\ \bibnamefont {Hung}}, \ and\
  \bibinfo {author} {\bibfnamefont {H.~J.}\ \bibnamefont {Kimble}},\ }\bibfield
   {title} {\enquote {\bibinfo {title} {{Two-dimensional photonic crystals for
  engineering atom–light interactions}},}\ }\href {\doibase
  10.1073/pnas.1822110116} {\bibfield  {journal} {\bibinfo  {journal}
  {Proceedings of the National Academy of Sciences}\ }\textbf {\bibinfo
  {volume} {116}},\ \bibinfo {pages} {12743} (\bibinfo {year}
  {2019})}\BibitemShut {NoStop}%
\bibitem [{\citenamefont {Feiguin}\ \emph {et~al.}(2020)\citenamefont
  {Feiguin}, \citenamefont {Garc{\'{i}}a-Ripoll},\ and\ \citenamefont
  {Gonz{\'{a}}lez-Tudela}}]{Feiguin2020}%
  \BibitemOpen
  \bibfield  {author} {\bibinfo {author} {\bibfnamefont {A.}~\bibnamefont
  {Feiguin}}, \bibinfo {author} {\bibfnamefont {J.~J.}\ \bibnamefont
  {Garc{\'{i}}a-Ripoll}}, \ and\ \bibinfo {author} {\bibfnamefont
  {A.}~\bibnamefont {Gonz{\'{a}}lez-Tudela}},\ }\bibfield  {title} {\enquote
  {\bibinfo {title} {{Qubit-photon corner states in all dimensions}},}\ }\href
  {\doibase 10.1103/PhysRevResearch.2.023082} {\bibfield  {journal} {\bibinfo
  {journal} {Physical Review Research}\ }\textbf {\bibinfo {volume} {2}},\
  \bibinfo {pages} {023082} (\bibinfo {year} {2020})}\BibitemShut {NoStop}%
\bibitem [{\citenamefont {Ruks}\ and\ \citenamefont {Busch}(2022)}]{Ruks2022}%
  \BibitemOpen
  \bibfield  {author} {\bibinfo {author} {\bibfnamefont {L.}~\bibnamefont
  {Ruks}}\ and\ \bibinfo {author} {\bibfnamefont {T.}~\bibnamefont {Busch}},\
  }\bibfield  {title} {\enquote {\bibinfo {title} {{Green's functions of and
  emission into discrete anisotropic and hyperbolic baths}},}\ }\href {\doibase
  10.1103/PhysRevResearch.4.023044} {\bibfield  {journal} {\bibinfo  {journal}
  {Physical Review Research}\ }\textbf {\bibinfo {volume} {4}},\ \bibinfo
  {pages} {023044} (\bibinfo {year} {2022})}\BibitemShut {NoStop}%
\bibitem [{\citenamefont {Vega}\ \emph {et~al.}(2023)\citenamefont {Vega},
  \citenamefont {Porras},\ and\ \citenamefont
  {Gonz{\'{a}}lez-Tudela}}]{Vega2023}%
  \BibitemOpen
  \bibfield  {author} {\bibinfo {author} {\bibfnamefont {C.}~\bibnamefont
  {Vega}}, \bibinfo {author} {\bibfnamefont {D.}~\bibnamefont {Porras}}, \ and\
  \bibinfo {author} {\bibfnamefont {A.}~\bibnamefont {Gonz{\'{a}}lez-Tudela}},\
  }\bibfield  {title} {\enquote {\bibinfo {title} {{Topological multimode
  waveguide QED}},}\ }\href {\doibase 10.1103/PhysRevResearch.5.023031}
  {\bibfield  {journal} {\bibinfo  {journal} {Physical Review Research}\
  }\textbf {\bibinfo {volume} {5}},\ \bibinfo {pages} {023031} (\bibinfo {year}
  {2023})}\BibitemShut {NoStop}%
\bibitem [{\citenamefont {Windt}\ \emph {et~al.}(2024)\citenamefont {Windt},
  \citenamefont {Bello}, \citenamefont {Demler},\ and\ \citenamefont
  {Cirac}}]{Windt2024}%
  \BibitemOpen
  \bibfield  {author} {\bibinfo {author} {\bibfnamefont {B.}~\bibnamefont
  {Windt}}, \bibinfo {author} {\bibfnamefont {M.}~\bibnamefont {Bello}},
  \bibinfo {author} {\bibfnamefont {E.}~\bibnamefont {Demler}}, \ and\ \bibinfo
  {author} {\bibfnamefont {J.~I.}\ \bibnamefont {Cirac}},\ }\bibfield  {title}
  {\enquote {\bibinfo {title} {{Fermionic matter-wave quantum optics with
  cold-atom impurity models}},}\ }\href {\doibase 10.1103/PhysRevA.109.023306}
  {\bibfield  {journal} {\bibinfo  {journal} {Physical Review A}\ }\textbf
  {\bibinfo {volume} {109}},\ \bibinfo {pages} {023306} (\bibinfo {year}
  {2024})}\BibitemShut {NoStop}%
\bibitem [{\citenamefont {Gonz{\'{a}}lez-Tudela}\ and\ \citenamefont
  {Cirac}(2018{\natexlab{a}})}]{Gonzalez-Tudela2018a}%
  \BibitemOpen
  \bibfield  {author} {\bibinfo {author} {\bibfnamefont {A.}~\bibnamefont
  {Gonz{\'{a}}lez-Tudela}}\ and\ \bibinfo {author} {\bibfnamefont {J.~I.}\
  \bibnamefont {Cirac}},\ }\bibfield  {title} {\enquote {\bibinfo {title}
  {{Exotic quantum dynamics and purely long-range coherent interactions in
  Dirac conelike baths}},}\ }\href {\doibase 10.1103/PhysRevA.97.043831}
  {\bibfield  {journal} {\bibinfo  {journal} {Physical Review A}\ }\textbf
  {\bibinfo {volume} {97}},\ \bibinfo {pages} {043831} (\bibinfo {year}
  {2018}{\natexlab{a}})}\BibitemShut {NoStop}%
\bibitem [{\citenamefont {Bienias}\ \emph {et~al.}(2022)\citenamefont
  {Bienias}, \citenamefont {Boettcher}, \citenamefont {Belyansky},
  \citenamefont {Koll{\'{a}}r},\ and\ \citenamefont {Gorshkov}}]{Bienias2022}%
  \BibitemOpen
  \bibfield  {author} {\bibinfo {author} {\bibfnamefont {P.}~\bibnamefont
  {Bienias}}, \bibinfo {author} {\bibfnamefont {I.}~\bibnamefont {Boettcher}},
  \bibinfo {author} {\bibfnamefont {R.}~\bibnamefont {Belyansky}}, \bibinfo
  {author} {\bibfnamefont {A.~J.}\ \bibnamefont {Koll{\'{a}}r}}, \ and\
  \bibinfo {author} {\bibfnamefont {A.~V.}\ \bibnamefont {Gorshkov}},\
  }\bibfield  {title} {\enquote {\bibinfo {title} {{Circuit Quantum
  Electrodynamics in Hyperbolic Space: From Photon Bound States to Frustrated
  Spin Models}},}\ }\href {\doibase 10.1103/PhysRevLett.128.013601} {\bibfield
  {journal} {\bibinfo  {journal} {Physical Review Letters}\ }\textbf {\bibinfo
  {volume} {128}},\ \bibinfo {pages} {013601} (\bibinfo {year}
  {2022})}\BibitemShut {NoStop}%
\bibitem [{\citenamefont {{De Bernardis}}\ \emph {et~al.}(2023)\citenamefont
  {{De Bernardis}}, \citenamefont {Piccioli}, \citenamefont {Rabl},\ and\
  \citenamefont {Carusotto}}]{DeBernardis2023}%
  \BibitemOpen
  \bibfield  {author} {\bibinfo {author} {\bibfnamefont {D.}~\bibnamefont {{De
  Bernardis}}}, \bibinfo {author} {\bibfnamefont {F.~S.}\ \bibnamefont
  {Piccioli}}, \bibinfo {author} {\bibfnamefont {P.}~\bibnamefont {Rabl}}, \
  and\ \bibinfo {author} {\bibfnamefont {I.}~\bibnamefont {Carusotto}},\
  }\bibfield  {title} {\enquote {\bibinfo {title} {{Chiral Quantum Optics in
  the Bulk of Photonic Quantum Hall Systems}},}\ }\href {\doibase
  10.1103/PRXQuantum.4.030306} {\bibfield  {journal} {\bibinfo  {journal} {PRX
  Quantum}\ }\textbf {\bibinfo {volume} {4}},\ \bibinfo {pages} {030306}
  (\bibinfo {year} {2023})}\BibitemShut {NoStop}%
\bibitem [{\citenamefont {Te{\v{c}}er}\ \emph {et~al.}(2023)\citenamefont
  {Te{\v{c}}er}, \citenamefont {{Di Liberto}}, \citenamefont {Silvi},
  \citenamefont {Montangero}, \citenamefont {Romanato},\ and\ \citenamefont
  {Calaj{\`{o}}}}]{Tecer2023}%
  \BibitemOpen
  \bibfield  {author} {\bibinfo {author} {\bibfnamefont {M.}~\bibnamefont
  {Te{\v{c}}er}}, \bibinfo {author} {\bibfnamefont {M.}~\bibnamefont {{Di
  Liberto}}}, \bibinfo {author} {\bibfnamefont {P.}~\bibnamefont {Silvi}},
  \bibinfo {author} {\bibfnamefont {S.}~\bibnamefont {Montangero}}, \bibinfo
  {author} {\bibfnamefont {F.}~\bibnamefont {Romanato}}, \ and\ \bibinfo
  {author} {\bibfnamefont {G.}~\bibnamefont {Calaj{\`{o}}}},\ }\href@noop {}
  {\enquote {\bibinfo {title} {{Strongly interacting photons in 2D waveguide
  QED}},}\ } (\bibinfo {year} {2023}),\ \Eprint
  {http://arxiv.org/abs/2312.07668} {arXiv:2312.07668} \BibitemShut {NoStop}%
\bibitem [{\citenamefont {Gonz{\'{a}}lez-Tudela}\ and\ \citenamefont
  {Cirac}(2018{\natexlab{b}})}]{Gonzalez-Tudela2018}%
  \BibitemOpen
  \bibfield  {author} {\bibinfo {author} {\bibfnamefont {A.}~\bibnamefont
  {Gonz{\'{a}}lez-Tudela}}\ and\ \bibinfo {author} {\bibfnamefont {J.~I.}\
  \bibnamefont {Cirac}},\ }\bibfield  {title} {\enquote {\bibinfo {title}
  {{Non-Markovian Quantum Optics with Three-Dimensional State-Dependent Optical
  Lattices}},}\ }\href {\doibase 10.22331/q-2018-10-01-97} {\bibfield
  {journal} {\bibinfo  {journal} {Quantum}\ }\textbf {\bibinfo {volume} {2}},\
  \bibinfo {pages} {97} (\bibinfo {year} {2018}{\natexlab{b}})}\BibitemShut
  {NoStop}%
\bibitem [{\citenamefont {Garc{\'{i}}a-Elcano}\ \emph
  {et~al.}(2020)\citenamefont {Garc{\'{i}}a-Elcano}, \citenamefont
  {Gonz{\'{a}}lez-Tudela},\ and\ \citenamefont
  {Bravo-Abad}}]{Garcia-Elcano2020}%
  \BibitemOpen
  \bibfield  {author} {\bibinfo {author} {\bibfnamefont {I.}~\bibnamefont
  {Garc{\'{i}}a-Elcano}}, \bibinfo {author} {\bibfnamefont {A.}~\bibnamefont
  {Gonz{\'{a}}lez-Tudela}}, \ and\ \bibinfo {author} {\bibfnamefont
  {J.}~\bibnamefont {Bravo-Abad}},\ }\bibfield  {title} {\enquote {\bibinfo
  {title} {{Tunable and Robust Long-Range Coherent Interactions between Quantum
  Emitters Mediated by Weyl Bound States}},}\ }\href {\doibase
  10.1103/PhysRevLett.125.163602} {\bibfield  {journal} {\bibinfo  {journal}
  {Physical Review Letters}\ }\textbf {\bibinfo {volume} {125}},\ \bibinfo
  {pages} {163602} (\bibinfo {year} {2020})}\BibitemShut {NoStop}%
\bibitem [{\citenamefont {Garc{\'{i}}a-Elcano}\ \emph
  {et~al.}(2023)\citenamefont {Garc{\'{i}}a-Elcano}, \citenamefont {Merino},
  \citenamefont {Bravo-Abad},\ and\ \citenamefont
  {Gonz{\'{a}}lez-Tudela}}]{Garcia-Elcano2023}%
  \BibitemOpen
  \bibfield  {author} {\bibinfo {author} {\bibfnamefont {I.}~\bibnamefont
  {Garc{\'{i}}a-Elcano}}, \bibinfo {author} {\bibfnamefont {J.}~\bibnamefont
  {Merino}}, \bibinfo {author} {\bibfnamefont {J.}~\bibnamefont {Bravo-Abad}},
  \ and\ \bibinfo {author} {\bibfnamefont {A.}~\bibnamefont
  {Gonz{\'{a}}lez-Tudela}},\ }\bibfield  {title} {\enquote {\bibinfo {title}
  {{Probing and harnessing photonic Fermi arc surface states using light-matter
  interactions}},}\ }\href {\doibase 10.1126/sciadv.adf8257} {\bibfield
  {journal} {\bibinfo  {journal} {Science Advances}\ }\textbf {\bibinfo
  {volume} {9}},\ \bibinfo {pages} {adf8257} (\bibinfo {year}
  {2023})}\BibitemShut {NoStop}%
\bibitem [{\citenamefont {von Neumann}\ and\ \citenamefont
  {Wigner}(1929)}]{vonNeumann1929}%
  \BibitemOpen
  \bibfield  {author} {\bibinfo {author} {\bibfnamefont {J.}~\bibnamefont {von
  Neumann}}\ and\ \bibinfo {author} {\bibfnamefont {E.~P.}\ \bibnamefont
  {Wigner}},\ }\bibfield  {title} {\enquote {\bibinfo {title} {{{\"{U}}ber
  merkw{\"{u}}rdige diskrete Eigenwerte}},}\ }\href@noop {} {\bibfield
  {journal} {\bibinfo  {journal} {Physikalische Zeitschrift}\ }\textbf
  {\bibinfo {volume} {30}},\ \bibinfo {pages} {467} (\bibinfo {year}
  {1929})}\BibitemShut {NoStop}%
\bibitem [{\citenamefont {Hsu}\ \emph {et~al.}(2016)\citenamefont {Hsu},
  \citenamefont {Zhen}, \citenamefont {Stone}, \citenamefont {Joannopoulos},\
  and\ \citenamefont {Solja{\v{c}}i{\'{c}}}}]{Hsu2016}%
  \BibitemOpen
  \bibfield  {author} {\bibinfo {author} {\bibfnamefont {C.~W.}\ \bibnamefont
  {Hsu}}, \bibinfo {author} {\bibfnamefont {B.}~\bibnamefont {Zhen}}, \bibinfo
  {author} {\bibfnamefont {A.~D.}\ \bibnamefont {Stone}}, \bibinfo {author}
  {\bibfnamefont {J.~D.}\ \bibnamefont {Joannopoulos}}, \ and\ \bibinfo
  {author} {\bibfnamefont {M.}~\bibnamefont {Solja{\v{c}}i{\'{c}}}},\
  }\bibfield  {title} {\enquote {\bibinfo {title} {{Bound states in the
  continuum}},}\ }\href {\doibase 10.1038/natrevmats.2016.48} {\bibfield
  {journal} {\bibinfo  {journal} {Nature Reviews Materials}\ }\textbf {\bibinfo
  {volume} {1}},\ \bibinfo {pages} {16048} (\bibinfo {year}
  {2016})}\BibitemShut {NoStop}%
\bibitem [{\citenamefont {Kang}\ \emph {et~al.}(2023)\citenamefont {Kang},
  \citenamefont {Liu}, \citenamefont {Chan},\ and\ \citenamefont
  {Xiao}}]{Kang2023}%
  \BibitemOpen
  \bibfield  {author} {\bibinfo {author} {\bibfnamefont {M.}~\bibnamefont
  {Kang}}, \bibinfo {author} {\bibfnamefont {T.}~\bibnamefont {Liu}}, \bibinfo
  {author} {\bibfnamefont {C.~T.}\ \bibnamefont {Chan}}, \ and\ \bibinfo
  {author} {\bibfnamefont {M.}~\bibnamefont {Xiao}},\ }\bibfield  {title}
  {\enquote {\bibinfo {title} {{Applications of bound states in the continuum
  in photonics}},}\ }\href {\doibase 10.1038/s42254-023-00642-8} {\bibfield
  {journal} {\bibinfo  {journal} {Nature Reviews Physics}\ }\textbf {\bibinfo
  {volume} {5}},\ \bibinfo {pages} {659} (\bibinfo {year} {2023})}\BibitemShut
  {NoStop}%
\bibitem [{\citenamefont {Xu}\ \emph {et~al.}(2023)\citenamefont {Xu},
  \citenamefont {Xing}, \citenamefont {Xue}, \citenamefont {Lu}, \citenamefont
  {Fan}, \citenamefont {Fan}, \citenamefont {Shum},\ and\ \citenamefont
  {Cong}}]{Xu2023}%
  \BibitemOpen
  \bibfield  {author} {\bibinfo {author} {\bibfnamefont {G.}~\bibnamefont
  {Xu}}, \bibinfo {author} {\bibfnamefont {H.}~\bibnamefont {Xing}}, \bibinfo
  {author} {\bibfnamefont {Z.}~\bibnamefont {Xue}}, \bibinfo {author}
  {\bibfnamefont {D.}~\bibnamefont {Lu}}, \bibinfo {author} {\bibfnamefont
  {J.}~\bibnamefont {Fan}}, \bibinfo {author} {\bibfnamefont {J.}~\bibnamefont
  {Fan}}, \bibinfo {author} {\bibfnamefont {P.~P.}\ \bibnamefont {Shum}}, \
  and\ \bibinfo {author} {\bibfnamefont {L.}~\bibnamefont {Cong}},\ }\bibfield
  {title} {\enquote {\bibinfo {title} {{Recent Advances and Perspective of
  Photonic Bound States in the Continuum}},}\ }\href {\doibase
  10.34133/ultrafastscience.0033} {\bibfield  {journal} {\bibinfo  {journal}
  {Ultrafast Science}\ }\textbf {\bibinfo {volume} {3}},\ \bibinfo {pages}
  {0033} (\bibinfo {year} {2023})}\BibitemShut {NoStop}%
\bibitem [{\citenamefont {Roushan}\ \emph {et~al.}(2017)\citenamefont
  {Roushan}, \citenamefont {Neill}, \citenamefont {Megrant}, \citenamefont
  {Chen}, \citenamefont {Babbush}, \citenamefont {Barends}, \citenamefont
  {Campbell}, \citenamefont {Chen}, \citenamefont {Chiaro}, \citenamefont
  {Dunsworth}, \citenamefont {Fowler}, \citenamefont {Jeffrey}, \citenamefont
  {Kelly}, \citenamefont {Lucero}, \citenamefont {Mutus}, \citenamefont
  {O'Malley}, \citenamefont {Neeley}, \citenamefont {Quintana}, \citenamefont
  {Sank}, \citenamefont {Vainsencher}, \citenamefont {Wenner}, \citenamefont
  {White}, \citenamefont {Kapit}, \citenamefont {Neven},\ and\ \citenamefont
  {Martinis}}]{Roushan2017}%
  \BibitemOpen
  \bibfield  {author} {\bibinfo {author} {\bibfnamefont {P.}~\bibnamefont
  {Roushan}}, \bibinfo {author} {\bibfnamefont {C.}~\bibnamefont {Neill}},
  \bibinfo {author} {\bibfnamefont {A.}~\bibnamefont {Megrant}}, \bibinfo
  {author} {\bibfnamefont {Y.}~\bibnamefont {Chen}}, \bibinfo {author}
  {\bibfnamefont {R.}~\bibnamefont {Babbush}}, \bibinfo {author} {\bibfnamefont
  {R.}~\bibnamefont {Barends}}, \bibinfo {author} {\bibfnamefont
  {B.}~\bibnamefont {Campbell}}, \bibinfo {author} {\bibfnamefont
  {Z.}~\bibnamefont {Chen}}, \bibinfo {author} {\bibfnamefont {B.}~\bibnamefont
  {Chiaro}}, \bibinfo {author} {\bibfnamefont {A.}~\bibnamefont {Dunsworth}},
  \bibinfo {author} {\bibfnamefont {A.}~\bibnamefont {Fowler}}, \bibinfo
  {author} {\bibfnamefont {E.}~\bibnamefont {Jeffrey}}, \bibinfo {author}
  {\bibfnamefont {J.}~\bibnamefont {Kelly}}, \bibinfo {author} {\bibfnamefont
  {E.}~\bibnamefont {Lucero}}, \bibinfo {author} {\bibfnamefont
  {J.}~\bibnamefont {Mutus}}, \bibinfo {author} {\bibfnamefont {P.~J.~J.}\
  \bibnamefont {O'Malley}}, \bibinfo {author} {\bibfnamefont {M.}~\bibnamefont
  {Neeley}}, \bibinfo {author} {\bibfnamefont {C.}~\bibnamefont {Quintana}},
  \bibinfo {author} {\bibfnamefont {D.}~\bibnamefont {Sank}}, \bibinfo {author}
  {\bibfnamefont {A.}~\bibnamefont {Vainsencher}}, \bibinfo {author}
  {\bibfnamefont {J.}~\bibnamefont {Wenner}}, \bibinfo {author} {\bibfnamefont
  {T.}~\bibnamefont {White}}, \bibinfo {author} {\bibfnamefont
  {E.}~\bibnamefont {Kapit}}, \bibinfo {author} {\bibfnamefont
  {H.}~\bibnamefont {Neven}}, \ and\ \bibinfo {author} {\bibfnamefont
  {J.}~\bibnamefont {Martinis}},\ }\bibfield  {title} {\enquote {\bibinfo
  {title} {Chiral ground-state currents of interacting photons in a synthetic
  magnetic field},}\ }\href {\doibase 10.1038/nphys3930} {\bibfield  {journal}
  {\bibinfo  {journal} {Nature Physics}\ }\textbf {\bibinfo {volume} {13}},\
  \bibinfo {pages} {146} (\bibinfo {year} {2017})}\BibitemShut {NoStop}%
\bibitem [{\citenamefont {Lodahl}\ \emph {et~al.}(2017)\citenamefont {Lodahl},
  \citenamefont {Mahmoodian}, \citenamefont {Stobbe}, \citenamefont
  {Rauschenbeutel}, \citenamefont {Schneeweiss}, \citenamefont {Volz},
  \citenamefont {Pichler},\ and\ \citenamefont {Zoller}}]{Lodahl2017}%
  \BibitemOpen
  \bibfield  {author} {\bibinfo {author} {\bibfnamefont {P.}~\bibnamefont
  {Lodahl}}, \bibinfo {author} {\bibfnamefont {S.}~\bibnamefont {Mahmoodian}},
  \bibinfo {author} {\bibfnamefont {S.}~\bibnamefont {Stobbe}}, \bibinfo
  {author} {\bibfnamefont {A.}~\bibnamefont {Rauschenbeutel}}, \bibinfo
  {author} {\bibfnamefont {P.}~\bibnamefont {Schneeweiss}}, \bibinfo {author}
  {\bibfnamefont {J.}~\bibnamefont {Volz}}, \bibinfo {author} {\bibfnamefont
  {H.}~\bibnamefont {Pichler}}, \ and\ \bibinfo {author} {\bibfnamefont
  {P.}~\bibnamefont {Zoller}},\ }\bibfield  {title} {\enquote {\bibinfo {title}
  {{Chiral quantum optics}},}\ }\href {\doibase 10.1038/nature21037} {\bibfield
   {journal} {\bibinfo  {journal} {Nature}\ }\textbf {\bibinfo {volume}
  {541}},\ \bibinfo {pages} {473} (\bibinfo {year} {2017})}\BibitemShut
  {NoStop}%
\bibitem [{\citenamefont {Hood}\ \emph {et~al.}(2016)\citenamefont {Hood},
  \citenamefont {Goban}, \citenamefont {Asenjo-Garcia}, \citenamefont {Lu},
  \citenamefont {Yu}, \citenamefont {Chang},\ and\ \citenamefont
  {Kimble}}]{Hood2016}%
  \BibitemOpen
  \bibfield  {author} {\bibinfo {author} {\bibfnamefont {J.~D.}\ \bibnamefont
  {Hood}}, \bibinfo {author} {\bibfnamefont {A.}~\bibnamefont {Goban}},
  \bibinfo {author} {\bibfnamefont {A.}~\bibnamefont {Asenjo-Garcia}}, \bibinfo
  {author} {\bibfnamefont {M.}~\bibnamefont {Lu}}, \bibinfo {author}
  {\bibfnamefont {S.~P.}\ \bibnamefont {Yu}}, \bibinfo {author} {\bibfnamefont
  {D.~E.}\ \bibnamefont {Chang}}, \ and\ \bibinfo {author} {\bibfnamefont
  {H.~J.}\ \bibnamefont {Kimble}},\ }\bibfield  {title} {\enquote {\bibinfo
  {title} {{Atom-atom interactions around the band edge of a photonic crystal
  waveguide}},}\ }\href {\doibase 10.1073/pnas.1603788113} {\bibfield
  {journal} {\bibinfo  {journal} {Proceedings of the National Academy of
  Sciences of the United States of America}\ }\textbf {\bibinfo {volume}
  {113}},\ \bibinfo {pages} {10507} (\bibinfo {year} {2016})}\BibitemShut
  {NoStop}%
\bibitem [{\citenamefont {Krinner}\ \emph {et~al.}(2018)\citenamefont
  {Krinner}, \citenamefont {Stewart}, \citenamefont {Pazmi{\~{n}}o},
  \citenamefont {Kwon},\ and\ \citenamefont {Schneble}}]{Krinner2018}%
  \BibitemOpen
  \bibfield  {author} {\bibinfo {author} {\bibfnamefont {L.}~\bibnamefont
  {Krinner}}, \bibinfo {author} {\bibfnamefont {M.}~\bibnamefont {Stewart}},
  \bibinfo {author} {\bibfnamefont {A.}~\bibnamefont {Pazmi{\~{n}}o}}, \bibinfo
  {author} {\bibfnamefont {J.}~\bibnamefont {Kwon}}, \ and\ \bibinfo {author}
  {\bibfnamefont {D.}~\bibnamefont {Schneble}},\ }\bibfield  {title} {\enquote
  {\bibinfo {title} {{Spontaneous emission of matter waves from a tunable open
  quantum system}},}\ }\href {\doibase 10.1038/s41586-018-0348-z} {\bibfield
  {journal} {\bibinfo  {journal} {Nature}\ }\textbf {\bibinfo {volume} {559}},\
  \bibinfo {pages} {589} (\bibinfo {year} {2018})}\BibitemShut {NoStop}%
\bibitem [{\citenamefont {Stewart}\ \emph {et~al.}(2020)\citenamefont
  {Stewart}, \citenamefont {Kwon}, \citenamefont {Lanuza},\ and\ \citenamefont
  {Schneble}}]{Stewart2020}%
  \BibitemOpen
  \bibfield  {author} {\bibinfo {author} {\bibfnamefont {M.}~\bibnamefont
  {Stewart}}, \bibinfo {author} {\bibfnamefont {J.}~\bibnamefont {Kwon}},
  \bibinfo {author} {\bibfnamefont {A.}~\bibnamefont {Lanuza}}, \ and\ \bibinfo
  {author} {\bibfnamefont {D.}~\bibnamefont {Schneble}},\ }\bibfield  {title}
  {\enquote {\bibinfo {title} {{Dynamics of matter-wave quantum emitters in a
  structured vacuum}},}\ }\href {\doibase 10.1103/physrevresearch.2.043307}
  {\bibfield  {journal} {\bibinfo  {journal} {Physical Review Research}\
  }\textbf {\bibinfo {volume} {2}},\ \bibinfo {pages} {043307} (\bibinfo {year}
  {2020})}\BibitemShut {NoStop}%
\bibitem [{\citenamefont {Liu}\ and\ \citenamefont {Houck}(2017)}]{Liu2017}%
  \BibitemOpen
  \bibfield  {author} {\bibinfo {author} {\bibfnamefont {Y.}~\bibnamefont
  {Liu}}\ and\ \bibinfo {author} {\bibfnamefont {A.~A.}\ \bibnamefont
  {Houck}},\ }\bibfield  {title} {\enquote {\bibinfo {title} {{Quantum
  electrodynamics near a photonic bandgap}},}\ }\href {\doibase
  10.1038/nphys3834} {\bibfield  {journal} {\bibinfo  {journal} {Nature
  Physics}\ }\textbf {\bibinfo {volume} {13}},\ \bibinfo {pages} {48} (\bibinfo
  {year} {2017})}\BibitemShut {NoStop}%
\bibitem [{\citenamefont {Sundaresan}\ \emph {et~al.}(2019)\citenamefont
  {Sundaresan}, \citenamefont {Lundgren}, \citenamefont {Zhu}, \citenamefont
  {Gorshkov},\ and\ \citenamefont {Houck}}]{Sundaresan2019}%
  \BibitemOpen
  \bibfield  {author} {\bibinfo {author} {\bibfnamefont {N.~M.}\ \bibnamefont
  {Sundaresan}}, \bibinfo {author} {\bibfnamefont {R.}~\bibnamefont
  {Lundgren}}, \bibinfo {author} {\bibfnamefont {G.}~\bibnamefont {Zhu}},
  \bibinfo {author} {\bibfnamefont {A.~V.}\ \bibnamefont {Gorshkov}}, \ and\
  \bibinfo {author} {\bibfnamefont {A.~A.}\ \bibnamefont {Houck}},\ }\bibfield
  {title} {\enquote {\bibinfo {title} {{Interacting Qubit-Photon Bound States
  with Superconducting Circuits}},}\ }\href {\doibase
  10.1103/PhysRevX.9.011021} {\bibfield  {journal} {\bibinfo  {journal}
  {Physical Review X}\ }\textbf {\bibinfo {volume} {9}},\ \bibinfo {pages}
  {011021} (\bibinfo {year} {2019})}\BibitemShut {NoStop}%
\bibitem [{\citenamefont {Harrington}\ \emph {et~al.}(2019)\citenamefont
  {Harrington}, \citenamefont {Naghiloo}, \citenamefont {Tan},\ and\
  \citenamefont {Murch}}]{Harrington2019}%
  \BibitemOpen
  \bibfield  {author} {\bibinfo {author} {\bibfnamefont {P.~M.}\ \bibnamefont
  {Harrington}}, \bibinfo {author} {\bibfnamefont {M.}~\bibnamefont
  {Naghiloo}}, \bibinfo {author} {\bibfnamefont {D.}~\bibnamefont {Tan}}, \
  and\ \bibinfo {author} {\bibfnamefont {K.~W.}\ \bibnamefont {Murch}},\
  }\bibfield  {title} {\enquote {\bibinfo {title} {{Bath engineering of a
  fluorescing artificial atom with a photonic crystal}},}\ }\href {\doibase
  10.1103/PhysRevA.99.052126} {\bibfield  {journal} {\bibinfo  {journal}
  {Physical Review A}\ }\textbf {\bibinfo {volume} {99}},\ \bibinfo {pages}
  {052126} (\bibinfo {year} {2019})}\BibitemShut {NoStop}%
\bibitem [{\citenamefont {Koll{\'{a}}r}\ \emph {et~al.}(2019)\citenamefont
  {Koll{\'{a}}r}, \citenamefont {Fitzpatrick},\ and\ \citenamefont
  {Houck}}]{Kollar2019}%
  \BibitemOpen
  \bibfield  {author} {\bibinfo {author} {\bibfnamefont {A.~J.}\ \bibnamefont
  {Koll{\'{a}}r}}, \bibinfo {author} {\bibfnamefont {M.}~\bibnamefont
  {Fitzpatrick}}, \ and\ \bibinfo {author} {\bibfnamefont {A.~A.}\ \bibnamefont
  {Houck}},\ }\bibfield  {title} {\enquote {\bibinfo {title} {{Hyperbolic
  lattices in circuit quantum electrodynamics}},}\ }\href {\doibase
  10.1038/s41586-019-1348-3} {\bibfield  {journal} {\bibinfo  {journal}
  {Nature}\ }\textbf {\bibinfo {volume} {571}},\ \bibinfo {pages} {45}
  (\bibinfo {year} {2019})}\BibitemShut {NoStop}%
\bibitem [{\citenamefont {Carusotto}\ \emph {et~al.}(2020)\citenamefont
  {Carusotto}, \citenamefont {Houck}, \citenamefont {Koll{\'{a}}r},
  \citenamefont {Roushan}, \citenamefont {Schuster},\ and\ \citenamefont
  {Simon}}]{Carusotto2020}%
  \BibitemOpen
  \bibfield  {author} {\bibinfo {author} {\bibfnamefont {I.}~\bibnamefont
  {Carusotto}}, \bibinfo {author} {\bibfnamefont {A.~A.}\ \bibnamefont
  {Houck}}, \bibinfo {author} {\bibfnamefont {A.~J.}\ \bibnamefont
  {Koll{\'{a}}r}}, \bibinfo {author} {\bibfnamefont {P.}~\bibnamefont
  {Roushan}}, \bibinfo {author} {\bibfnamefont {D.~I.}\ \bibnamefont
  {Schuster}}, \ and\ \bibinfo {author} {\bibfnamefont {J.}~\bibnamefont
  {Simon}},\ }\bibfield  {title} {\enquote {\bibinfo {title} {{Photonic
  materials in circuit quantum electrodynamics}},}\ }\href {\doibase
  10.1038/s41567-020-0815-y} {\bibfield  {journal} {\bibinfo  {journal} {Nature
  Physics}\ }\textbf {\bibinfo {volume} {16}},\ \bibinfo {pages} {268}
  (\bibinfo {year} {2020})}\BibitemShut {NoStop}%
\bibitem [{\citenamefont {Mirhosseini}\ \emph {et~al.}(2018)\citenamefont
  {Mirhosseini}, \citenamefont {Kim}, \citenamefont {Ferreira}, \citenamefont
  {Kalaee}, \citenamefont {Sipahigil}, \citenamefont {Keller},\ and\
  \citenamefont {Painter}}]{Mirhosseini2018}%
  \BibitemOpen
  \bibfield  {author} {\bibinfo {author} {\bibfnamefont {M.}~\bibnamefont
  {Mirhosseini}}, \bibinfo {author} {\bibfnamefont {E.}~\bibnamefont {Kim}},
  \bibinfo {author} {\bibfnamefont {V.~S.}\ \bibnamefont {Ferreira}}, \bibinfo
  {author} {\bibfnamefont {M.}~\bibnamefont {Kalaee}}, \bibinfo {author}
  {\bibfnamefont {A.}~\bibnamefont {Sipahigil}}, \bibinfo {author}
  {\bibfnamefont {A.~J.}\ \bibnamefont {Keller}}, \ and\ \bibinfo {author}
  {\bibfnamefont {O.}~\bibnamefont {Painter}},\ }\bibfield  {title} {\enquote
  {\bibinfo {title} {{Superconducting metamaterials for waveguide quantum
  electrodynamics}},}\ }\href {\doibase 10.1038/s41467-018-06142-z} {\bibfield
  {journal} {\bibinfo  {journal} {Nature Communications}\ }\textbf {\bibinfo
  {volume} {9}},\ \bibinfo {pages} {3706} (\bibinfo {year} {2018})}\BibitemShut
  {NoStop}%
\bibitem [{\citenamefont {Indrajeet}\ \emph {et~al.}(2020)\citenamefont
  {Indrajeet}, \citenamefont {Wang}, \citenamefont {Hutchings}, \citenamefont
  {Taketani}, \citenamefont {Wilhelm}, \citenamefont {Lahaye},\ and\
  \citenamefont {Plourde}}]{Indrajeet2020}%
  \BibitemOpen
  \bibfield  {author} {\bibinfo {author} {\bibfnamefont {S.}~\bibnamefont
  {Indrajeet}}, \bibinfo {author} {\bibfnamefont {H.}~\bibnamefont {Wang}},
  \bibinfo {author} {\bibfnamefont {M.~D.}\ \bibnamefont {Hutchings}}, \bibinfo
  {author} {\bibfnamefont {B.~G.}\ \bibnamefont {Taketani}}, \bibinfo {author}
  {\bibfnamefont {F.~K.}\ \bibnamefont {Wilhelm}}, \bibinfo {author}
  {\bibfnamefont {M.~D.}\ \bibnamefont {Lahaye}}, \ and\ \bibinfo {author}
  {\bibfnamefont {B.~L.}\ \bibnamefont {Plourde}},\ }\bibfield  {title}
  {\enquote {\bibinfo {title} {{Coupling a Superconducting Qubit to a
  Left-Handed Metamaterial Resonator}},}\ }\href {\doibase
  10.1103/PhysRevApplied.14.064033} {\bibfield  {journal} {\bibinfo  {journal}
  {Physical Review Applied}\ }\textbf {\bibinfo {volume} {14}},\ \bibinfo
  {pages} {064033} (\bibinfo {year} {2020})}\BibitemShut {NoStop}%
\bibitem [{\citenamefont {Kim}\ \emph {et~al.}(2021)\citenamefont {Kim},
  \citenamefont {Zhang}, \citenamefont {Ferreira}, \citenamefont {Banker},
  \citenamefont {Iverson}, \citenamefont {Sipahigil}, \citenamefont {Bello},
  \citenamefont {Gonz{\'{a}}lez-Tudela}, \citenamefont {Mirhosseini},\ and\
  \citenamefont {Painter}}]{Kim2021}%
  \BibitemOpen
  \bibfield  {author} {\bibinfo {author} {\bibfnamefont {E.}~\bibnamefont
  {Kim}}, \bibinfo {author} {\bibfnamefont {X.}~\bibnamefont {Zhang}}, \bibinfo
  {author} {\bibfnamefont {V.~S.}\ \bibnamefont {Ferreira}}, \bibinfo {author}
  {\bibfnamefont {J.}~\bibnamefont {Banker}}, \bibinfo {author} {\bibfnamefont
  {J.~K.}\ \bibnamefont {Iverson}}, \bibinfo {author} {\bibfnamefont
  {A.}~\bibnamefont {Sipahigil}}, \bibinfo {author} {\bibfnamefont
  {M.}~\bibnamefont {Bello}}, \bibinfo {author} {\bibfnamefont
  {A.}~\bibnamefont {Gonz{\'{a}}lez-Tudela}}, \bibinfo {author} {\bibfnamefont
  {M.}~\bibnamefont {Mirhosseini}}, \ and\ \bibinfo {author} {\bibfnamefont
  {O.}~\bibnamefont {Painter}},\ }\bibfield  {title} {\enquote {\bibinfo
  {title} {{Quantum Electrodynamics in a Topological Waveguide}},}\ }\href
  {\doibase 10.1103/PhysRevX.11.011015} {\bibfield  {journal} {\bibinfo
  {journal} {Physical Review X}\ }\textbf {\bibinfo {volume} {11}},\ \bibinfo
  {pages} {011015} (\bibinfo {year} {2021})}\BibitemShut {NoStop}%
\bibitem [{\citenamefont {Ferreira}\ \emph {et~al.}(2021)\citenamefont
  {Ferreira}, \citenamefont {Banker}, \citenamefont {Sipahigil}, \citenamefont
  {Matheny}, \citenamefont {Keller}, \citenamefont {Kim}, \citenamefont
  {Mirhosseini},\ and\ \citenamefont {Painter}}]{Ferreira2021}%
  \BibitemOpen
  \bibfield  {author} {\bibinfo {author} {\bibfnamefont {V.~S.}\ \bibnamefont
  {Ferreira}}, \bibinfo {author} {\bibfnamefont {J.}~\bibnamefont {Banker}},
  \bibinfo {author} {\bibfnamefont {A.}~\bibnamefont {Sipahigil}}, \bibinfo
  {author} {\bibfnamefont {M.~H.}\ \bibnamefont {Matheny}}, \bibinfo {author}
  {\bibfnamefont {A.~J.}\ \bibnamefont {Keller}}, \bibinfo {author}
  {\bibfnamefont {E.}~\bibnamefont {Kim}}, \bibinfo {author} {\bibfnamefont
  {M.}~\bibnamefont {Mirhosseini}}, \ and\ \bibinfo {author} {\bibfnamefont
  {O.}~\bibnamefont {Painter}},\ }\bibfield  {title} {\enquote {\bibinfo
  {title} {{Collapse and Revival of an Artificial Atom Coupled to a Structured
  Photonic Reservoir}},}\ }\href {\doibase 10.1103/PhysRevX.11.041043}
  {\bibfield  {journal} {\bibinfo  {journal} {Physical Review X}\ }\textbf
  {\bibinfo {volume} {11}},\ \bibinfo {pages} {41043} (\bibinfo {year}
  {2021})}\BibitemShut {NoStop}%
\bibitem [{\citenamefont {Scigliuzzo}\ \emph {et~al.}(2022)\citenamefont
  {Scigliuzzo}, \citenamefont {Calaj{\`{o}}}, \citenamefont {Ciccarello},
  \citenamefont {{Perez Lozano}}, \citenamefont {Bengtsson}, \citenamefont
  {Scarlino}, \citenamefont {Wallraff}, \citenamefont {Chang}, \citenamefont
  {Delsing},\ and\ \citenamefont {Gasparinetti}}]{Scigliuzzo2022}%
  \BibitemOpen
  \bibfield  {author} {\bibinfo {author} {\bibfnamefont {M.}~\bibnamefont
  {Scigliuzzo}}, \bibinfo {author} {\bibfnamefont {G.}~\bibnamefont
  {Calaj{\`{o}}}}, \bibinfo {author} {\bibfnamefont {F.}~\bibnamefont
  {Ciccarello}}, \bibinfo {author} {\bibfnamefont {D.}~\bibnamefont {{Perez
  Lozano}}}, \bibinfo {author} {\bibfnamefont {A.}~\bibnamefont {Bengtsson}},
  \bibinfo {author} {\bibfnamefont {P.}~\bibnamefont {Scarlino}}, \bibinfo
  {author} {\bibfnamefont {A.}~\bibnamefont {Wallraff}}, \bibinfo {author}
  {\bibfnamefont {D.}~\bibnamefont {Chang}}, \bibinfo {author} {\bibfnamefont
  {P.}~\bibnamefont {Delsing}}, \ and\ \bibinfo {author} {\bibfnamefont
  {S.}~\bibnamefont {Gasparinetti}},\ }\bibfield  {title} {\enquote {\bibinfo
  {title} {{Controlling Atom-Photon Bound States in an Array of
  Josephson-Junction Resonators}},}\ }\href {\doibase
  10.1103/physrevx.12.031036} {\bibfield  {journal} {\bibinfo  {journal}
  {Physical Review X}\ }\textbf {\bibinfo {volume} {12}},\ \bibinfo {pages}
  {31036} (\bibinfo {year} {2022})}\BibitemShut {NoStop}%
\bibitem [{\citenamefont {Zhang}\ \emph
  {et~al.}(2023{\natexlab{a}})\citenamefont {Zhang}, \citenamefont {Kim},
  \citenamefont {Mark}, \citenamefont {Choi},\ and\ \citenamefont
  {Painter}}]{Zhang2023}%
  \BibitemOpen
  \bibfield  {author} {\bibinfo {author} {\bibfnamefont {X.}~\bibnamefont
  {Zhang}}, \bibinfo {author} {\bibfnamefont {E.}~\bibnamefont {Kim}}, \bibinfo
  {author} {\bibfnamefont {D.~K.}\ \bibnamefont {Mark}}, \bibinfo {author}
  {\bibfnamefont {S.}~\bibnamefont {Choi}}, \ and\ \bibinfo {author}
  {\bibfnamefont {O.}~\bibnamefont {Painter}},\ }\bibfield  {title} {\enquote
  {\bibinfo {title} {{A superconducting quantum simulator based on a
  photonic-bandgap metamaterial}},}\ }\href {\doibase 10.1126/science.ade7651}
  {\bibfield  {journal} {\bibinfo  {journal} {Science}\ }\textbf {\bibinfo
  {volume} {379}},\ \bibinfo {pages} {278} (\bibinfo {year}
  {2023}{\natexlab{a}})}\BibitemShut {NoStop}%
\bibitem [{\citenamefont {Jouanny}\ \emph {et~al.}(2024)\citenamefont
  {Jouanny}, \citenamefont {Frasca}, \citenamefont {Weibel}, \citenamefont
  {Peyruchat}, \citenamefont {Scigliuzzo}, \citenamefont {Oppliger},
  \citenamefont {Palma}, \citenamefont {Sbroggio}, \citenamefont {Beaulieu},
  \citenamefont {Zilberberg},\ and\ \citenamefont {Scarlino}}]{Jouanny2024}%
  \BibitemOpen
  \bibfield  {author} {\bibinfo {author} {\bibfnamefont {V.}~\bibnamefont
  {Jouanny}}, \bibinfo {author} {\bibfnamefont {S.}~\bibnamefont {Frasca}},
  \bibinfo {author} {\bibfnamefont {V.~J.}\ \bibnamefont {Weibel}}, \bibinfo
  {author} {\bibfnamefont {L.}~\bibnamefont {Peyruchat}}, \bibinfo {author}
  {\bibfnamefont {M.}~\bibnamefont {Scigliuzzo}}, \bibinfo {author}
  {\bibfnamefont {F.}~\bibnamefont {Oppliger}}, \bibinfo {author}
  {\bibfnamefont {F.~D.}\ \bibnamefont {Palma}}, \bibinfo {author}
  {\bibfnamefont {D.}~\bibnamefont {Sbroggio}}, \bibinfo {author}
  {\bibfnamefont {G.}~\bibnamefont {Beaulieu}}, \bibinfo {author}
  {\bibfnamefont {O.}~\bibnamefont {Zilberberg}}, \ and\ \bibinfo {author}
  {\bibfnamefont {P.}~\bibnamefont {Scarlino}},\ }\href@noop {} {\enquote
  {\bibinfo {title} {{Band engineering and study of disorder using topology in
  compact high kinetic inductance cavity arrays}},}\ } (\bibinfo {year}
  {2024}),\ \Eprint {http://arxiv.org/abs/2403.18150} {arXiv:2403.18150}
  \BibitemShut {NoStop}%
\bibitem [{\citenamefont {Rosenberg}\ \emph {et~al.}(2017)\citenamefont
  {Rosenberg}, \citenamefont {Kim}, \citenamefont {Das}, \citenamefont {Yost},
  \citenamefont {Gustavsson}, \citenamefont {Hover}, \citenamefont {Krantz},
  \citenamefont {Melville}, \citenamefont {Racz}, \citenamefont {Samach},
  \citenamefont {Weber}, \citenamefont {Yan}, \citenamefont {Yoder},
  \citenamefont {Kerman},\ and\ \citenamefont {Oliver}}]{Rosenberg2017}%
  \BibitemOpen
  \bibfield  {author} {\bibinfo {author} {\bibfnamefont {D.}~\bibnamefont
  {Rosenberg}}, \bibinfo {author} {\bibfnamefont {D.}~\bibnamefont {Kim}},
  \bibinfo {author} {\bibfnamefont {R.}~\bibnamefont {Das}}, \bibinfo {author}
  {\bibfnamefont {D.}~\bibnamefont {Yost}}, \bibinfo {author} {\bibfnamefont
  {S.}~\bibnamefont {Gustavsson}}, \bibinfo {author} {\bibfnamefont
  {D.}~\bibnamefont {Hover}}, \bibinfo {author} {\bibfnamefont
  {P.}~\bibnamefont {Krantz}}, \bibinfo {author} {\bibfnamefont
  {A.}~\bibnamefont {Melville}}, \bibinfo {author} {\bibfnamefont
  {L.}~\bibnamefont {Racz}}, \bibinfo {author} {\bibfnamefont {G.~O.}\
  \bibnamefont {Samach}}, \bibinfo {author} {\bibfnamefont {S.~J.}\
  \bibnamefont {Weber}}, \bibinfo {author} {\bibfnamefont {F.}~\bibnamefont
  {Yan}}, \bibinfo {author} {\bibfnamefont {J.~L.}\ \bibnamefont {Yoder}},
  \bibinfo {author} {\bibfnamefont {A.~J.}\ \bibnamefont {Kerman}}, \ and\
  \bibinfo {author} {\bibfnamefont {W.~D.}\ \bibnamefont {Oliver}},\ }\bibfield
   {title} {\enquote {\bibinfo {title} {{3D integrated superconducting
  qubits}},}\ }\href {\doibase 10.1038/s41534-017-0044-0} {\bibfield  {journal}
  {\bibinfo  {journal} {npj Quantum Information}\ }\textbf {\bibinfo {volume}
  {3}},\ \bibinfo {pages} {42} (\bibinfo {year} {2017})}\BibitemShut {NoStop}%
\bibitem [{\citenamefont {Rahamim}\ \emph {et~al.}(2017)\citenamefont
  {Rahamim}, \citenamefont {Behrle}, \citenamefont {Peterer}, \citenamefont
  {Patterson}, \citenamefont {Spring}, \citenamefont {Tsunoda}, \citenamefont
  {Manenti}, \citenamefont {Tancredi},\ and\ \citenamefont
  {Leek}}]{Rahamim2017}%
  \BibitemOpen
  \bibfield  {author} {\bibinfo {author} {\bibfnamefont {J.}~\bibnamefont
  {Rahamim}}, \bibinfo {author} {\bibfnamefont {T.}~\bibnamefont {Behrle}},
  \bibinfo {author} {\bibfnamefont {M.~J.}\ \bibnamefont {Peterer}}, \bibinfo
  {author} {\bibfnamefont {A.}~\bibnamefont {Patterson}}, \bibinfo {author}
  {\bibfnamefont {P.~A.}\ \bibnamefont {Spring}}, \bibinfo {author}
  {\bibfnamefont {T.}~\bibnamefont {Tsunoda}}, \bibinfo {author} {\bibfnamefont
  {R.}~\bibnamefont {Manenti}}, \bibinfo {author} {\bibfnamefont
  {G.}~\bibnamefont {Tancredi}}, \ and\ \bibinfo {author} {\bibfnamefont
  {P.~J.}\ \bibnamefont {Leek}},\ }\bibfield  {title} {\enquote {\bibinfo
  {title} {{Double-sided coaxial circuit QED with out-of-plane wiring}},}\
  }\href {\doibase 10.1063/1.4984299} {\bibfield  {journal} {\bibinfo
  {journal} {Applied Physics Letters}\ }\textbf {\bibinfo {volume} {110}},\
  \bibinfo {pages} {222602} (\bibinfo {year} {2017})}\BibitemShut {NoStop}%
\bibitem [{\citenamefont {Kosen}\ \emph {et~al.}(2022)\citenamefont {Kosen},
  \citenamefont {Li}, \citenamefont {Rommel}, \citenamefont {Shiri},
  \citenamefont {Warren}, \citenamefont {Gr{\"{o}}nberg}, \citenamefont
  {Salonen}, \citenamefont {Abad}, \citenamefont {Bizn{\'{a}}rov{\'{a}}},
  \citenamefont {Caputo}, \citenamefont {Chen}, \citenamefont {Grigoras},
  \citenamefont {Johansson}, \citenamefont {Kockum}, \citenamefont
  {Kri{\v{z}}an}, \citenamefont {Lozano}, \citenamefont {Norris}, \citenamefont
  {Osman}, \citenamefont {Fern{\'{a}}ndez-Pend{\'{a}}s}, \citenamefont
  {Ronzani}, \citenamefont {Roudsari}, \citenamefont {Simbierowicz},
  \citenamefont {Tancredi}, \citenamefont {Wallraff}, \citenamefont {Eichler},
  \citenamefont {Govenius},\ and\ \citenamefont {Bylander}}]{Kosen2022}%
  \BibitemOpen
  \bibfield  {author} {\bibinfo {author} {\bibfnamefont {S.}~\bibnamefont
  {Kosen}}, \bibinfo {author} {\bibfnamefont {H.-X.}\ \bibnamefont {Li}},
  \bibinfo {author} {\bibfnamefont {M.}~\bibnamefont {Rommel}}, \bibinfo
  {author} {\bibfnamefont {D.}~\bibnamefont {Shiri}}, \bibinfo {author}
  {\bibfnamefont {C.}~\bibnamefont {Warren}}, \bibinfo {author} {\bibfnamefont
  {L.}~\bibnamefont {Gr{\"{o}}nberg}}, \bibinfo {author} {\bibfnamefont
  {J.}~\bibnamefont {Salonen}}, \bibinfo {author} {\bibfnamefont
  {T.}~\bibnamefont {Abad}}, \bibinfo {author} {\bibfnamefont {J.}~\bibnamefont
  {Bizn{\'{a}}rov{\'{a}}}}, \bibinfo {author} {\bibfnamefont {M.}~\bibnamefont
  {Caputo}}, \bibinfo {author} {\bibfnamefont {L.}~\bibnamefont {Chen}},
  \bibinfo {author} {\bibfnamefont {K.}~\bibnamefont {Grigoras}}, \bibinfo
  {author} {\bibfnamefont {G.}~\bibnamefont {Johansson}}, \bibinfo {author}
  {\bibfnamefont {A.~F.}\ \bibnamefont {Kockum}}, \bibinfo {author}
  {\bibfnamefont {C.}~\bibnamefont {Kri{\v{z}}an}}, \bibinfo {author}
  {\bibfnamefont {D.~P.}\ \bibnamefont {Lozano}}, \bibinfo {author}
  {\bibfnamefont {G.~J.}\ \bibnamefont {Norris}}, \bibinfo {author}
  {\bibfnamefont {A.}~\bibnamefont {Osman}}, \bibinfo {author} {\bibfnamefont
  {J.}~\bibnamefont {Fern{\'{a}}ndez-Pend{\'{a}}s}}, \bibinfo {author}
  {\bibfnamefont {A.}~\bibnamefont {Ronzani}}, \bibinfo {author} {\bibfnamefont
  {A.~F.}\ \bibnamefont {Roudsari}}, \bibinfo {author} {\bibfnamefont
  {S.}~\bibnamefont {Simbierowicz}}, \bibinfo {author} {\bibfnamefont
  {G.}~\bibnamefont {Tancredi}}, \bibinfo {author} {\bibfnamefont
  {A.}~\bibnamefont {Wallraff}}, \bibinfo {author} {\bibfnamefont
  {C.}~\bibnamefont {Eichler}}, \bibinfo {author} {\bibfnamefont
  {J.}~\bibnamefont {Govenius}}, \ and\ \bibinfo {author} {\bibfnamefont
  {J.}~\bibnamefont {Bylander}},\ }\bibfield  {title} {\enquote {\bibinfo
  {title} {{Building blocks of a flip-chip integrated superconducting quantum
  processor}},}\ }\href {\doibase 10.1088/2058-9565/ac734b} {\bibfield
  {journal} {\bibinfo  {journal} {Quantum Science and Technology}\ }\textbf
  {\bibinfo {volume} {7}},\ \bibinfo {pages} {035018} (\bibinfo {year}
  {2022})}\BibitemShut {NoStop}%
\bibitem [{\citenamefont {Press}\ \emph {et~al.}(2007)\citenamefont {Press},
  \citenamefont {Teukolsky}, \citenamefont {Vettering},\ and\ \citenamefont
  {Flannery}}]{Press2007}%
  \BibitemOpen
  \bibfield  {author} {\bibinfo {author} {\bibfnamefont {W.~H.}\ \bibnamefont
  {Press}}, \bibinfo {author} {\bibfnamefont {S.~A.}\ \bibnamefont
  {Teukolsky}}, \bibinfo {author} {\bibfnamefont {W.~T.}\ \bibnamefont
  {Vettering}}, \ and\ \bibinfo {author} {\bibfnamefont {B.~P.}\ \bibnamefont
  {Flannery}},\ }\href@noop {} {\emph {\bibinfo {title} {Numerical Recipes: The
  Art of Scientific Computing}}},\ \bibinfo {edition} {3rd}\ ed.\ (\bibinfo
  {publisher} {Cambridge University Press},\ \bibinfo {year}
  {2007})\BibitemShut {NoStop}%
\bibitem [{\citenamefont {Hatano}\ and\ \citenamefont
  {Suzuki}(2005)}]{Hatano2005}%
  \BibitemOpen
  \bibfield  {author} {\bibinfo {author} {\bibfnamefont {N.}~\bibnamefont
  {Hatano}}\ and\ \bibinfo {author} {\bibfnamefont {M.}~\bibnamefont
  {Suzuki}},\ }\enquote {\bibinfo {title} {Finding exponential product formulas
  of higher orders},}\ in\ \href {\doibase 10.1007/11526216_2} {\emph {\bibinfo
  {booktitle} {Quantum Annealing and Other Optimization Methods}}},\ \bibinfo
  {editor} {edited by\ \bibinfo {editor} {\bibfnamefont {A.}~\bibnamefont
  {Das}}\ and\ \bibinfo {editor} {\bibfnamefont {B.}~\bibnamefont
  {K.~Chakrabarti}}}\ (\bibinfo  {publisher} {Springer Berlin Heidelberg},\
  \bibinfo {address} {Berlin, Heidelberg},\ \bibinfo {year} {2005})\ pp.\
  \bibinfo {pages} {37--68}\BibitemShut {NoStop}%
\bibitem [{\citenamefont {Al-Mohy}\ and\ \citenamefont
  {Higham}(2010)}]{AlMohy2010}%
  \BibitemOpen
  \bibfield  {author} {\bibinfo {author} {\bibfnamefont {A.}~\bibnamefont
  {Al-Mohy}}\ and\ \bibinfo {author} {\bibfnamefont {N.}~\bibnamefont
  {Higham}},\ }\bibfield  {title} {\enquote {\bibinfo {title} {{A New Scaling
  and Squaring Algorithm for the Matrix Exponential}},}\ }\href {\doibase
  10.1137/09074721X} {\bibfield  {journal} {\bibinfo  {journal} {SIAM Journal
  on Matrix Analysis and Applications}\ }\textbf {\bibinfo {volume} {31}},\
  \bibinfo {pages} {970} (\bibinfo {year} {2010})}\BibitemShut {NoStop}%
\bibitem [{\citenamefont {Moler}\ and\ \citenamefont {Loan}(2003)}]{Moler2003}%
  \BibitemOpen
  \bibfield  {author} {\bibinfo {author} {\bibfnamefont {C.}~\bibnamefont
  {Moler}}\ and\ \bibinfo {author} {\bibfnamefont {C.}~\bibnamefont {Loan}},\
  }\bibfield  {title} {\enquote {\bibinfo {title} {Nineteen dubious ways to
  compute the exponential of a matrix, twenty-five years later},}\ }\href
  {\doibase 10.1137/S00361445024180} {\bibfield  {journal} {\bibinfo  {journal}
  {Society for Industrial and Applied Mathematics}\ }\textbf {\bibinfo {volume}
  {45}},\ \bibinfo {pages} {3} (\bibinfo {year} {2003})}\BibitemShut {NoStop}%
\bibitem [{\citenamefont {Cooley}\ and\ \citenamefont
  {Tukey}(1965)}]{Cooley1965}%
  \BibitemOpen
  \bibfield  {author} {\bibinfo {author} {\bibfnamefont {J.~W.}\ \bibnamefont
  {Cooley}}\ and\ \bibinfo {author} {\bibfnamefont {J.~W.}\ \bibnamefont
  {Tukey}},\ }\bibfield  {title} {\enquote {\bibinfo {title} {An algorithm for
  the machine calculation of complex fourier series},}\ }\href
  {http://www.jstor.org/stable/2003354} {\bibfield  {journal} {\bibinfo
  {journal} {Mathematics of Computation}\ }\textbf {\bibinfo {volume} {19}},\
  \bibinfo {pages} {297} (\bibinfo {year} {1965})}\BibitemShut {NoStop}%
\bibitem [{\citenamefont {Dicke}(1954)}]{Dicke1954}%
  \BibitemOpen
  \bibfield  {author} {\bibinfo {author} {\bibfnamefont {R.~H.}\ \bibnamefont
  {Dicke}},\ }\bibfield  {title} {\enquote {\bibinfo {title} {{Coherence in
  Spontaneous Radiation Processes}},}\ }\href {\doibase 10.1103/PhysRev.93.99}
  {\bibfield  {journal} {\bibinfo  {journal} {Physical Review}\ }\textbf
  {\bibinfo {volume} {93}},\ \bibinfo {pages} {99} (\bibinfo {year}
  {1954})}\BibitemShut {NoStop}%
\bibitem [{\citenamefont {Gross}\ and\ \citenamefont
  {Haroche}(1982)}]{Gross1982}%
  \BibitemOpen
  \bibfield  {author} {\bibinfo {author} {\bibfnamefont {M.}~\bibnamefont
  {Gross}}\ and\ \bibinfo {author} {\bibfnamefont {S.}~\bibnamefont
  {Haroche}},\ }\bibfield  {title} {\enquote {\bibinfo {title} {{Superradiance:
  An essay on the theory of collective spontaneous emission}},}\ }\href
  {\doibase 10.1016/0370-1573(82)90102-8} {\bibfield  {journal} {\bibinfo
  {journal} {Physics Reports}\ }\textbf {\bibinfo {volume} {93}},\ \bibinfo
  {pages} {301} (\bibinfo {year} {1982})}\BibitemShut {NoStop}%
\bibitem [{\citenamefont {Zhang}\ \emph
  {et~al.}(2023{\natexlab{b}})\citenamefont {Zhang}, \citenamefont {Liu},
  \citenamefont {Gong},\ and\ \citenamefont {Wang}}]{Zhang2023a}%
  \BibitemOpen
  \bibfield  {author} {\bibinfo {author} {\bibfnamefont {X.}~\bibnamefont
  {Zhang}}, \bibinfo {author} {\bibfnamefont {C.}~\bibnamefont {Liu}}, \bibinfo
  {author} {\bibfnamefont {Z.}~\bibnamefont {Gong}}, \ and\ \bibinfo {author}
  {\bibfnamefont {Z.}~\bibnamefont {Wang}},\ }\bibfield  {title} {\enquote
  {\bibinfo {title} {{Quantum interference and controllable magic cavity QED
  via a giant atom in a coupled resonator waveguide}},}\ }\href {\doibase
  10.1103/PhysRevA.108.013704} {\bibfield  {journal} {\bibinfo  {journal}
  {Physical Review A}\ }\textbf {\bibinfo {volume} {108}},\ \bibinfo {pages}
  {013704} (\bibinfo {year} {2023}{\natexlab{b}})}\BibitemShut {NoStop}%
\bibitem [{\citenamefont {Du}\ \emph {et~al.}(2021)\citenamefont {Du},
  \citenamefont {Cai}, \citenamefont {Wu}, \citenamefont {Wang},\ and\
  \citenamefont {Li}}]{Du2021a}%
  \BibitemOpen
  \bibfield  {author} {\bibinfo {author} {\bibfnamefont {L.}~\bibnamefont
  {Du}}, \bibinfo {author} {\bibfnamefont {M.-R.}\ \bibnamefont {Cai}},
  \bibinfo {author} {\bibfnamefont {J.-H.}\ \bibnamefont {Wu}}, \bibinfo
  {author} {\bibfnamefont {Z.}~\bibnamefont {Wang}}, \ and\ \bibinfo {author}
  {\bibfnamefont {Y.}~\bibnamefont {Li}},\ }\bibfield  {title} {\enquote
  {\bibinfo {title} {{Single-photon nonreciprocal excitation transfer with
  non-Markovian retarded effects}},}\ }\href {\doibase
  10.1103/PhysRevA.103.053701} {\bibfield  {journal} {\bibinfo  {journal}
  {Physical Review A}\ }\textbf {\bibinfo {volume} {103}},\ \bibinfo {pages}
  {053701} (\bibinfo {year} {2021})}\BibitemShut {NoStop}%
\bibitem [{\citenamefont {Calaj{\'{o}}}\ \emph {et~al.}(2016)\citenamefont
  {Calaj{\'{o}}}, \citenamefont {Ciccarello}, \citenamefont {Chang},\ and\
  \citenamefont {Rabl}}]{Calajo2016}%
  \BibitemOpen
  \bibfield  {author} {\bibinfo {author} {\bibfnamefont {G.}~\bibnamefont
  {Calaj{\'{o}}}}, \bibinfo {author} {\bibfnamefont {F.}~\bibnamefont
  {Ciccarello}}, \bibinfo {author} {\bibfnamefont {D.}~\bibnamefont {Chang}}, \
  and\ \bibinfo {author} {\bibfnamefont {P.}~\bibnamefont {Rabl}},\ }\bibfield
  {title} {\enquote {\bibinfo {title} {{Atom-field dressed states in slow-light
  waveguide QED}},}\ }\href {\doibase 10.1103/PhysRevA.93.033833} {\bibfield
  {journal} {\bibinfo  {journal} {Physical Review A}\ }\textbf {\bibinfo
  {volume} {93}},\ \bibinfo {pages} {033833} (\bibinfo {year}
  {2016})}\BibitemShut {NoStop}%
\bibitem [{\citenamefont {Bay}\ \emph {et~al.}(1997)\citenamefont {Bay},
  \citenamefont {Lambropoulos},\ and\ \citenamefont {M{\o}lmer}}]{Bay1997}%
  \BibitemOpen
  \bibfield  {author} {\bibinfo {author} {\bibfnamefont {S.}~\bibnamefont
  {Bay}}, \bibinfo {author} {\bibfnamefont {P.}~\bibnamefont {Lambropoulos}}, \
  and\ \bibinfo {author} {\bibfnamefont {K.}~\bibnamefont {M{\o}lmer}},\
  }\bibfield  {title} {\enquote {\bibinfo {title} {{Atom-atom interaction in
  strongly modified reservoirs}},}\ }\href {\doibase 10.1103/PhysRevA.55.1485}
  {\bibfield  {journal} {\bibinfo  {journal} {Physical Review A}\ }\textbf
  {\bibinfo {volume} {55}},\ \bibinfo {pages} {1485} (\bibinfo {year}
  {1997})}\BibitemShut {NoStop}%
\bibitem [{\citenamefont {Lambropoulos}\ \emph {et~al.}(2000)\citenamefont
  {Lambropoulos}, \citenamefont {Nikolopoulos}, \citenamefont {Nielsen},\ and\
  \citenamefont {Bay}}]{Lambropoulos2000}%
  \BibitemOpen
  \bibfield  {author} {\bibinfo {author} {\bibfnamefont {P.}~\bibnamefont
  {Lambropoulos}}, \bibinfo {author} {\bibfnamefont {G.~M.}\ \bibnamefont
  {Nikolopoulos}}, \bibinfo {author} {\bibfnamefont {T.~R.}\ \bibnamefont
  {Nielsen}}, \ and\ \bibinfo {author} {\bibfnamefont {S.}~\bibnamefont
  {Bay}},\ }\bibfield  {title} {\enquote {\bibinfo {title} {{Fundamental
  quantum optics in structured reservoirs}},}\ }\href {\doibase
  10.1088/0034-4885/63/4/201} {\bibfield  {journal} {\bibinfo  {journal}
  {Reports on Progress in Physics}\ }\textbf {\bibinfo {volume} {63}},\
  \bibinfo {pages} {455} (\bibinfo {year} {2000})}\BibitemShut {NoStop}%
\bibitem [{\citenamefont {Shahmoon}\ and\ \citenamefont
  {Kurizki}(2013)}]{Shahmoon2013}%
  \BibitemOpen
  \bibfield  {author} {\bibinfo {author} {\bibfnamefont {E.}~\bibnamefont
  {Shahmoon}}\ and\ \bibinfo {author} {\bibfnamefont {G.}~\bibnamefont
  {Kurizki}},\ }\bibfield  {title} {\enquote {\bibinfo {title} {{Nonradiative
  interaction and entanglement between distant atoms}},}\ }\href {\doibase
  10.1103/PhysRevA.87.033831} {\bibfield  {journal} {\bibinfo  {journal}
  {Physical Review A}\ }\textbf {\bibinfo {volume} {87}},\ \bibinfo {pages}
  {033831} (\bibinfo {year} {2013})}\BibitemShut {NoStop}%
\bibitem [{\citenamefont {Korenblit}\ \emph {et~al.}(2012)\citenamefont
  {Korenblit}, \citenamefont {Kafri}, \citenamefont {Campbell}, \citenamefont
  {Islam}, \citenamefont {Edwards}, \citenamefont {Gong}, \citenamefont {Lin},
  \citenamefont {Duan}, \citenamefont {Kim}, \citenamefont {Kim},\ and\
  \citenamefont {Monroe}}]{Korenblit2012}%
  \BibitemOpen
  \bibfield  {author} {\bibinfo {author} {\bibfnamefont {S.}~\bibnamefont
  {Korenblit}}, \bibinfo {author} {\bibfnamefont {D.}~\bibnamefont {Kafri}},
  \bibinfo {author} {\bibfnamefont {W.~C.}\ \bibnamefont {Campbell}}, \bibinfo
  {author} {\bibfnamefont {R.}~\bibnamefont {Islam}}, \bibinfo {author}
  {\bibfnamefont {E.~E.}\ \bibnamefont {Edwards}}, \bibinfo {author}
  {\bibfnamefont {Z.~X.}\ \bibnamefont {Gong}}, \bibinfo {author}
  {\bibfnamefont {G.~D.}\ \bibnamefont {Lin}}, \bibinfo {author} {\bibfnamefont
  {L.~M.}\ \bibnamefont {Duan}}, \bibinfo {author} {\bibfnamefont
  {J.}~\bibnamefont {Kim}}, \bibinfo {author} {\bibfnamefont {K.}~\bibnamefont
  {Kim}}, \ and\ \bibinfo {author} {\bibfnamefont {C.}~\bibnamefont {Monroe}},\
  }\bibfield  {title} {\enquote {\bibinfo {title} {{Quantum simulation of spin
  models on an arbitrary lattice with trapped ions}},}\ }\href {\doibase
  10.1088/1367-2630/14/9/095024} {\bibfield  {journal} {\bibinfo  {journal}
  {New Journal of Physics}\ }\textbf {\bibinfo {volume} {14}},\ \bibinfo
  {pages} {095024} (\bibinfo {year} {2012})}\BibitemShut {NoStop}%
\bibitem [{\citenamefont {Qiao}\ \emph {et~al.}(2024)\citenamefont {Qiao},
  \citenamefont {Cai}, \citenamefont {Wang}, \citenamefont {Du}, \citenamefont
  {Jin}, \citenamefont {Chen}, \citenamefont {Wang}, \citenamefont {Luan},
  \citenamefont {Gao}, \citenamefont {Sun}, \citenamefont {Tian}, \citenamefont
  {Zhang},\ and\ \citenamefont {Kim}}]{Qiao2024}%
  \BibitemOpen
  \bibfield  {author} {\bibinfo {author} {\bibfnamefont {M.}~\bibnamefont
  {Qiao}}, \bibinfo {author} {\bibfnamefont {Z.}~\bibnamefont {Cai}}, \bibinfo
  {author} {\bibfnamefont {Y.}~\bibnamefont {Wang}}, \bibinfo {author}
  {\bibfnamefont {B.}~\bibnamefont {Du}}, \bibinfo {author} {\bibfnamefont
  {N.}~\bibnamefont {Jin}}, \bibinfo {author} {\bibfnamefont {W.}~\bibnamefont
  {Chen}}, \bibinfo {author} {\bibfnamefont {P.}~\bibnamefont {Wang}}, \bibinfo
  {author} {\bibfnamefont {C.}~\bibnamefont {Luan}}, \bibinfo {author}
  {\bibfnamefont {E.}~\bibnamefont {Gao}}, \bibinfo {author} {\bibfnamefont
  {X.}~\bibnamefont {Sun}}, \bibinfo {author} {\bibfnamefont {H.}~\bibnamefont
  {Tian}}, \bibinfo {author} {\bibfnamefont {J.}~\bibnamefont {Zhang}}, \ and\
  \bibinfo {author} {\bibfnamefont {K.}~\bibnamefont {Kim}},\ }\bibfield
  {title} {\enquote {\bibinfo {title} {Tunable quantum simulation of spin
  models with a two-dimensional ion crystal},}\ }\href {\doibase
  10.1038/s41567-023-02378-9} {\bibfield  {journal} {\bibinfo  {journal}
  {Nature Physics}\ }\textbf {\bibinfo {volume} {20}},\ \bibinfo {pages} {623}
  (\bibinfo {year} {2024})}\BibitemShut {NoStop}%
\bibitem [{\citenamefont {Fauseweh}(2024)}]{Fauseweh2024}%
  \BibitemOpen
  \bibfield  {author} {\bibinfo {author} {\bibfnamefont {B.}~\bibnamefont
  {Fauseweh}},\ }\bibfield  {title} {\enquote {\bibinfo {title} {{Quantum
  many-body simulations on digital quantum computers: State-of-the-art and
  future challenges}},}\ }\href {\doibase 10.1038/s41467-024-46402-9}
  {\bibfield  {journal} {\bibinfo  {journal} {Nature Communications}\ }\textbf
  {\bibinfo {volume} {15}},\ \bibinfo {pages} {2123} (\bibinfo {year}
  {2024})}\BibitemShut {NoStop}%
\bibitem [{\citenamefont {Chen}\ and\ \citenamefont {Kockum}(2024)}]{Chen2024}%
  \BibitemOpen
  \bibfield  {author} {\bibinfo {author} {\bibfnamefont {G.}~\bibnamefont
  {Chen}}\ and\ \bibinfo {author} {\bibfnamefont {A.~F.}\ \bibnamefont
  {Kockum}},\ }\bibfield  {title} {\enquote {\bibinfo {title} {{Simulating open
  quantum systems with giant atoms}},}\ }\href@noop {} {\bibfield  {journal}
  {\bibinfo  {journal} {in preparation}\ } (\bibinfo {year}
  {2024})}\BibitemShut {NoStop}%
\bibitem [{\citenamefont {Leonforte}\ \emph {et~al.}(2024)\citenamefont
  {Leonforte}, \citenamefont {Sun}, \citenamefont {Valenti}, \citenamefont
  {Spagnolo}, \citenamefont {Illuminati}, \citenamefont {Carollo},\ and\
  \citenamefont {Ciccarello}}]{Leonforte2024}%
  \BibitemOpen
  \bibfield  {author} {\bibinfo {author} {\bibfnamefont {L.}~\bibnamefont
  {Leonforte}}, \bibinfo {author} {\bibfnamefont {X.}~\bibnamefont {Sun}},
  \bibinfo {author} {\bibfnamefont {D.}~\bibnamefont {Valenti}}, \bibinfo
  {author} {\bibfnamefont {B.}~\bibnamefont {Spagnolo}}, \bibinfo {author}
  {\bibfnamefont {F.}~\bibnamefont {Illuminati}}, \bibinfo {author}
  {\bibfnamefont {A.}~\bibnamefont {Carollo}}, \ and\ \bibinfo {author}
  {\bibfnamefont {F.}~\bibnamefont {Ciccarello}},\ }\href@noop {} {\enquote
  {\bibinfo {title} {{Quantum optics with giant atoms in a structured photonic
  bath}},}\ } (\bibinfo {year} {2024}),\ \Eprint
  {http://arxiv.org/abs/2402.10275} {arXiv:2402.10275} \BibitemShut {NoStop}%
\bibitem [{\citenamefont {Cohen-Tannoudji}\ \emph {et~al.}(1998)\citenamefont
  {Cohen-Tannoudji}, \citenamefont {Dupont-Roc},\ and\ \citenamefont
  {Grynberg}}]{Cohen-Tannoudji}%
  \BibitemOpen
  \bibfield  {author} {\bibinfo {author} {\bibfnamefont {C.}~\bibnamefont
  {Cohen-Tannoudji}}, \bibinfo {author} {\bibfnamefont {J.}~\bibnamefont
  {Dupont-Roc}}, \ and\ \bibinfo {author} {\bibfnamefont {G.}~\bibnamefont
  {Grynberg}},\ }\enquote {\bibinfo {title} {{Nonperturbative Calculation of
  Transition Amplitudes}},}\ in\ \href {\doibase
  https://doi.org/10.1002/9783527617197.ch3} {\emph {\bibinfo {booktitle}
  {Atom—Photon Interactions}}}\ (\bibinfo  {publisher} {John Wiley \& Sons,
  Ltd},\ \bibinfo {year} {1998})\ Chap.~\bibinfo {chapter} {3}, pp.\ \bibinfo
  {pages} {165--255}\BibitemShut {NoStop}%
\bibitem [{\citenamefont {Morita}(1971)}]{Morita1971}%
  \BibitemOpen
  \bibfield  {author} {\bibinfo {author} {\bibfnamefont {T.}~\bibnamefont
  {Morita}},\ }\bibfield  {title} {\enquote {\bibinfo {title} {{Useful
  Procedure for Computing the Lattice Green's Function‐Square, Tetragonal,
  and bcc Lattices}},}\ }\href {\doibase 10.1063/1.1665800} {\bibfield
  {journal} {\bibinfo  {journal} {Journal of Mathematical Physics}\ }\textbf
  {\bibinfo {volume} {12}},\ \bibinfo {pages} {1744} (\bibinfo {year}
  {1971})}\BibitemShut {NoStop}%
\end{thebibliography}%

\end{document}